\documentclass{cernrep}

\usepackage{varwidth}
\usepackage{xcolor}

\begin{document}

\title{Cyclotrons: Magnetic Design and Beam Dynamics}
\author{W. Kleeven and S. Zaremba}
\institute{Ion Beam Applications, Louvain-La-Neuve, Belgium}
\maketitle

\begin{abstract}

Classical, isochronous, and synchro-cyclotrons are introduced. Transverse and longitudinal beam dynamics in these accelerators
are covered. The problem of vertical focusing and iscochronism in compact isochronous cyclotrons is treated in some detail. Different methods
for isochronization of the cyclotron magnetic field are discussed. The limits of the classical cyclotron are explained.
Typical features of the synchro-cyclotron, such as the beam cap\-ture problem, stable phase motion, and the extraction
problem are discussed.
The main design goals for beam injection are explained and special prob\-lems related to a central region with an internal ion
source are considered. The principle of a Penning ion gauge source is addressed. The issue of ver\-tical focusing in the cyclotron centre
is briefly discussed. Several examples of numerical simulations are given. Different methods of (axial) injection are
briefly outlined.
Different solutions for beam extraction are described. These include the internal target, extraction by stripping, resonant
extraction using a deflector, regenerative extraction, and self-extraction.
Different methods of creating a turn separation are explained.
Different types of extraction device, such as harmonic coils, deflectors, and gradient corrector channels,
are outlined. Some general considerations for cyclotron magnetic design are given and the use of modern magnetic modelling tools is discussed, with a few illustrative examples.
An approach is chosen where the accent is less on completeness and rigorousness
(because this has already been done) and more on explaining and illustrating the main principles that are used in medical cyclotrons.
Sometimes a more industrial viewpoint is taken. The use of complicated formulae is limited. 
\end{abstract}

\section{Different types of cyclotron}

\subsection{The basic equation of the cyclotron---the classical cyclotron}

Consider a particle with charge $q$ and mass $m$ that moves with constant velocity $v$ in a uniform
magnetic field $B$. Such a particle moves in a circle with radius $r$; the centripetal force is provided by the Lorentz force
acting on the particle:

\begin{equation}
\frac{mv^2}{r} = qvB\ .
\end{equation}

The angular velocity is given by

\begin{equation}\label{eq:freq}
 \omega =\frac{v}{r} =\frac{qB}{m}\ .
\end{equation}

\begin{figure}
\begin{center}
\includegraphics[width=7cm]{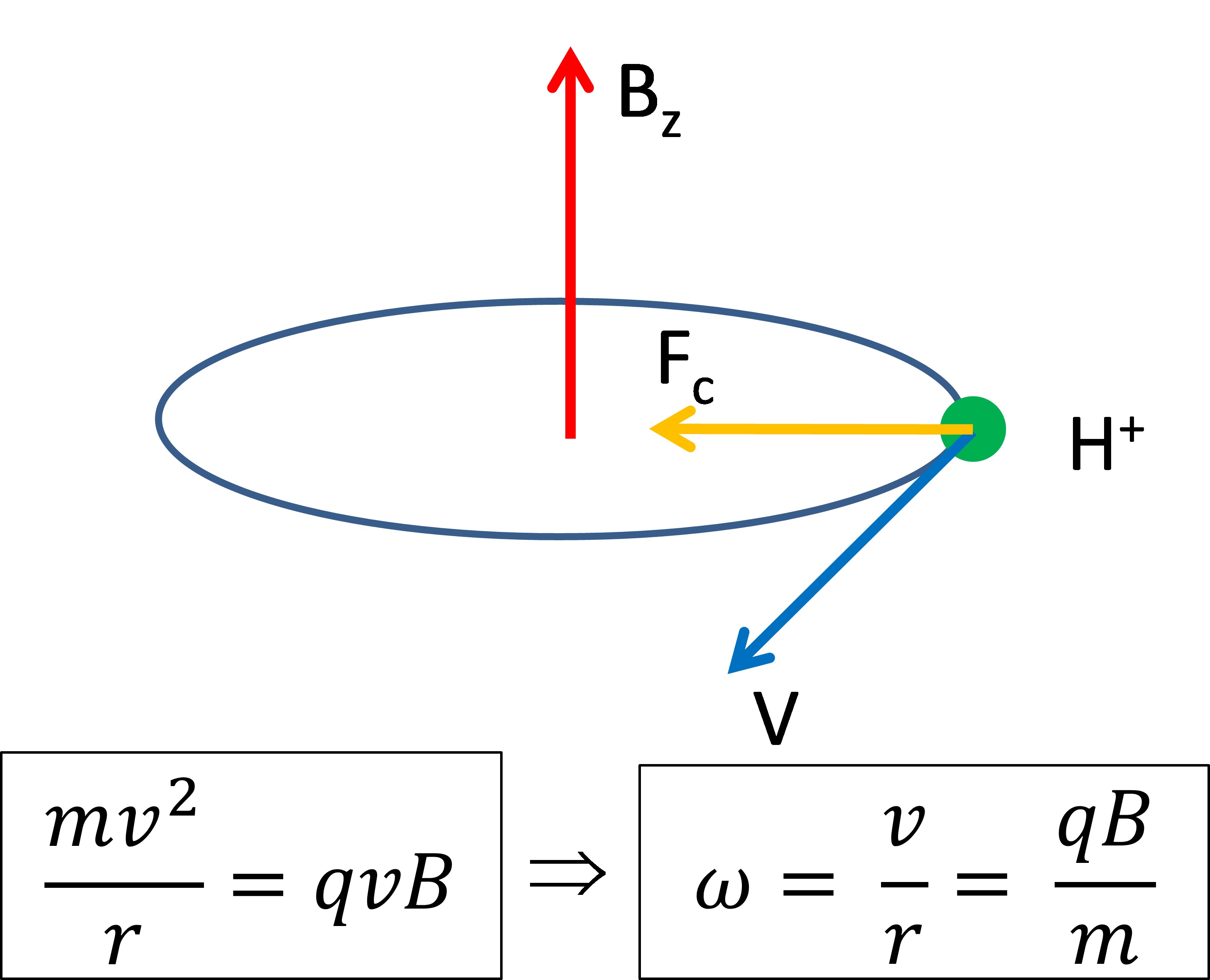}
\caption{At first order, a particle in a cyclotron rotates at constant frequency, independent of radius or energy}
\label{fig:clascycl1}
\end{center}
\end{figure}

This is illustrated in \Fref{fig:clascycl1}. Thus, the angular velocity is constant: it is independent of radius, velocity,
energy (in the non-relativistic limit), or time.

There are a few very important consequences of this feature:

\begin{enumerate}
\item particles can be accelerated with an RF system that operates at constant frequency;
\item the orbits start their path in the centre (injection) and spiral outward to the pole radius (extraction);
\item the magnetic field is constant in time;
\item the RF structure and the magnetic structure are completely integrated: the same RF structure will accelerate the beam many times
      (allowing for a compact, cost-effective accelerator);
\item the operation of the accelerator and thus the beam is a fully continuous wave.
\end{enumerate}

\begin{figure}
\begin{center}
\includegraphics[width=12cm]{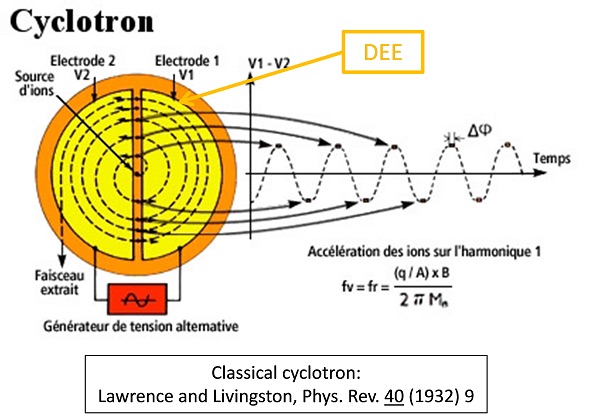}
\caption{Synchronism in a cyclotron between the particle rotation and the RF accelerating wave (courtesy of Frederic Chautard.)}
\label{fig:clascycl2}
\end{center}
\end{figure}

The cyclotron was invented in 1932 by Lawrence and Livingston\cite{Lawrence1932}.  This type
(quasi-uniform magnetic field) is called  the classical cyclotron. The principle of the cyclotron is illustrated in
\Fref{fig:clascycl2}. The frequency of the RF-structure and the magnetic field are related as

\begin{equation}
f_{\mathrm{RF}} \approx 15.2\:h\:\frac{Z}{A}\: B\ .
\end{equation}

\noindent Here $Z$ and $A$ are the charge number and mass number of the particle, $h$ is the harmonic mode of the acceleration
$h\:=\:f_{\mathrm{RF}}\:/\:f_{\mathrm{ion}}\:$;
$f_{\mathrm{RF}}$ is expressed in megahertz and $B$ in tesla.

There is a fundamental problem with the classical cyclotron, which can be seen as follows.

\begin{enumerate}
\item In a uniform magnetic field there is no vertical focusing (the motion is meta-stable).
\item During acceleration, the relativistic mass increases; therefore, the angular velocity is actually not constant
      but gradually decreases (see Eqs.~(\ref{eq:freq}) and (\ref{eq:mass})). A loss of synchronism occurs between the RF and the beam (loss of isochronism).
\item Simply increasing the magnetic field with radius is not possible, because the motion then becomes vertically unstable.
\end{enumerate}

The particle angular velocity taking into account the relativistic mass increase is given by

\begin{equation} \label{eq:mass}
\omega = \frac{qB}{m_0}\:\sqrt{1-\left(\frac{v}{c}\right)^2}\ .
\end{equation}

\noindent Here $m_0$ is the particle rest-mass and $c$ is the speed of light.

Let us make a small sidestep and see how much energy can be achieved with the classical cyclo\-tron. Assume a magnetic field
with a small negative gradient, such that some vertical focusing is pro\-vided.
The magnetic field as a function of radius is given by

\begin{equation}
B(r) = B_0 \left(\frac{r}{r_0}\right)^{-n}\ .
\end{equation}

\begin{figure}
\begin{center}
\includegraphics[width=7cm]{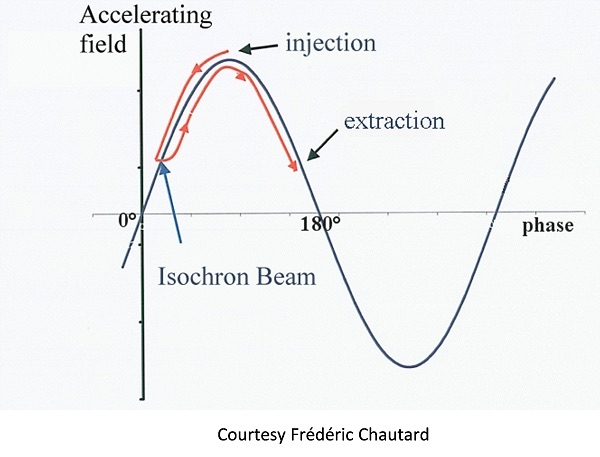}
\caption{Owing to the relativistic mass increase and the small negative magnetic field gradient, the particle in a classical cyclotron
runs steadily out of phase with respect to the RF. Nevertheless, acceleration can be obtained during a limited number of turns (courtesy of Frederic Chautard.)}
\label{fig:clascycl3}
\end{center}
\end{figure}

\noindent Here, $r_0$ is some reference radius and $B_0$ is the field at that radius.
The field index $n$ is defined as
\begin{equation*}
 n \: = \: -\frac{\mathrm{d}B}{\mathrm{d}r} \cdot \frac{r}{B} \ .
\end{equation*}
Vertical tuning is related to the field index as $\nu_z=\sqrt{n}$. During the acceleration, the particles
gradually run out of phase with respect to the RF. However, the RF frequency can be tuned such that, in the cyclotron centre,
the magnetic field is too high. Here the particles are extracted from the ion source at an RF phase of
approximately 90$^\circ$, but then the phase will decrease because the RF frequency is too low. Since the
magnetic field decreases with radius, there will be, after some number of turns, a moment where the revolution frequency and
RF frequency are exactly the same. Beyond that point, the RF phase will start to increase, because the RF frequency is now too high.
This RF phase motion is illustrated in \Fref{fig:clascycl3}.

The longitudinal motion can be studied using a simple Excel  model.
The energy and phase of the accelerated particle are
found by integrating the following equations:

\begin{eqnarray}
\Delta E_n &=& qNV_{\mathrm{dee}}\cos\Phi_n\ , \\
\Delta \Phi_n &=& 2\pi h \frac{B-B_{\mathrm{iso}}}{B_{\mathrm{iso}}}\ .
\end{eqnarray}

\noindent Here, $\Delta E_n$ is the energy gain at turn $n$, $V_{\mathrm{dee}}$ is the maximum dee voltage, $N$ is the number of accelerating gaps,
$\Phi_n$ is the RF phase at turn $n$,
$\Delta\Phi_n$ is the RF phase slip in the turn $n$, $h$ is the harmonic mode, and $B_{\mathrm{iso}}$ is the isochronous magnetic field
corresponding to the given RF frequency and taking into account the relativistic mass increase.

\begin{figure}
\begin{center}
\includegraphics[width=10cm]{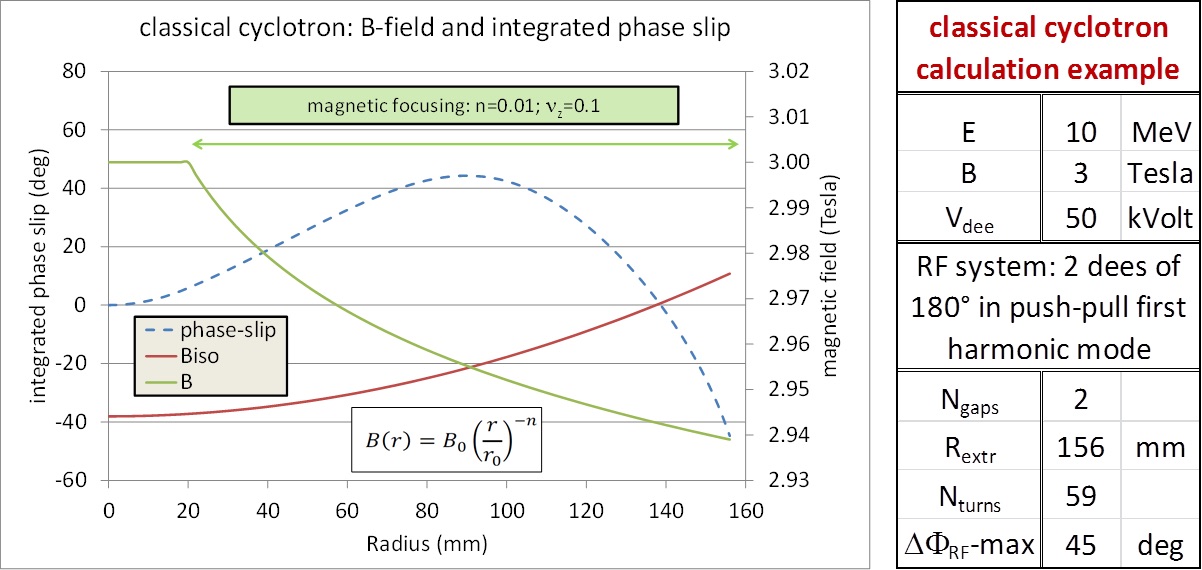}
\caption{A simple Excel model can show how much energy can be reached in a weak-focusing rotational symmetric cyclotron, depending on the
dee voltage, the field index, and the magnetic field value.}
\label{fig:clascycl4}
\end{center}
\end{figure}

An example of such a calculation is shown in \Fref{fig:clascycl4}. This cyclotron is a candidate
for a small super\-conducting machine for isotope production for positron emission tomography. The main parameters are shown in the same figure. It
can be seen that quite a high dee voltage is needed to limit the number of turns and thus the RF phase slip. In this case, an
RF system with two 180$^\circ$ dees in push--pull mode was assumed. In such a system, the two opposite dees oscillate 180$^\circ$
out of phase, such that the total maximum energy gain per turn is four times the dee voltage. It is seen that protons of $10\UMeV$
can be obtained in a field of $3\UT$, at an extraction radius of $156\Umm$, with a dee voltage of $50\UkV$, and about 60 turns in the
cyclotron. At such a low energy, it is possible to accelerate \EH$^-$ without substantial losses by magnetic stripping. Such a cyclotron
is currently under construction in the CIEMAT Institute in Madrid, Spain\cite{Oliver2013}.

\subsection{Another solution: the synchro-cyclotron}\label{another}

A solution for the energy and vertical focusing limitations of the classical cyclotron has been intro\-duced independently by
Veksler\cite{Veksler1945} and McMillan\cite{McMillan1945}. (Note that the synchrotron
was also invented independently by Veksler and McMillan and is described in the same papers.) This solution, the synchro-cyclotron, differs in the following ways from the classical cyclotron:

\begin{enumerate}
\item the magnetic field gradually decreases with radius in order to obtain weak vertical focusing:
\begin{equation}
 n=-\frac{r}{B}\frac{\mathrm{d}B}{\mathrm{d}r} \Rightarrow\ \nu_z = \sqrt{n}\ ;
\end{equation}
\item the RF frequency gradually decreases with time, to compensate for the decrease in magnetic field
and the increase in particle mass (see \Eref{eq:freq}).
\end{enumerate}

\noindent This type of cyclotron brings about several important consequences.

\begin{enumerate}
\item Much higher energies can be obtained, in the range $100\UMeV$ to $1\UGeV$.
\item The RF is pulsed but the magnetic field is constant in time (which is not the case in a synchrotron).
\item The beam is no longer continuous wave but is modulated (pulsed) in time.
\item The average beam current is much lower than for a continuous-wave machine (OK for proton therapy).
\item There is a longitudinal beam dynamics similar to that of the synchrotron.
\item The beam can only be captured in the cyclotron centre during a short time-window.
\item The timing between the RF frequency, RF voltage, and ion source needs to be well defined and controlled.
\item A more complicated (but not necessarily more expensive) RF system is needed to obtain the re\-quired frequency modulation.
\item The RF frequency cannot be varied very quickly (rotating capacitor) and therefore the acceleration must be slow.  This implies the
following:
\begin{enumerate}
\item low energy gain per turn;
\item many turns up to extraction;
\item low RF voltage and low RF power needed.
\end{enumerate}
\item There is only a very small turn separation at extraction. Therefore a special extraction method (called a
regenerative extraction) is needed to get the beam out of the machine.
\end{enumerate}

\begin{figure}
\begin{center}
\includegraphics[width=\textwidth]{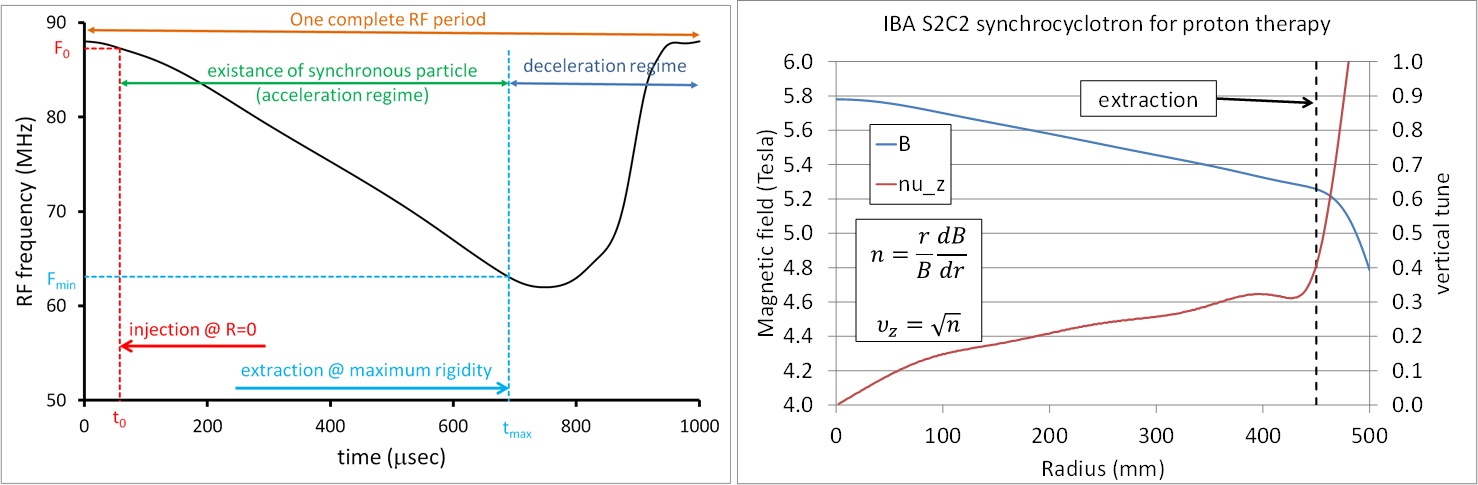}
\caption{Some properties of the IBA superconducting synchro-cyclotron (S2C2). Left: modulation of the RF frequency as a function of time within one pulse. Right: average magnetic field
and vertical tuning as a function of radius. }
\label{fig:s2c21}
\end{center}
\end{figure}

Recently, IBA has developed a $230\UMeV$ superconducting synchro-cyclotron (S2C2) for proton therapy. The advantage of such a solution,
as compared with compact superconducting isochronous cyclotrons, is that the average magnetic field can be increased to substantially
higher values, because there is no concern about lack of vertical focusing. \Figure[b]~\ref{fig:s2c21} shows some properties
of this cyclotron. The graph on the right shows the average magnetic field and the vertical focusing frequency (the passive extraction
system was not installed in this case). The magnetic field in the centre is about $5.8\UT$ and the extraction radius is about $450\Umm$;
the pole radius is $500\Umm$. Also shown is the vertical focusing frequency; in this weak-focusing machine, the vertical focusing is
produced solely by the negative field gradient. The graph on the left illustrates the time structure of the RF. The pulse length is
$1\Ums$ and the corresponding pulse rate is $1\UkHz$. The RF frequency varies from about $88\UMHz$ (when the beam is captured
in the cyclotron centre) to about $63\UMHz$ (when the beam begins to be extracted at $r=450\Umm$). The total acceleration time is about
$600\Uus$, and the number of turns in this cyclotron is greater than $45\,000$.

\begin{figure}
\begin{center}
\includegraphics[width=12cm]{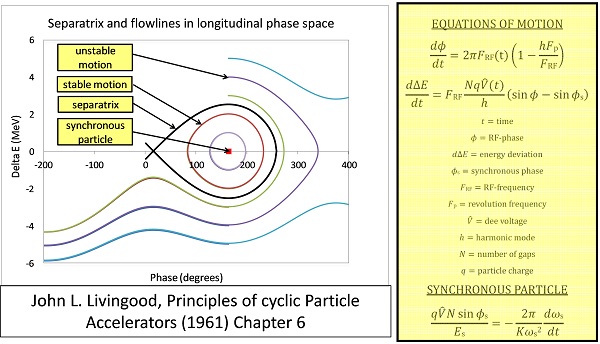}
\caption{Longitudinal beam dynamics in a synchro-cyclotron. Left: flow lines and separatrix
in longitudinal phase space. Right: equations of motion that govern this phase space.}
\label{fig:longps}
\end{center}
\end{figure}

\Figure[b]~\ref{fig:longps} illustrates the longitudinal beam dynamics in a synchro-cyclotron such as
the S2C2. The graph on the left shows the longitudinal
phase space. For the synchronous particle, the angular velocity is (by definition) always the same as the RF frequency at all
radii in the machine. A non-synchronous particle executes an oscillation around this synchronous particle. The horizontal
axis is the RF phase and the vertical axis is the energy difference of the particle with respect to the synchronous particle.
For small excursions, particles execute elliptical (symmetric) oscillations around the synchronous point. For larger excursions,
owing to the non-linear character of the dynamics, the flow lines start to deform.
The separatrix separates the stable zone from the unstable zone. Inside the separatrix, there
remains, on average, a resonance between the RF frequency and the particle revolution frequency, and the particle will be
accelerated. Outside the separatrix, there is no longer a resonant acceleration of the particle and it will stay close to
a fixed radius in the cyclotron. The right panel of \Fref{fig:longps} shows the equations of motion that govern the
longitudinal phase space. More explanations of this can be found elsewhere, \eg in the textbook by Livingood\cite{Livingood1961}.

\subsection{The isochronous cyclotron}

In the isochronous cyclotron, an additional resource of vertical focusing is introduced by allowing the magnetic field to vary
with azimuth along a circle. This additional focusing is so strong that it dominates the vertical defocusing arising
from a radially increasing field. The radial increase can be made such that the revolution frequency of
the particles remains constant in the machine, even for relativistic energies (for which the mass increase is significant).
This new resource of vertical focusing was invented by Thomas\cite{Thomas1938}. In the next section, this type of cyclotron is
discussed in more detail.

\section{More about compact azimuthally varying field cyclotrons}

\subsection{Vertical focusing in cyclotrons}

To better understand the vertical focusing in a cyclotron, consider the vertical component of the Lorentz force, $F_z$:

\begin{equation}
F_z =q(\vec{v}\times\vec{B})_z =-q\left(v_\theta B_r-v_r B_\theta\right)\ .
\end{equation}

The first term, $v_\theta B_r$, is obtained in a radially decreasing, rotationally symmetric magnetic field, such as for
the classical cyclotron or the synchro-cyclotron. If only this term is present, this would correspond to the case of weak focusing.
The second term, $v_r B_\theta$, requires an azimuthal modulation of the magnetic field. If such a modulation exists,  it will, by itself,
also generate a radial component of the velocity.

\begin{figure}
\begin{center}
\includegraphics[width=12cm]{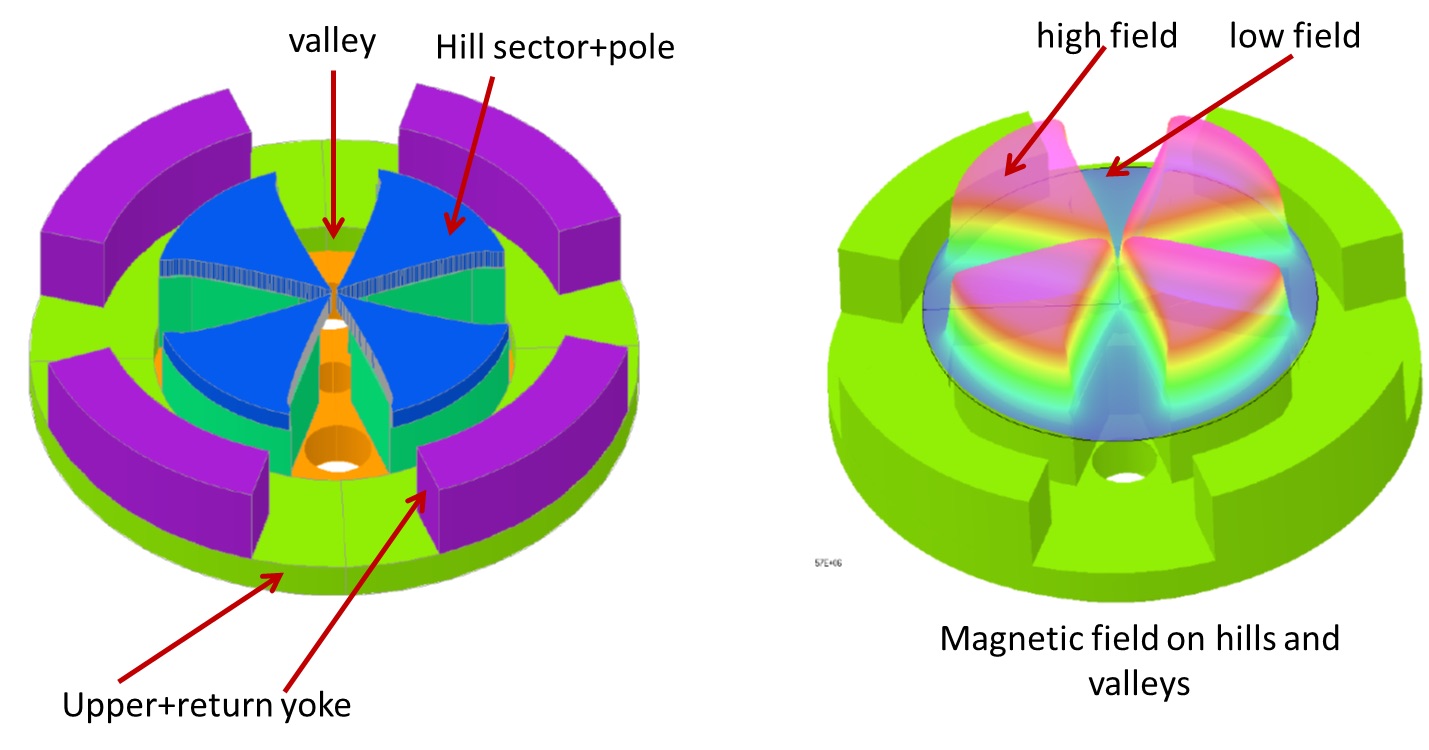}
\caption{Magnet of a compact azimuthally varying field cyclotron. Left: the hill sectors and poles, the valleys, and the
different parts of the yoke. Right: histogram of the magnetic field.}
\label{fig:cycloavf}
\end{center}
\end{figure}

The azimuthal field modulation can be produced by introducing high-field sectors (hills), separated by low-field
regions (valleys). This is illustrated in \Fref{fig:cycloavf}, which shows the magnet of a compact four-fold symmetrical azimuthally varying field (AVF) cyclotron with four hills and four valleys. The hill sectors are mounted on upper and lower
plates of the yoke and surrounded by a return yoke placed in between the upper and lower plates. The plates contain
circular holes in the valleys, which are used for vacuum pumping or installation of RF cavities. The right
panel shows a histogram of the magnetic field in the median plane superimposed on the geometry. This field map was
computed using the 3D finite-element software package Opera-3d from Vector Fields Cobham Technical Services\cite{Cobham2015}.

\begin{figure}
\begin{center}
\includegraphics[width=12cm]{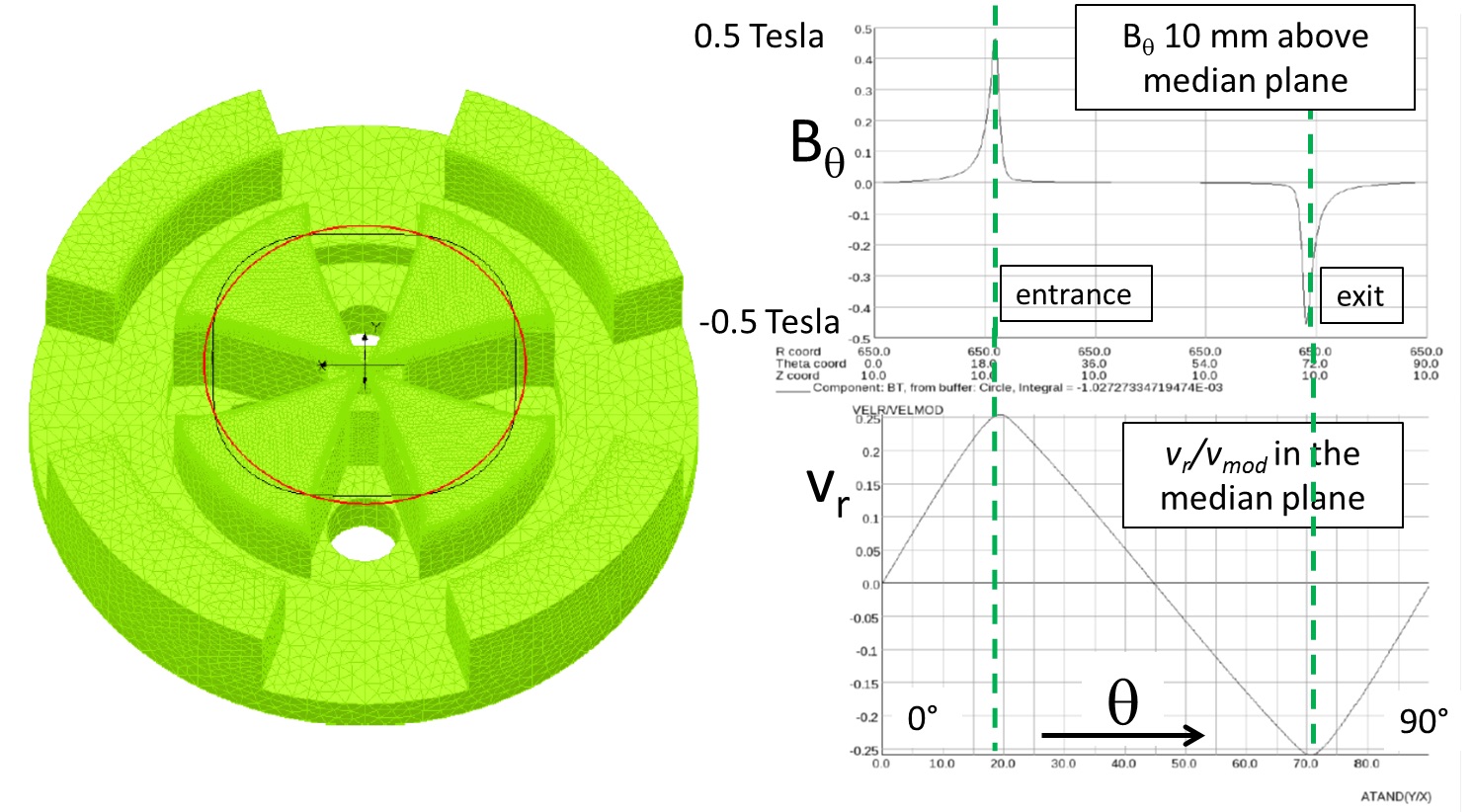}
\caption{Vertical focusing in an azimuthally varying field cyclotron. Left: scalloping of the orbit with respect
to the geometrical circle. Right: the radial velocity component and the azimuthal field component  $10\Umm$ above
the median plane. The product of these gives the stabilizing vertical Lorentz forces directed towards the median plane.}
\label{fig:vertfoc1}
\end{center}
\end{figure}

\Figure[b]~\ref{fig:vertfoc1} illustrates the vertical focusing in such an AVF cyclotron. The drawing on the left shows
how a computed closed orbit oscillates around a reference circle to produce a scalloped orbit. The
upper graph on the right shows the azimuthal component of the field in a circle $10\Umm$ from the median plane. It can be
seen that $B_\theta$ is strongly peaked at the entrance and exit of the sector. The lower graph on the right shows the (normalized)
radial component of the velocity $v_r$. The maximum of this component is also at the entrance and exit of the sector.
The product of both terms is positive at both the sector entrance and exit, indicating that the vertical focusing is
concentrated at these azimuthal locations and is always positive (not alternating).

\begin{figure}
\begin{center}
\includegraphics[width=7cm]{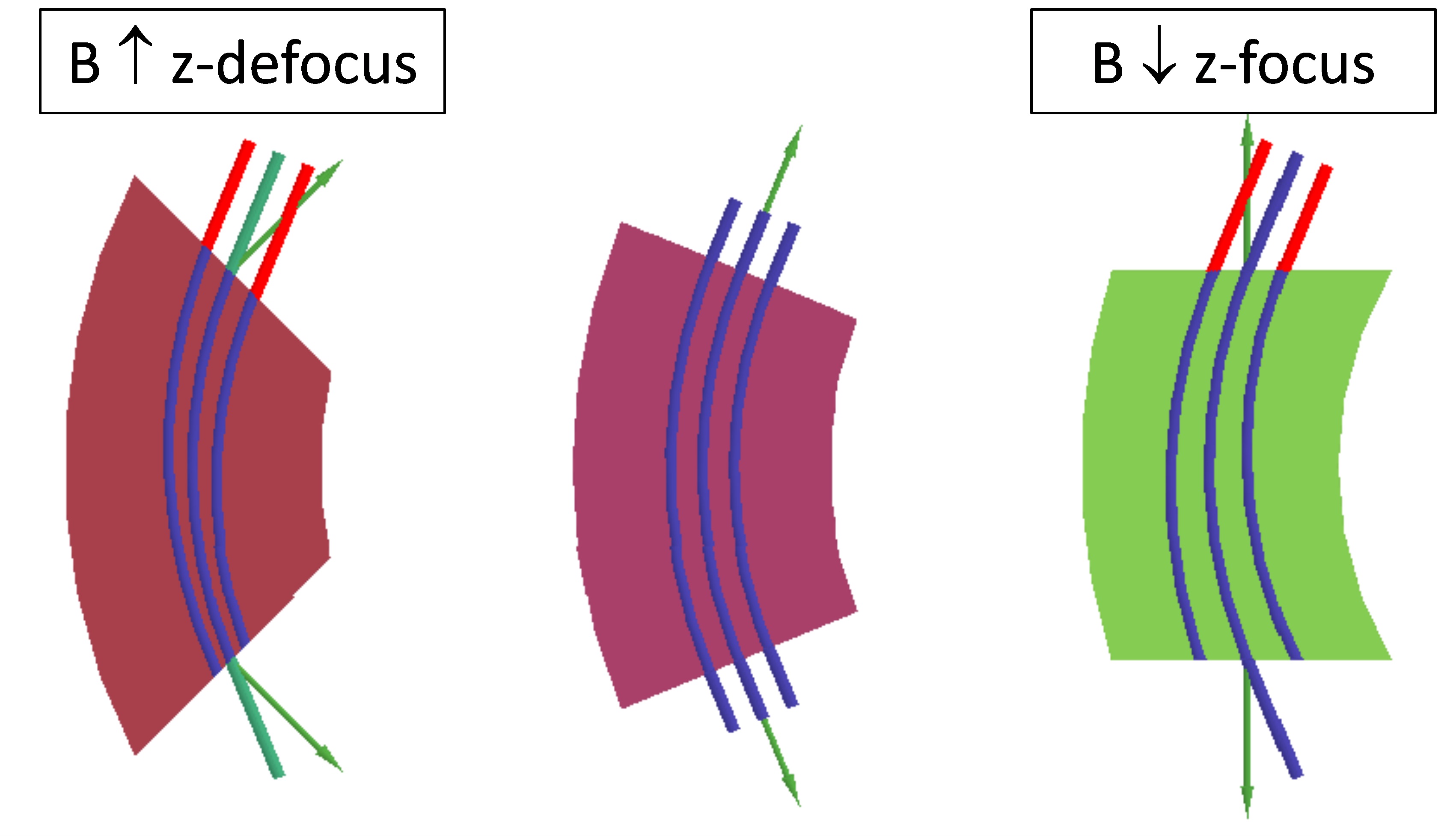}
\caption{Vertical focusing in a cyclotron relates closely to vertical focusing at the entrance and exit of a dipole magnet.}
\label{fig:vertfoc2}
\end{center}
\end{figure}

The vertical focusing in a cyclotron with straight sectors is of the same nature as the edge focusing that occurs at
the entrance and exit of the dipole bending magnets. This is shown in \Fref{fig:vertfoc2}. To find the
sign of the vertical focusing at an edge, one should draw the normal vector on the orbit, pointing away from the orbit centre.
If the magnetic field along this direction is decreasing, then the edge will be vertically focusing. Otherwise, it will be
vertically defocusing (see, for example, the TRANSPORT manual\cite{Trans1980}).

It is interesting to note that Thomas\cite{Thomas1938} invented sector focusing (Thomas focusing) in 1938,
several years before the invention of the synchro-cyclotron and the synchrotron. However, his solution could not be applied immediately, owing to the
increased complexity of the magnetic structure. This is why synchro-cyclotrons have been used at the birth of proton therapy.

\begin{figure}
\begin{center}
\includegraphics[width=9cm]{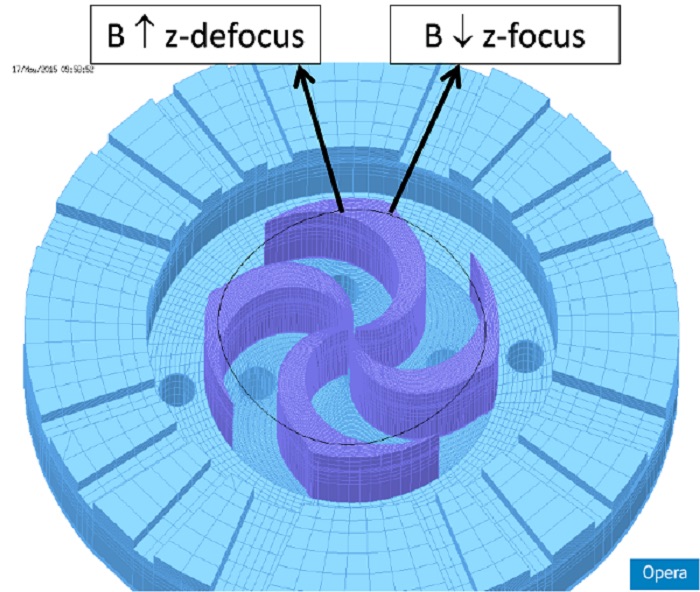}
\caption{Vertical focusing in an azimuthally varying field cyclotron can be strongly increased by spiralling the hill sectors.}
\label{fig:vertfoc3}
\end{center}
\end{figure}

The vertical focusing created in an AVF cyclotron can be strongly increased if the shape of the sectors is changed from
straight to spiral.  \Figure[b]~\ref{fig:vertfoc3} shows an Opera-3d preprocessor model of a compact cyclotron with spiralled sectors.
By drawing the normal vector on the closed orbit at the sector edges, it can be seen that the angle between the orbit and the edge
can be made rather large (choosing a large spiral); thus, generating a strong vertical (de-)focusing effect. However,
it can also be seen that the direction of the vertical force changes sign between entrance and exit of the sector. Thus, the spiralling
of the sectors creates a sequence of alternating focusing, which can become relatively strong. This strong (alternating) focusing was
invented by Christofilos\cite{Christ1950} and Courant \textit{et al.}\cite{Courant1952}.

We note that in many compact cyclotrons, the vertical focusing is not only concentrated at the sector edges, but can be more
distributed along the closed orbit:

\begin{enumerate}
\item edge focusing occurs at the entrance and exit of the hill sectors;
\item for spiral sectors, this focusing starts to alternate and can be made stronger;
\item in the middle of a hill sector, there can be a positive field gradient (\eg by application of an elliptical pole gap),
creating vertical defocusing;
\item in the middle of the valley, there is often a negative field gradient, creating vertical focusing.
\end{enumerate}

The strength of the azimuthal field variation in a cyclotron is expressed in the flutter function $F$. This function is defined
as

\begin{equation}
F(r) = \frac{\overline{B^2}-(\, \overline{B}\, )^2}{(\, \overline{B}\, )^2}\ .
\end{equation}

\noindent Here, $\overline{B}$ is the average of the median magnetic field over the azimuthal range from 0$^\circ$ to 360$^\circ$
and $\overline{B^2}$ is the average over the square of this field.
The median plane magnetic field can be represented in a Fourier series as

\begin{equation}
B(r,\theta) = \overline{B}(r)\left[1+\sum_{n=1}^\infty A_n(r)\cos n\theta + B_n(r)\sin n\theta\right]\ ,
\end{equation}

\noindent where $A_n$ and $B_n$ are the normalized Fourier harmonics of the field.
With this representation of the field, the flutter can be written as

\begin{equation}
F=\sum_{n=1}^\infty \frac{A_n^2+B_n^2}{2}\ .
\end{equation}

\begin{figure}
\begin{center}
\includegraphics[width=12cm]{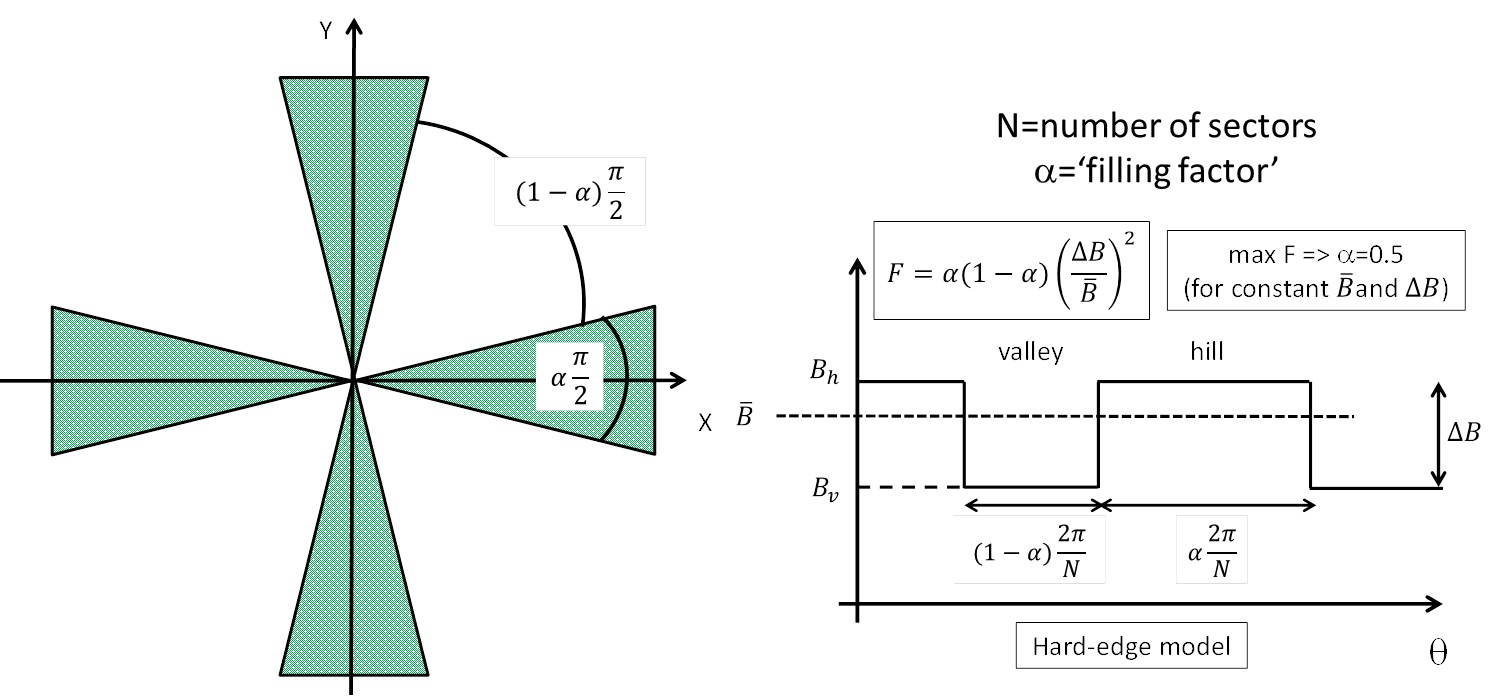}
\caption{Simple model to estimate flutter in a hard-edge approximation}
\label{fig:flut}
\end{center}
\end{figure}

Often, in a compact cyclotron, a hard-edge model of the magnetic field can be used to estimate the flutter.
This is illustrated in \Fref{fig:flut} for a cyclotron with four-fold symmetry. The drawing on the left defines the hill
angle $\alpha\pi/2$ and the valley angle $(1-\alpha)\pi/2$. The parameter $\alpha$ is a kind of filling factor.
The drawing on the right shows the hard-edge field approximation, with $B_\mathrm{v}$ the field in the valley and $B_\mathrm{h}$ the field in
the hill. The parameter $N$ is the number of the symmetry periods in the magnet.
It is easily seen that for such a model, the flutter takes the
form

\begin{equation} \label{flutje}
F=\alpha(1-\alpha)\left(\frac{\Delta B}{B}\right)^2\ ,
\end{equation}

\noindent where $\Delta B=B_h-B_v$. Thus, the maximum flutter is obtained for $\alpha=0.5$, where the hills and the valleys have the same width.

The flutter is a useful quantity, because the betatron oscillation frequencies can be expressed quite precisely
in terms of $F$. The expressions for the vertical ($\nu_z$) and the radial ($\nu_r$) tunings are given by:

\begin{eqnarray}
\nu_z^2 = k&+&\frac{N^2}{N^2-1}F(1+2\tan^2\xi)\ , \label{nuz} \\
\nu_r^2 = (1-k)&+&\frac{3N^2}{(N^2-1)(N^2-4)}F(1+\tan^2\xi)\ . \label{nur}
\end{eqnarray}

\begin{figure}
\begin{center}
\includegraphics[width=6cm]{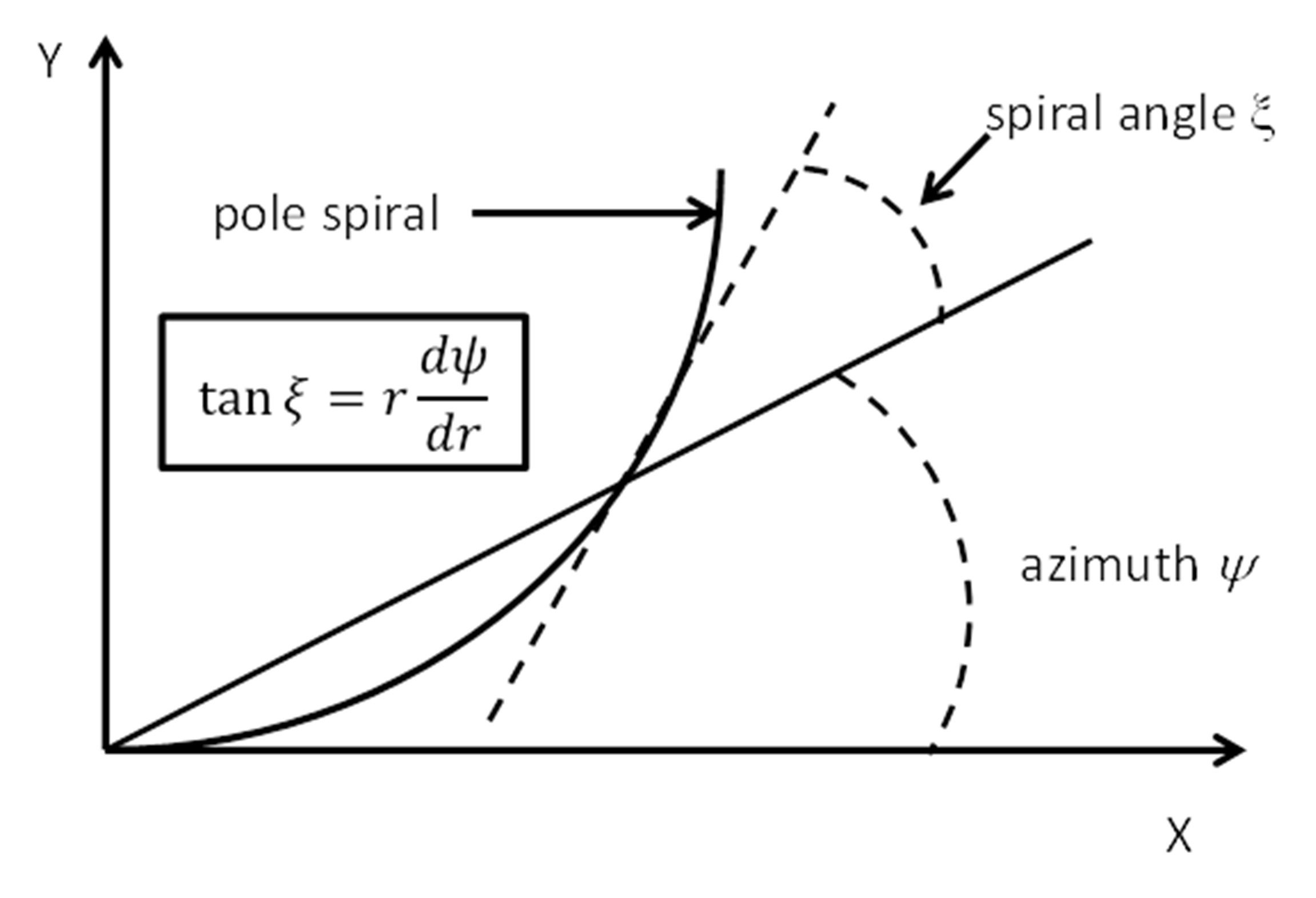}
\caption{Spiral angle $\xi$ of the pole}
\label{fig:spiral}
\end{center}
\end{figure}

\noindent Here, $k$ is the field index and $\xi$ is the spiral angle of the pole. This angle is defined in
\Fref{fig:spiral}

We note that Eqs.~(\ref{nuz}) and (\ref{nur}) are approximations.
A better approximation has been obtained by Hagedoorn
and Verster\cite{Hagedoorn1962}; they express the tuning functions in terms of the Fourier components $A_n$ and $B_n$
of the magnetic field.

\subsection{Major milestones in cyclotron development}

We have seen the main differences between the three types of cyclotron that have been invented, starting in
the 1920s. Now, we can make a brief overview of the major milestones that have been
achieved in the development of the cyclotron. Note that some of the features in this list will be discussed
later on in this course.

\begin{enumerate}
\item Classical cyclotron (Lawrence and Livingston\cite{Lawrence1932}):
        \begin{enumerate}
        \item uniform magnetic field $\Rightarrow$ loss of isochronism due to relativistic mass increase $\Rightarrow$ limited energy;
        \item continuous wave but weak focusing $\Rightarrow$ low currents.        \end{enumerate}
\item Synchro-cyclotron (McMillan--Veksler\cite{Veksler1945,McMillan1945}):
        \begin{enumerate}
        \item decreasing $B(r)$ but time-varying RF frequency $\Rightarrow$ high energies achievable;
        \item pulsed operation and weak focusing $\Rightarrow$ very low currents.        \end{enumerate}
\item The isochronous AVF cyclotron (Thomas focusing):        \begin{enumerate}
        \item azimuthally varying magnetic fields with hills and valleys;
        \item allows both isochronism and vertical stability;
        \item continuous-wave operation, high energies, and high currents;
        \item radial sectors $\Rightarrow$ edge focusing\cite{Thomas1938};        \item spiral sectors $\Rightarrow$ alternating focusing\cite{Christ1950,Courant1952}.        \end{enumerate}
\item The separate sector cyclotron (Willax\cite{Willax1963}):
        \begin{enumerate}
        \item no more valleys $\Rightarrow$ hills constructed from separate dipole magnets;
        \item more space for accelerating cavities and injection elements;        \item example: PSI cyclotron at Villigen, Switzerland;
        \item very high energy ($590\UMeV$) and very high current ($2.5\UmA$) $\Rightarrow$ $1.5\UMW$ beam power.        \end{enumerate}
\item \EH$^-$ cyclotron (TRIUMF, Richardson\cite{Richardson1972}):
        \begin{enumerate}
        \item easy extraction by \EH$^-$ stripping;
        \item low magnetic field (centre $3\UkG$) because of electromagnetic stripping;
        \item TRIUMF is the largest cyclotron in the world ($17\Um$ pole diameter).        \end{enumerate}
\item Superconducting cyclotron: Fraser,  Chalk River, Blosser, MSU\cite{Bigham1973,Blosser1975}:
        \begin{enumerate}
        \item high magnetic field (up to $5\UT$) $\Rightarrow$ high energies at compact design.        \end{enumerate}
\item Superconducting synchro-cyclotrons (Wu--Blosser--Antaya\cite{Wu1990}):
        \begin{enumerate}
        \item very high average magnetic fields ($9\UT$ (Mevion) and almost $6\UT$ (IBA));
        \item very compact $\Rightarrow$ cost reduction $\Rightarrow$ future proton therapy machines.
 \end{enumerate}
\end{enumerate}

\subsection{Commercial cyclotrons for medical applications}

The company IBA was founded in 1986. Since then, more than 300 cyclotrons for isotope production have been sold by IBA, as well as about 40 cyclotrons for proton therapy for cancer treatment. There are a few reasons why cyclotrons are so successful in the medical
market, for radiopharmaceutical applications such as isotope productions and also for proton therapy applications:

\begin{enumerate}
\item cyclotrons are very cost-effective machines for achieving:
        \begin{enumerate}
        \item the required energies (<$100\UMeV$ for radiopharmaceuticals and <$250\UMeV$ for proton ther\-apy);
  \item sufficient beam current (up to 1-$2\UmA$ for radiopharmaceuticals and <$1\UuA$ for proton ther\-apy).
        \end{enumerate}
\item efficient use of RF power $\Rightarrow$ the same RF cavities are used many ($N_\mathrm{turns}$) times;
\item very compact design:
        \begin{enumerate}
        \item the magnetic and RF-structures are fully integrated into system;
        \item the system is single stage $\Rightarrow$ no injector accelerator is needed.
        \end{enumerate}
\item for radiopharmaceuticals, the required energies can be achieved with moderate magnetic fields (1--$2\UT$), allowing
        for conventional magnet technology;
\item simple RF system:
         \begin{enumerate}
        \item constant RF frequency (10--$100\UMHz$), allowing for continuous-wave operation;
        \item moderate RF voltages (10--$100\UkV$).
        \end{enumerate}
\item easy injection into the cyclotron (internal ion source or by axial injection);
\item for radiopharmaceuticals, simple extraction based on stripping of \EH$^-$ ions.
\end{enumerate}

\begin{figure}
\begin{center}
\includegraphics[width=12cm]{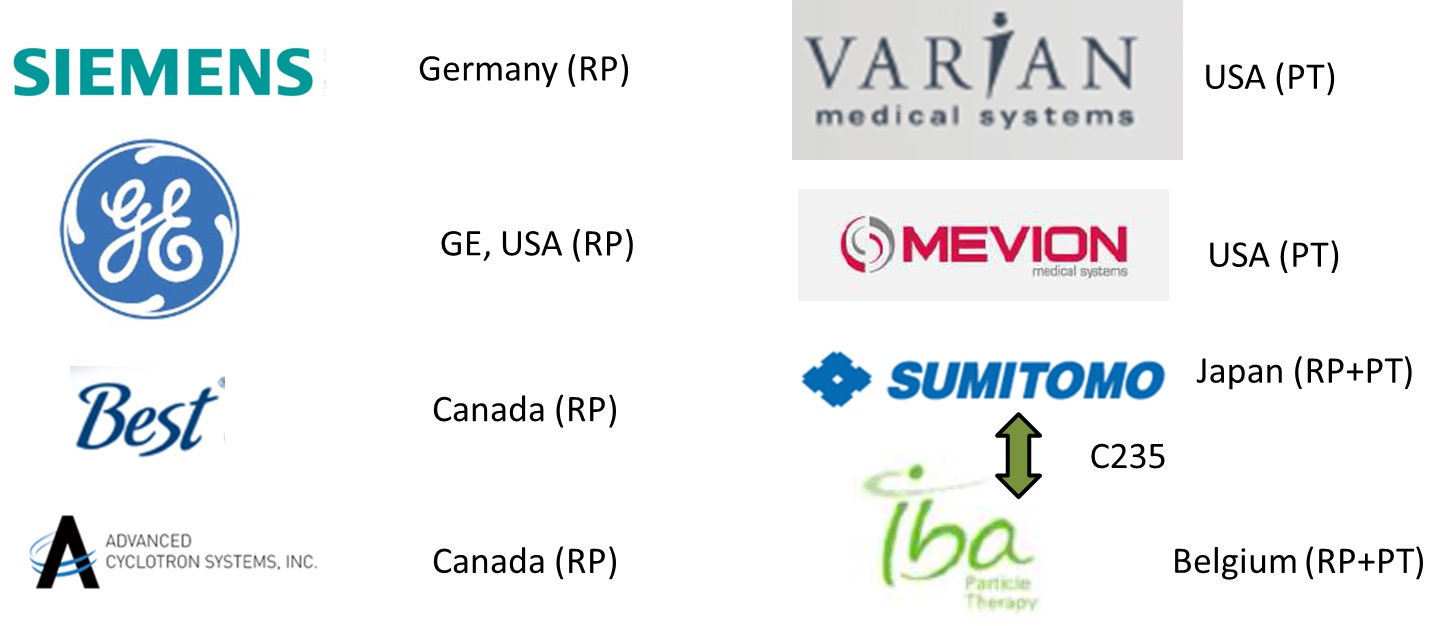}
\caption{The~most important cyclotron vendors or manufacturers. RP, radiopharmaceuticals (mostly isotope production);
PT, proton therapy.}
\label{fig:vendors}
\end{center}
\end{figure}

Globally, there are not so many cyclotron vendors and manufacturers. The largest are listed in \Fref{fig:vendors}.
Note that Sumitomo and IBA collaborated in the development of the C235 cyclotron.

\begin{figure}
\begin{center}
\includegraphics[width=12cm]{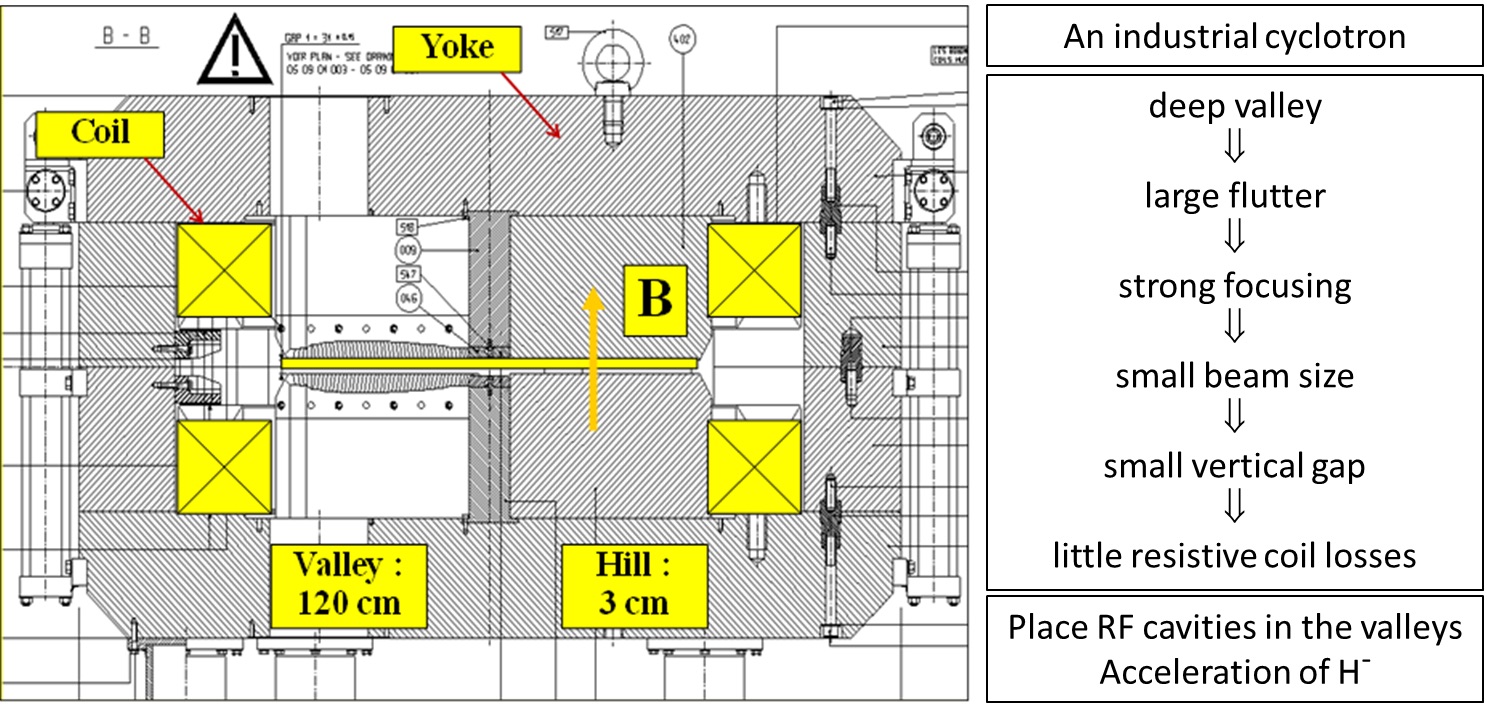}
\caption{Deep-valley cyclotron; the concept is used in many compact isotope production and proton therapy cyclotrons.}
\label{fig:deepvalley}
\end{center}
\end{figure}

\begin{figure}
\begin{center}
\includegraphics[width=9cm]{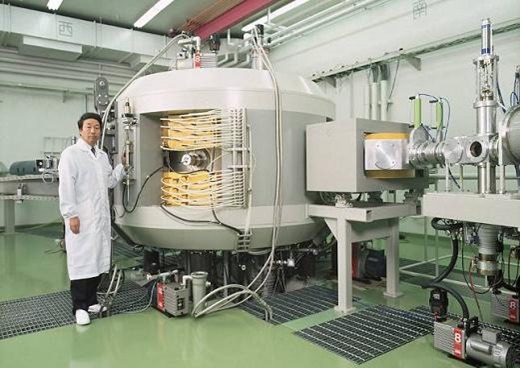}
\caption{ IBA C30 cyclotron}
\label{fig:c30}
\end{center}
\end{figure}

A classical example of a compact industrial isochronous cyclotron for medical isotope production
is the C30, developed by IBA in 1986. The magnet of this machine is shown in \Fref{fig:deepvalley} and the cyclotron
itself in \Fref{fig:c30}. It is characterized by the very large
ratio of valley pole gap to hill pole gap (so-called deep-valley design). This produces a large magnetic flutter and thus relatively
strong vertical focusing. As a consequence, the vertical beam size remains small, so the pole gap in the hills can be
small too ($30\Umm$). In a non-saturated magnet, the magnetic field is inversely proportional to the hill gap;
thus, with such a small hill gap, the ampere turns required in the main coil can be strongly reduced. The success of the
different types of cyclotron for isotope production stems from the following features, which were integrated in the design:

\begin{enumerate}
\item the deep-valley design, allowing for low electric power dissipation in the coils;
\item the four-fold rotational symmetry:
        \begin{enumerate}
        \item allowing a compact design with two RF cavities placed in two opposite valleys;
        \item two remaining valleys remaining for pumping and diagnostics, ion sources, \etc
        \end{enumerate}
\item acceleration of negative ions (\EH$^-$ or \ED$^-$):        \begin{enumerate}
        \item simple extraction by stripping, with almost 100\% extraction efficiency.
        \end{enumerate}
\item simple injection by the use of an internal Penning ionization gauge (PIG) source or external in\-jection with a spiral inflector.
\end{enumerate}

\subsection{Isochronization of the magnetic field}

\begin{figure}
\begin{center}
\includegraphics[width=9cm]{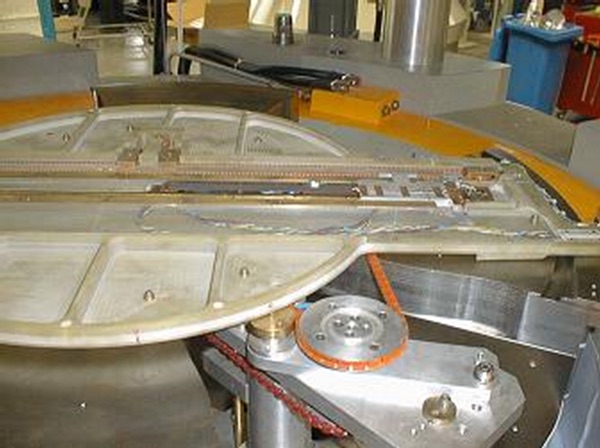}
\caption{Mapping system carrying a Hall probe that moves on a polar grid in the cyclotron median plane}
\label{fig:mapping}
\end{center}
\end{figure}

The cyclotron is perfectly isochronous if the particle angular velocity is constant everywhere in the cyclotron, independent of the
energy of the particle. To achieve this, the average magnetic field needs to be correctly shaped as a function of radius.
It is impossible to obtain perfect isochronism just from the design of the magnet. The required precision of the average
magnetic field is of the order of $10^{-4}$ to $10^{-5}$. To assess the field error, a precise mapping of the magnetic field
in the median plane of the cyclotron is needed. This is done with an automated and computer controlled mapping system,
such as shown in \Fref{fig:mapping}. The mapping system moves a Hall probe (or a search coil) on a 2D polar or Cartesian grid
in order to obtain a full field map. The probe positioning can be pneumatic (compressed air) or motorized. The Hall probes or search
coils need to be precisely calibrated against NMR, and possible temperature effects need to be compensated. The field map
is analysed by computing equilibrium orbits and determining the revolution frequency as a function of particle energy. The
iron of the hill sectors must be shimmed, to improve the isochronism.

Some essential information is obtained from a cyclotron field map.

\begin{enumerate}
\item The level of isochronism $\Rightarrow$ the RF phase slip (per turn and accumulated).
\item Information about transverse optical stability $\Rightarrow$ the tune functions of the betatron oscillations.
\item Possible crossing of dangerous resonances $\Rightarrow$ the tune operating diagram.
\item Magnetic field errors $\Rightarrow$ first- and second-harmonic errors may be drivers for resonances.\item The optical functions (Twiss parameters) of a closed orbit can also be obtained. This may be useful when matching the  extracted beam to a beam line or target.

\end{enumerate}

We note that besides the harmonic errors, median plane errors may also exist in a cyclotron. Such errors can push
the beam out of the median plane. These errors are very difficult to measure because a possible radial magnetic field component is
always much smaller than the main vertical field component; thus, an almost perfect alignment of the Hall probe with respect
to the median plane is needed.

An analysis of a magnetic field map can be done at different levels:

\begin{enumerate}
\item by Fourier analysis and inspection of the average magnetic field $\bar{B}(r)$ and harmonic field errors;
\item by a static orbit analysis $\Rightarrow$ acceleration is turned off and a series of closed orbits and their prop\-erties are determined
        at the relevant range of energies;
\item by computation of accelerated orbits as needed in specific cases, such as:
        \begin{enumerate}
        \item central region studies and design;
        \item extraction studies;
        \item studies of resonance crossing.
        \end{enumerate}
\end{enumerate}

Closed orbits in a cyclotron are computed by solving the static (non-accelerated) motion\cite{Verster1962,Gordon1984}
of the particle. Two types of closed orbit exist.

\begin{enumerate}
\item Equilibrium orbits have the same $N$-fold symmetry as the cyclotron. They are obtained in the ideal magnetic field map
        where errors have been removed.\item Periodic orbits have a periodicity of $2\pi$ and are obtained in a real (measured) field map with errors.
\end{enumerate}

Different dedicated programs are available, such as CYCLOPS\cite{Gordon1984} and EOMSU. At IBA, we use a custom-written program.
These programs solve the equations of motion and determine the proper initial conditions, such that the orbit closes in itself.

The closed orbit code computes the phase slip per turn on each orbit. However, the integrated (accumulated) phase slip
will depend on the energy gain per turn. For a larger dee voltage, there will be less turns and thus less integrated phase slip. Conversely, the energy gain per turn depends on the RF phase slip that was already accumulated. To take this into account,
a self-consistent formula is needed, as follows\cite{Gordon1984}:

\begin{equation}
\Phi(E) = \arcsin\left(\frac{2\pi h}{f_\mathrm{RF}}\int_0^E\frac{\Delta f(E^\prime)}{\Delta E_0(E^\prime)}\mathrm{d}E^\prime\right)\ .
\end{equation}

\noindent Here, $\Phi$ is the integrated phase slip, $h$ is the harmonic mode, $f_\mathrm{RF}$ is the RF frequency, $\Delta f$ is the
closed orbit frequency error, and $E_0$ is the nominal energy gain per turn.

\subsection{Different ways to isochronize a cyclotron}\label{iso_cyclo}

\begin{figure}
\begin{center}
\includegraphics[width=12cm]{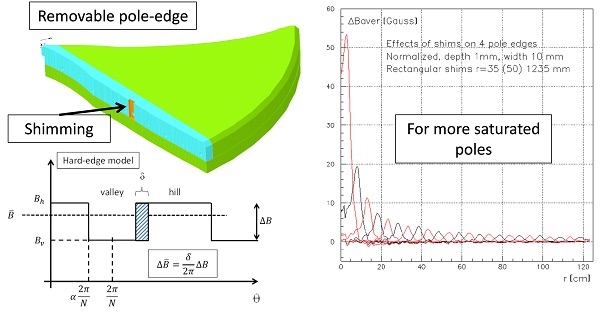}
\caption{Upper left: isochronization of a cyclotron magnetic field is achieved by machining the profile of a remov\-able pole edge.
Lower left: a rough estimate of the shimming effect (change of the average magnetic field)
can be obtained with a hard-edge model. Right: a better prediction of the shimming
can be obtained by calculating (with a 3D finite-element code) the change of the radial profile of the average magnetic field
due to a well-defined pole cut and repeating this for several cuts at varying radii.}
\label{fig:isochronisation1}
\end{center}
\end{figure}

\begin{figure}
\begin{center}
\includegraphics[width=12cm]{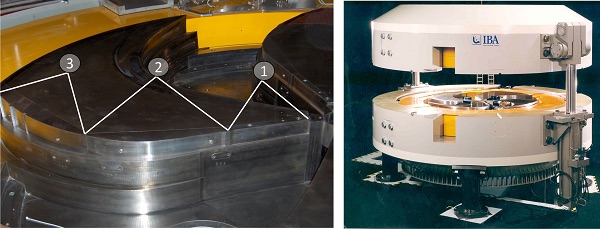}
\caption{Left:  IBA C235 cyclotron for proton therapy. Right: in this cyclotron there are three removable pole edges
for isochronization of the magnetic field.}
\label{fig:isochronisation2}
\end{center}
\end{figure}

Often, cyclotrons for medical isotope production or proton therapy are fixed-field, single-particle ma\-chines.
In such a case, the isochronization of the magnetic field can be achieved by shimming the iron of the pole sectors. This is illustrated
in \Fref{fig:isochronisation1}. Each pole contains an (easily) removable pole edge that can be shimmed.
For a rough estimate of the shimming $\delta$ that is needed to compensate a certain field error $\Delta B$, a hard-edge
model can be used, as illustrated in the lower left panel of \Fref{fig:isochronisation1}. This gives

\begin{equation}
\Delta \bar{B} = \frac{\delta}{2\pi}\Delta B\ ,
\end{equation}

\noindent  where $\Delta B$ is the difference between the hill field and the valley field. Care must be taken that not too much iron is removed. Several iterations are
often needed, with some safety margin applied each time. A hard-edge model is not so precise and does not take into account
the effect that magnetic flux is not completely removed together with the iron that has been cut, but may redistribute to radii other than where the cut was made. This may particularly occur when the iron is saturated. A better estimate can be obtained by
using a 3D finite-element code (such as Opera-3d) to calculate the effect of a multitude of individual small shims at gradually
increasing radius on the pole edge. Then, for each pole cut, the modification of the average magnetic field as a function of radius
is obtained. This is illustrated in the right panel of \Fref{fig:isochronisation1}. From this, a shimming matrix is obtained,
which relates the change of the average field $\Delta \bar{B}(r_1,r_2)$ at the radius $r_2$ due to a small cut at radius $r_1$.
Such a matrix needs to be calculated once for a given (prototype) cyclotron and can then be used to speed up
the isochronization of all following cyclotrons of the same type. \Figure[b]~\ref{fig:isochronisation2} shows, as an example, the IBA C235 cyclotron for proton therapy. There are three
removable pole edges on each pole.

\begin{figure}
\begin{center}
\includegraphics[width=12cm]{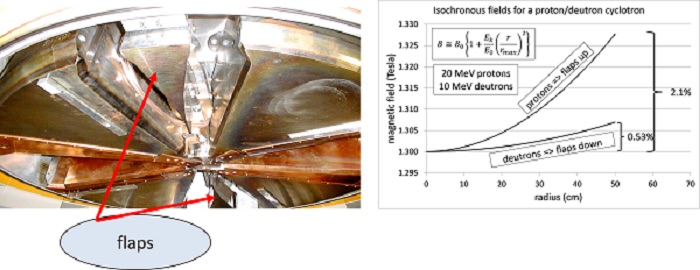}
\caption{Left: upper poles of the IBA C18/9 cyclotron. The movable iron wedges (flaps) are used to change the
average magnetic field profile from protons (flaps close to the median plane) to deuterons (flaps farther away from the median
plane). Right: relativistic correction of the magnetic field, as needed for protons (about 2.1\%)
and for deuterons (about 0.53\%).}
\label{fig:isochronisation3}
\end{center}
\end{figure}

\begin{figure}
\begin{center}
\includegraphics[width=11cm]{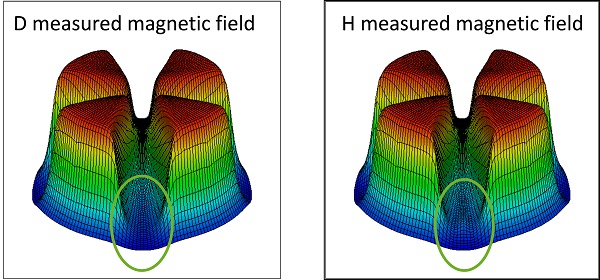}
\caption{Finite-element Opera-3d simulation of the effect of the flaps in the IBA C18/9 cyclotron. Left: the neutron field with the flaps farther away from the median plane. Right: the proton field
with the flaps close to the median plane.}
\label{fig:isochronisation4}
\end{center}
\end{figure}

Modern cyclotrons for the production of PET isotopes can often accelerate two types of particle, namely
\EH$^{-}$ and \ED$^{-}$ ions. For \ED$^{-}$ ions, about half the energy of \EH$^{-}$ ions can be obtained.
The extraction is achieved by stripping and the energy can be varied by moving the radial position of the stripping
foil; the magnetic field remains fixed. However, the relativistic field correction needed for
\EH$^{-}$ ions is about four times as large as for \ED$^{-}$ ions. Thus, two different isochronous field maps need to be
made in the machine. In the IBA cyclotrons, this is done with the so-called `flaps'; these are movable iron wedges that are placed
in two opposite valleys in the cyclotron. For the \EH$^{-}$ ion field, the flaps are moved vertically to a position close to the median
plane. In this configuration, the average magnetic field increases approximately 2\% (for the IBA C18/9 cyclotron) in order to
create the \EH$^{-}$ isochronous field shape. For \ED$^{-}$ ions, the flaps are moved farther away from the median plane, such that
their contribution to the field is strongly reduced.
In the cyclotron, there are still removable pole edges on the hills, which can be shimmed to create an isochronous field shape for the deuterons. The wedge shapes of the flaps are optimized, to create
the isochronous field shape for the protons. This optimization needs to be done only once (for the prototype cyclotron).
The geometry of the C18/9 cyclotron is shown in \Fref{fig:isochronisation3}, together with an illustration of both the
proton and deuteron isochronous field profiles.
Figure~\ref{fig:isochronisation4} shows a finite-element Opera-3d simulation of the effect of the flaps
in the IBA C18/9 cyclotron. The right panel shows the proton field, where the flaps are close to the median plane,
and the left panel shows the deuteron field, where the flaps are farther away from the median plane.

\begin{figure}
\begin{center}
\includegraphics[width=8cm]{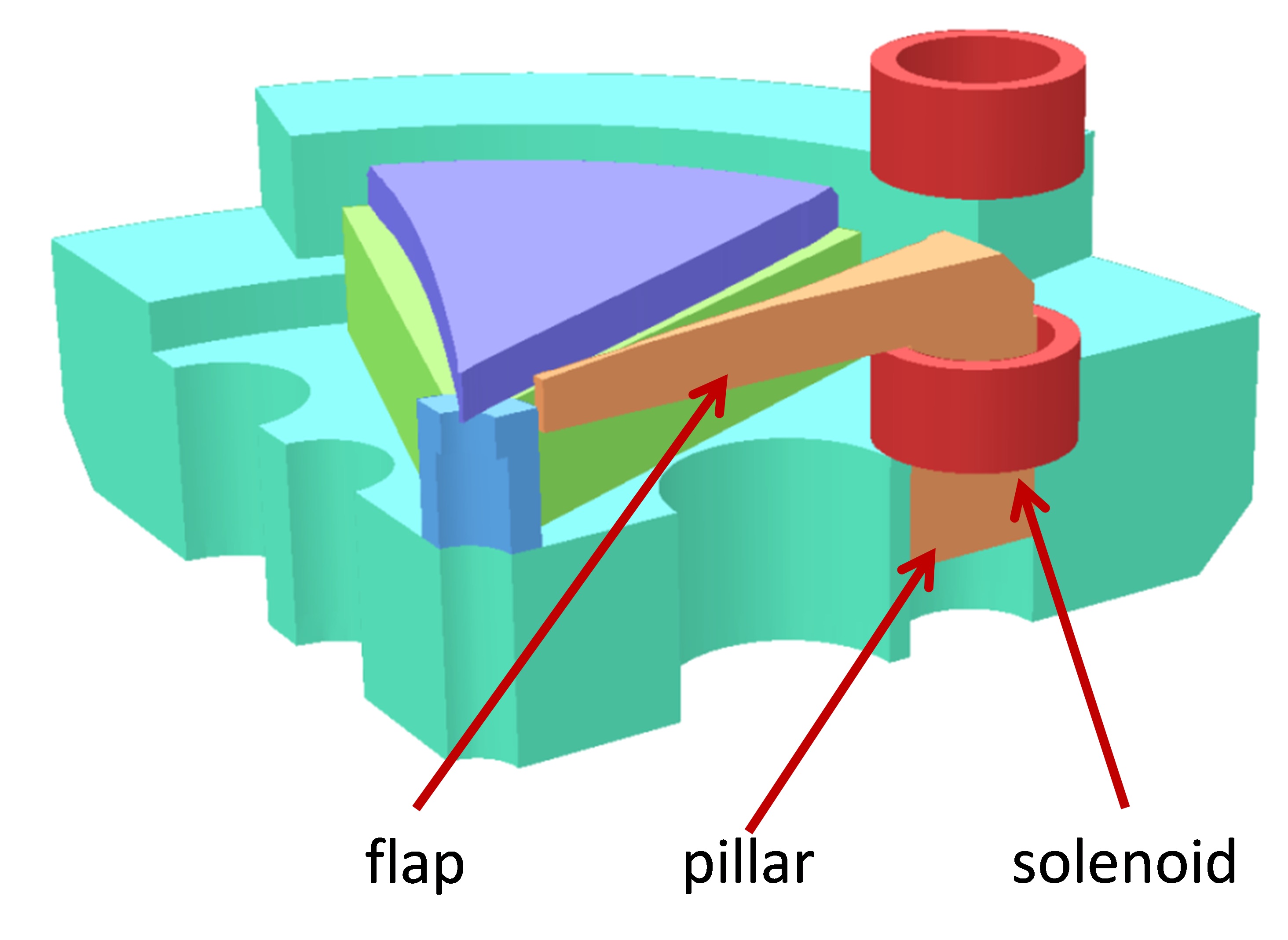}
\caption{Alternative method for isochronization of a dual-particle cyclotron. In this case, the flaps are in a fixed position
and connected magnetically to the base of the return yoke by a solid iron cylinder. The magnetic flux into the flaps is controlled
by the solenoidal coil around the cylinder. This method has not yet been realized but has been patented by IBA.}
\label{fig:isochronisation5}
\end{center}
\end{figure}

The flaps, as discussed, cannot so easily be applied to higher-energy cyclotrons (\eg $70\UMeV$ p and $35\UMeV$ d machine).
For such a cyclotron, a larger relativistic correction is needed (about~7\% for $70\UMeV$ p), which cannot be produced by the
`floating flaps' geometry. A way to solve this is to connect the flaps with an iron cylinder to the base of the valley, as illustrated
in \Fref{fig:isochronisation5}. Here, much more magnetic flux is guided towards the median plane, owing to
higher magnetization of the iron of the flaps. To produce both (p and d) field maps one could again move the flaps vertically,
close to (or away from) the median plane, but in this case the cylinder, attached to the flaps, also has to move into a circular
hole machined into the return yoke. A much simpler solution would be not to move the flaps at all but to place a solenoid coil
around the cylinder; in this way, the flux guidance from the base of the valley towards the median plane can be set (and optimized)
by the DC current in the solenoid (see \Fref{fig:isochronisation5}).

\begin{figure}
\begin{center}
\includegraphics[width=12cm]{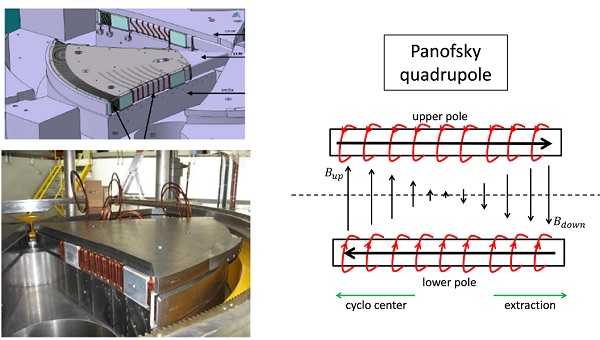}
\caption{Yet another method for isochronization of a dual-particle cyclotron is to split the pole into three layers and to wind coils
around the middle layers. As explained in the right figure, these coils push magnetic field from the outer pole region towards
the centre or vice versa, depending on the polarity of the coil current. Note that in the upper left figure, the cover has been removed.}
\label{fig:isochronisation6}
\end{center}
\end{figure}

Yet another method is applied in the IBA C70XP cyclotron\cite{Zaremba2007}.
This cyclotron accelerates four different particle types, namely
$\EH^-$ to $70\UMeV$ and ions of $\ED^-$, $\EH_2^+$, and $^{4}\EHe^{2+}$ to $35\UMeV$. There are two different isochronous field shapes: the first
is for the $q/m\,=\,1/1$ particle ($\EH^-$) and the other for the $q/m\,=\,1/2$ particles ($\ED^-$, $\EH_2^+$ and $^{4}\EHe^{2+}$).
The hill sectors are divided
into three layers (lower~=~sector, middle~=~pole and upper~=~cover) with an air gap above and below the middle layer).
This enables winding of a coil around this pole, as illustrated in \Fref{fig:isochronisation6}.  With this coil, magnetic
flux can be pushed from the extraction region towards the centre or vice versa, thus modifying the profile of the average magnetic
field. In the actual machine, there is not one coil but three independent coils wound around each pole, which are needed to shape
the average field with sufficient precision.

\begin{figure}
\begin{center}
\includegraphics[width=7.5cm]{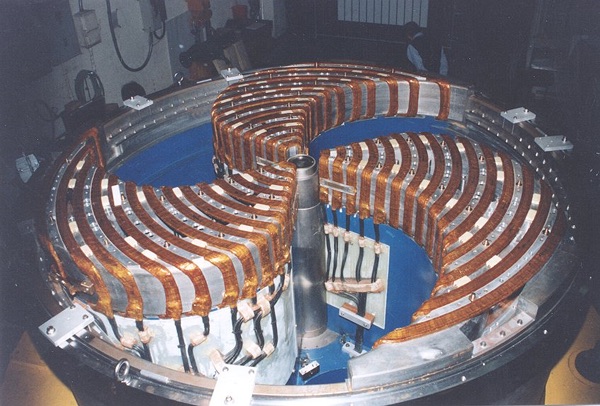}
\caption{A method similar to the one outlined in \Fref{fig:isochronisation6} is used in the AGOR superconducting cyclotron\cite{Gales1986,Schreuder1996}}
\label{fig:isochronisation7}
\end{center}
\end{figure}

A similar, more general, method is applied in the AGOR superconducting cyclotron\cite{Gales1986,Schreuder1996} (\Fref{fig:isochronisation7}). This is a
variable-energy, multiparticle, superconducting cyclotron that has been mainly used for nuclear physics research.
It requires a much broader range
of magnetic field maps and adjustments. To obtain this kind of flexibility, there are 15 independent Panofsky-type
correction coils placed around each pole.

\begin{figure}
\begin{center}
\includegraphics[width=\textwidth]{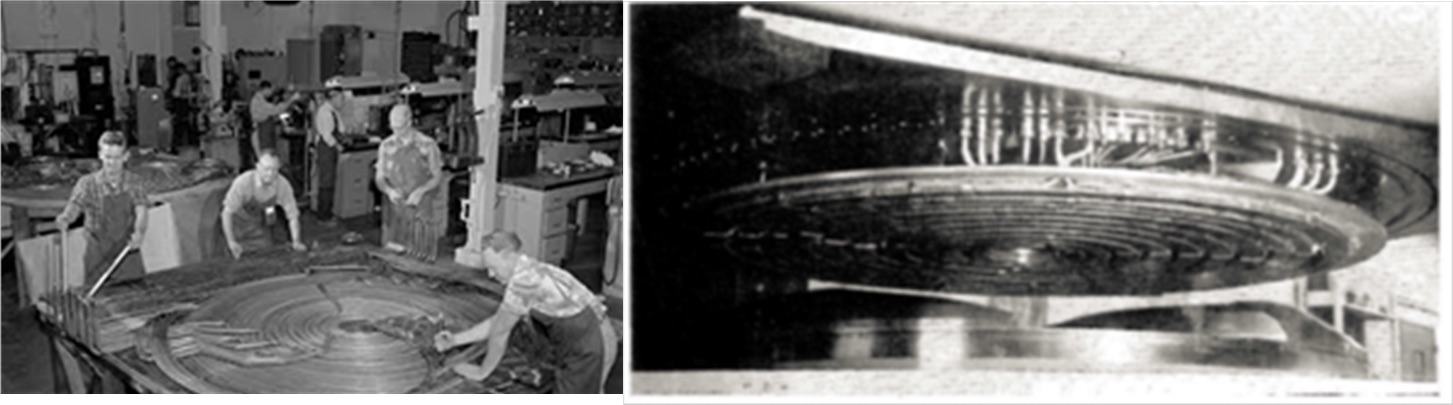}
\caption{A more general method is used to isochronize multiparticle variable-energy cyclotrons. Here, an array of concentric
circular coils is placed on the pole and each coil uses its own independent power supply. In this way, there is a lot
of flexibility in creating the required average magnetic field profile. Left: correction coils for the  Berkeley 88-inch cyclotron\cite{Berkeley}.
Right:  correction coils of the Philips 30 MeV azimuthally varying field cyclotron\cite{Verst1962}.}
\label{fig:isochronisation8}
\end{center}
\end{figure}

Another general method, used for variable-energy multiparticle cyclotrons is to place a set of independent circular coils on
the pole of the cyclotron. This is, for example, applied in the Berkeley 88-inch cyclotron\cite{Kelly1962}
and the Philips $30\UMeV$ variable-energy cyclotron\cite{Verst1962}, as
illustrated in \Fref{fig:isochronisation8}.

\subsection{Example: A $70\UMeV$ industrial cyclotron for isotope production}\label{c70_example}

\begin{figure}
\begin{center}
\includegraphics[width=\textwidth]{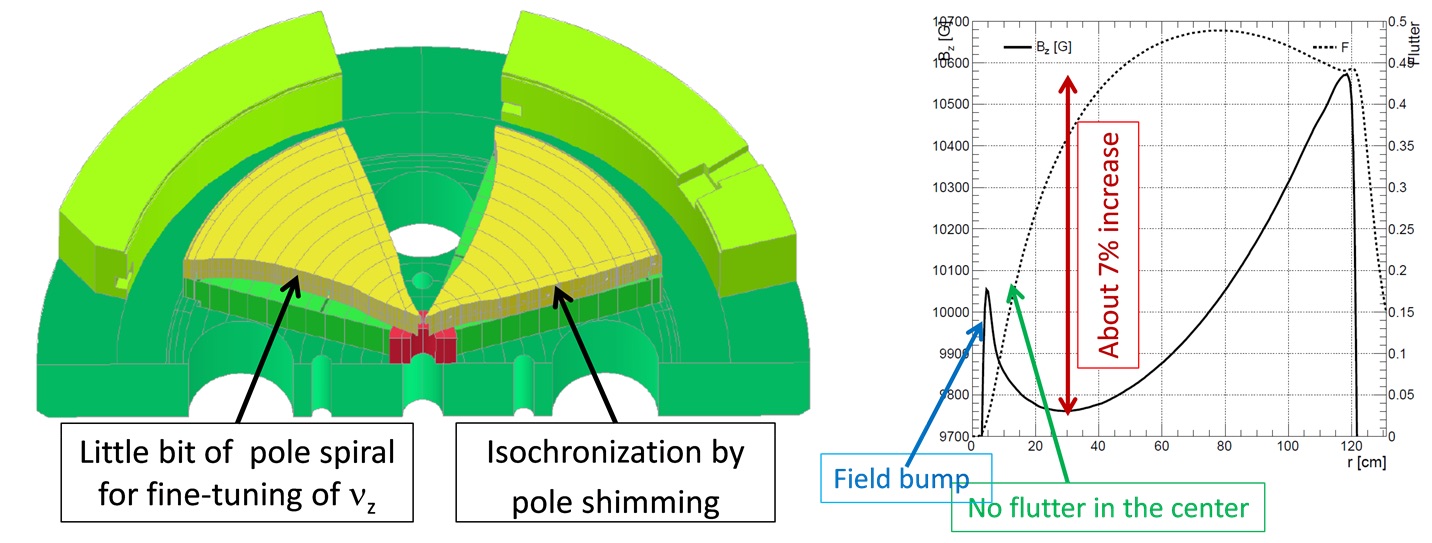}
\caption{Left: the IBA C70 as an example of an industrial cyclotron for medical isotope production,
showing the small spiral in the pole to adjust vertical focusing and the removable pole edge for isochronization.
Right: average magnetic field and flutter of this cyclotron.}
\label{fig:c70_example2}
\end{center}
\end{figure}

Recently at IBA, a $70\UMeV$ cyclotron for the production of medical radioisotopes has been designed and constructed.
This cyclotron is under commissioning at the time of writing (end of 2015).  The C70 is a high-intensity ($750\UuA$), four-fold symmetrical
$\EH^-$ cyclotron with axial injection and dual beam extraction by stripping. The magnetic structure is given in
\Fref{fig:c70_example2}. In this single-particle, fixed-field  machine, isochronization is ensured by shimming the
removable pole edges. It can be seen that a small amount of pole spiral has been applied  to increase and fine tune the
vertical betatron frequency $\nu_z$. The right panel of \Fref{fig:c70_example2} shows the shape of the average magnetic field and the flutter
in this cyclotron.

The average magnetic field increases by approximately 1\% per $10\UMeV$ up to 7\% at the maximum radius.
It is seen that the flutter goes to zero in the centre of the cyclotron. To provide some additional vertical magnetic focusing in this region, a field bump is provided, which generates a negative field index. This zone
is not isochronous and thus generates some RF phase slip, but not so much because it is crossed in a few turns. Some additional
vertical electrical focusing is obtained at the first few accelerating gaps (as discussed in \Sref{injection}).
The sharp drop of the magnetic field towards the centre of the cyclotron is due to the axial hole in the
return yoke needed for the axial injection.

\begin{figure}
\begin{center}
\includegraphics[width=\textwidth]{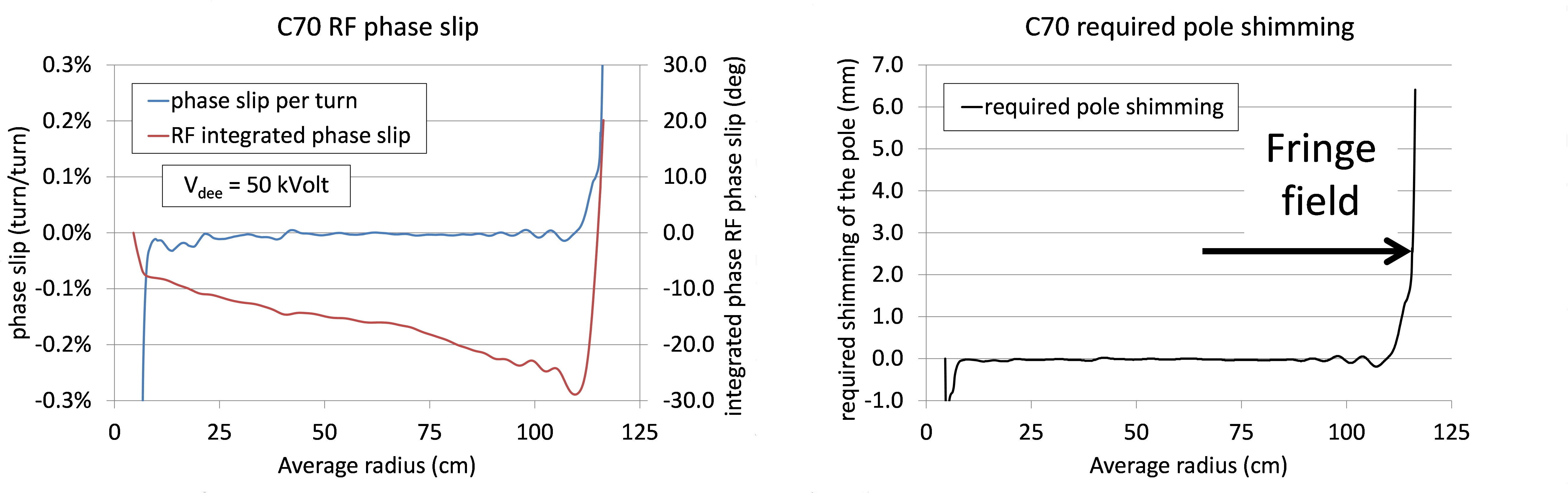}
\caption{Left: RF phase slip per turn and integrated RF phase slip in the IBA C70 cyclotron. Right: calculated phase slip
per turn can be related to the required shimming of the removable pole edges.}
\label{fig:c70_example4}
\end{center}
\end{figure}

\Figure[b]~\ref{fig:c70_example4} shows the phase slip per turn and the integrated RF phase slip in this cyclotron.
The horizontal axis is the average radius of the closed orbits. These orbits are found up to an energy of $71.4\UMeV$. The highest energy orbits enter the radial fringe area of the poles where a large
part of the accumulated phase slip is generated. The RF frequency is tuned such that the (negative) minimum and (positive) maximum of the
integrated phase slip are equal. In this case, the maximum phase slip of 30$^\circ$ is considered acceptable; the actual shift will be smaller because particles are extracted at energies less than $70\UMeV$.
The right panel of \Fref{fig:c70_example4} shows the amount of shimming of the pole that would be needed to
isochronize the field perfectly. The negative values correspond with cutting of the iron. The sharp rise at the highest
radii is due to the fringe field of the magnet. This error is intrinsic to the magnet design and cannot be corrected.

\begin{figure}
\begin{center}
\includegraphics[width=\textwidth]{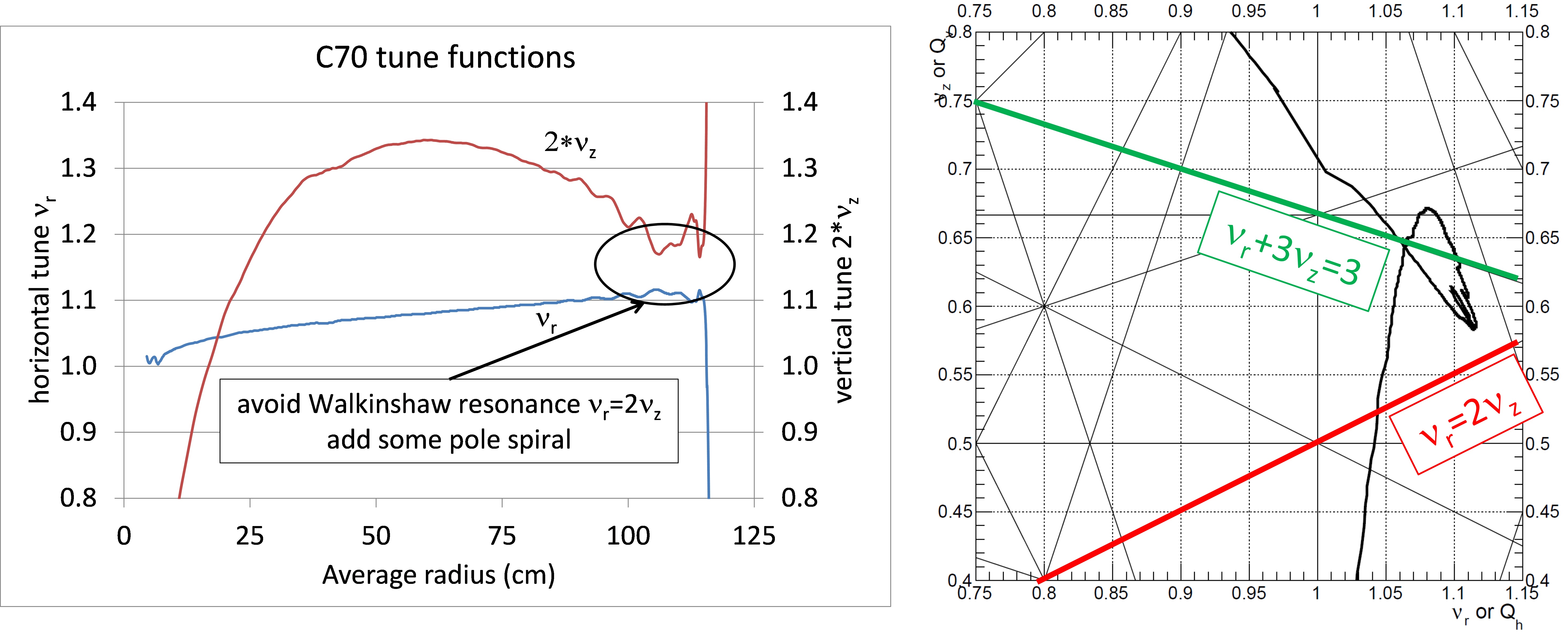}
\caption{Left: radial ($\nu_r$) and vertical ($\nu_z$) tune function of the IBA C70 cyclotron as a function of radius. Right:
related tuning diagram, showing the working curve and some important resonance lines.}
\label{fig:c70_example5}
\end{center}
\end{figure}

\Figure[b]~\ref{fig:c70_example5} shows the tuning functions and tuning operating diagram of the C70. In the left panel, the vertical
tuning, $\nu_z$, has been multiplied by a factor  of two, to clearly visualize the situation of the Walkinshaw resonance
$\nu_r=2\nu_z$. This is considered as a resonance that might be dangerous in the region of extraction, especially if the
radial beam size is large. In the design of the magnet,
the (slight) spiralling of the pole was introduced in an effort to avoid this resonance. This spiral increases the value of $\nu_z$ and lifts the red curve
in \Fref{fig:c70_example5} above the blue one. It is seen in the resonance diagram that besides the Walkinshaw, the
resonance $\nu_r+3\nu_z=3$ is crossed a few times. However, this resonance is considered much less dangerous, because it is
of higher order (four, as compared with three for the Walkinshaw) and also because it is not a structural resonance (it is driven by a
field error of harmonic three and not by some intrinsic harmonic of the magnetic field).

\subsection{The limit of a three-fold rotational symmetry cyclotron}

\begin{figure}
\begin{center}
\includegraphics[width=\textwidth]{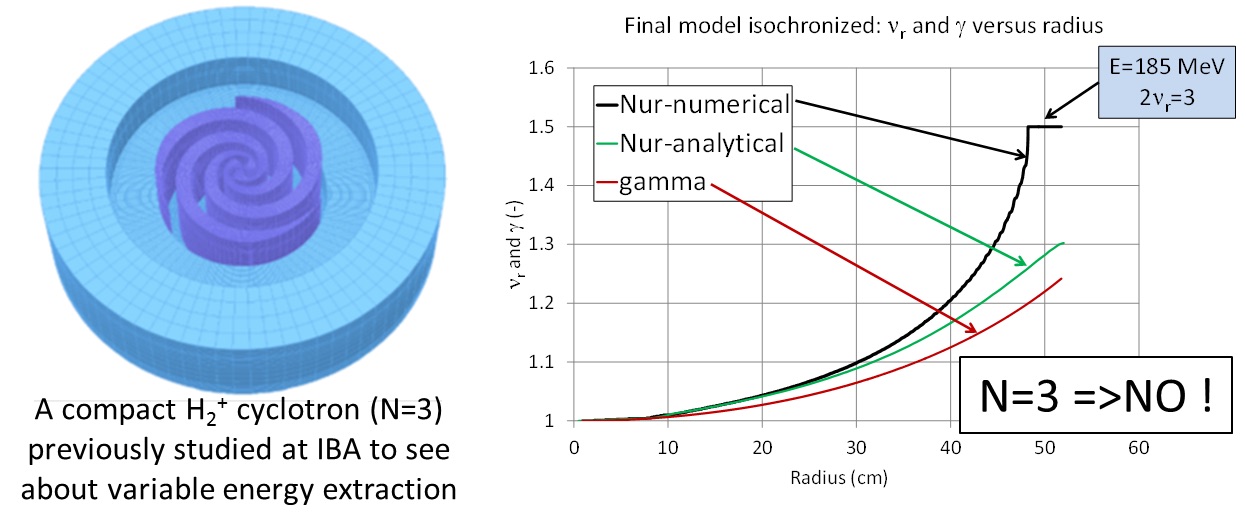}
\caption{Energy limitation of a cyclotron with three-fold rotational symmetry due to the $2\nu_r=3$ resonance.
The flutter is produced by iron poles and is therefore limited in a high-field cyclotron.  Left: Opera-3d model of the compact $\EH_2^+$ cyclotron used in the simulations. Right: $\nu_r$ curve as a function of radius, showing
the $2\nu_r=3$ stop band.}
\label{fig:resonance1}
\end{center}
\end{figure}

In this section, we make a small sidestep and ask whether a cyclotron with three-fold
symmetry can be used for proton therapy at about $230\UMeV$.
The simple formula ($\nu_r\approx\gamma$) saying that the radial betatron tuning, $\nu_r$, is about equal to the
relativistic $\gamma$ is not valid when the resonance

\begin{equation}
2\nu_r=3
\end{equation}

\noindent is approached. This resonance is driven by quadrupole terms (radial field gradients) of symmetry 3. For the
cyclotron considered, this is the basic symmetry of the magnetic field; therefore, this resonance is a structural resonance.
For a proton therapy cyclotron of $230\UMeV$, we have $\gamma\approx 1.25$, and one could think that this is still far enough away
from the resonance value $\nu_r=1.5$. However, this is not the case. At IBA, we conducted a small study, to see
whether variable-energy stripping extraction could be made in a compact $\EH_2^+$ cyclotron with symmetry $N=3$.
The left panel of \Fref{fig:resonance1} shows the Opera-3d magnetic model of this cyclotron. The flutter in the high-field
(assumed superconducting) cyclotron is limited by the fact that it is produced only by the iron poles.
The right panel of \Fref{fig:resonance1} shows the radial tune function as a function of the radius. The red curve corresponds
to the simple approximation $\nu_r\approx\gamma$. The green curve is obtained from the analytic formula derived by
Hagedoorn and Verster\cite{Hagedoorn1962}, which is closely related to \Eref{nur} (but more precise than it).
The black curve is obtained from the closed orbit calculations in the actual field map.  Here, it is clearly seen that the
theory is no longer valid when the resonance is approached. At an energy of $185\UMeV$, the beam optics enters the stop band
of the resonance. At this energy, the beam would no longer be stable and thus $185\UMeV$ is the maximum energy that can be achieved.

\begin{figure}
\begin{center}
\includegraphics[width=12cm]{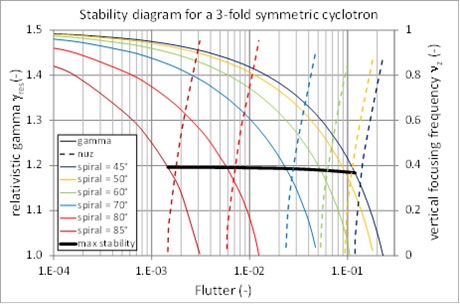}
\caption{Stability diagram of a cyclotron with three-fold rotational symmetry, assuming that the flutter is relatively small,
as it is produced by the iron poles. The solid black line shows the maximum kinetic energy (relativistic $\gamma$)
that can be obtained, taking into account the requirements of vertical focusing, as well as the $2\nu_r=3$ stop band.}
\label{fig:resonance2}
\end{center}
\end{figure}

To better understand this, we repeated the derivation of the Hamiltonian made by Hagedoorn and Verster\cite{Hagedoorn1962}, but instead of studying the radial motion in a coordinate frame rotating
at the (dimensionless) frequency equal to 1 (which is a good approximation when $\nu_r\approx 1$),
we studied the motion in a coordinate frame rotating at a frequency equal to 1.5 (which should give a better approximation when
$\nu_r\approx 1.5$). We note that Hagedoorn's theory is very precise, as long as the flutter is not too large.
The results of this analysis are summarized in \Fref{fig:resonance2}. The horizontal axis gives
the flutter of the magnetic field. The left vertical axis gives the value of the relativistic $\gamma$ at the $2\nu_r=3$ resonance. The right vertical axis gives the value of the vertical betatron frequency, $\nu_z$, in the
magnetic field. The magnetic field index is chosen such that the cyclotron is isochronous. Different lines in \Fref{fig:resonance2}
correspond to different values of the pole spiral angle. The coloured solid lines must be read on the left axis and show
at what value of the flutter and what value of the spiral angle, the $2\nu_r=3$ resonance sets in. It can be seen that, for a
given flutter, the maximum energy decreases with increasing spiral angle. Similarly, for a given spiral angle,
the maximum energy decreases with increasing flutter.  The dashed lines represent the vertical tuning $\nu_z$, which must be read
on the right axis. Here, it can be seen that $\nu_z$ increases with both increasing flutter and increasing spiral angle.
For a stable cyclotron, it is necessary that one remains below the inset of the resonance and also that the vertical tuning is positive.
This limit of stability, for both radial and vertical motion, is found by looking for the points on the solid coloured curves
for which the vertical tune is exactly zero.  Connecting these points gives the bold black line in \Fref{fig:resonance2}.
This line represents the stability limit of a compact three-fold symmetric AVF cyclotron (with small flutter).
It can be seen that in all cases, the maximum relativistic $\gamma$ is almost the same and corresponds to a maximum proton energy
of about $185\UMeV$.

\subsection{The notion of the orbit centre and the magnetic centre in a cyclotron}

\begin{figure}
\begin{center}
\includegraphics[width=\textwidth]{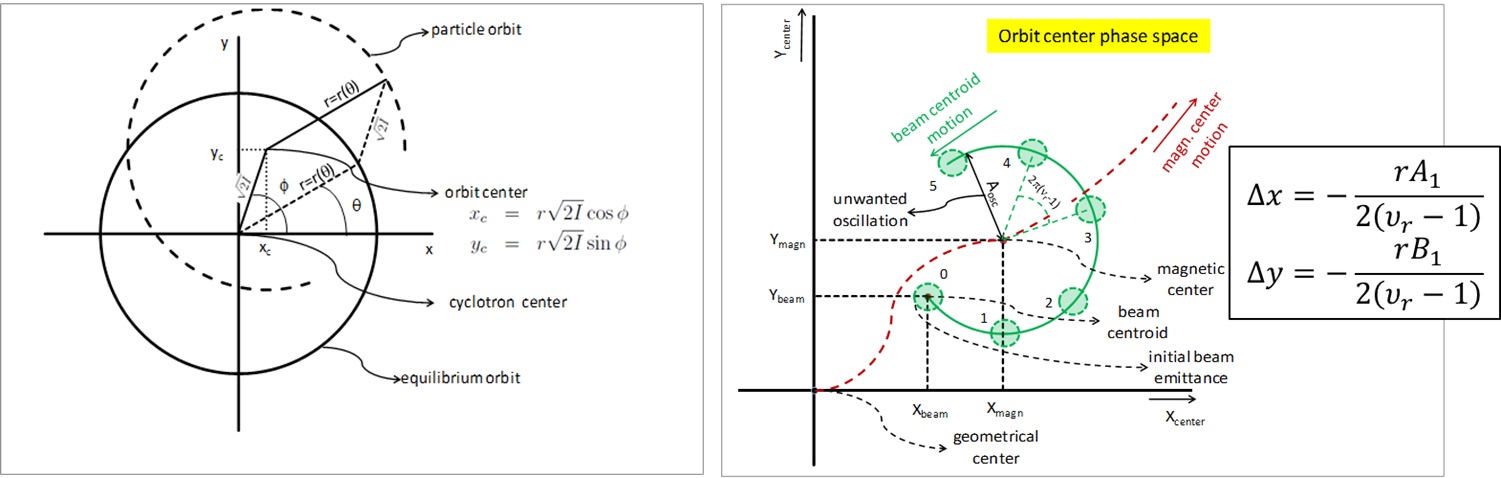}
\caption{Left: in a cyclotron, the radial betatron oscillation can be represented as a (slow) motion of the orbit centre.
Right: the betatron oscillations take place with respect to the magnetic centre of the cyclotron.
Owing to a first-harmonic field error, the magnetic centre shifts away from the geometric centre of the cyclotron.}
\label{fig:orbit_centre2}
\end{center}
\end{figure}

Betatron oscillations in a cyclotron can be represented by the usual amplitude and phase, but also by the coordinates
of the orbit centre. The latter can be more convenient, because the orbit centre oscillates slowly ($\nu_r-1$) compared with the betatron oscillation itself ($\nu_r$). In the orbit centre representation, the equations of motion can
be simplified using approximations that make use of the slowly varying character of this motion and the integration can be
done much faster. This may be especially useful in a synchro-cyclotron where the particle makes many turns
(approximately $50\,000$ for the S2C2)
and full orbit integration from the source to extraction is almost impossible.
The left panel of \Fref{fig:orbit_centre2} illustrates the radial betatron oscillation around the equilibrium orbit
in terms of the orbit centre. The real orbit can be reconstructed from the orbit centre coordinates and the equilibrium orbit
radius $r(\theta)$.
In this illustration, a circular equilibrium orbit (synchro-cyclotron) is shown.
For AVF cyclotrons, this will be a scalloped orbit.

Particles execute a betatron oscillation around the magnetic centre. A first-harmonic field error displaces the magnetic
centre of the cyclotron relative to the geometrical centre. This displacement is given by:

\begin{eqnarray}
  \Delta x &=& -\frac{rA_1}{2(\nu_r-1)}\ , \\
  \Delta y &=& -\frac{rB_1}{2(\nu_r-1)}\ .
\end{eqnarray}

When there is acceleration, the magnetic centre itself is also moving and the total motion is a superposition of the two
separate motions. This is illustrated in the right panel of \Fref{fig:orbit_centre2}.  The beam quality (emittance)
degrades when the beam centroid is not following the magnetic centre. This may occur in two ways.

\begin{enumerate}
\item A beam centring error at injection.
\item Accelerating through a region where the gradient of the first harmonic is large. This is a non-adiabatic effect (which
will not occur in a synchro-cyclotron where the acceleration is very slow).
\end{enumerate}
The coherent amplitude $A_\mathrm{osc}$ of the betatron oscillation is a good measure of the harmful effect of the centring error. The
numbers in \Fref{fig:orbit_centre2} indicate subsequent turns.
Hagedoorn and Verster\cite{Hagedoorn1962} have derived a Hamiltonian description of the dynamics of the orbit centre. Their
theory includes linear and non-linear motion (separatrix), as well as the influence of field errors.

\subsection{Harmonic field errors in a map}

As discussed in the previous section, a first-harmonic field error will de-centre the closed orbit. This effect becomes
large when $\nu_r\simeq 1$. This happens in the cyclotron centre and in the radial fringe region of the pole. An off-centred
orbit might become sensitive to other resonances. Excessively high harmonic errors must be avoided. However, a localized first-harmonic field bump may be used to create a coherent beam oscillation, enabling extraction of the beam from the cyclotron (precessional
extraction).

The gradient of the first harmonic can drive the $2\nu_z=1$ resonance. In the stop band of this resonance,
the motion becomes vertically unstable. This may lead to amplitude growth and emittance growth.
The stop band of this resonance is defined by\cite{Kleeven1992}:

\begin{equation}
 4\nu_z\left\lvert\nu_z-\frac{1}{2}\right\rvert < \sqrt{r\frac{\mathrm{d}C_1}{\mathrm{d}r}}\ ,
\end{equation}

\noindent where $C_n=\sqrt{A_n^2+B_n^2}$.

The gradient of a second harmonic can drive the $2\nu_r=2$ resonance. In the stop band of this resonance,
the horizontal motion becomes unstable. This problem may occur, for example, when second-harmonic iron shims are used to isochronize
a dual-particle (proton--deuteron) cyclotron. The stop band of this resonance is given by\cite{Kleeven1992}

\begin{equation}
 4|\nu_r-1| < \left\lvert 2C_2+r\frac{\mathrm{d}C_2}{\mathrm{d}r} \right\rvert \ .
\end{equation}

\begin{figure}
\begin{center}
\includegraphics[width=9cm]{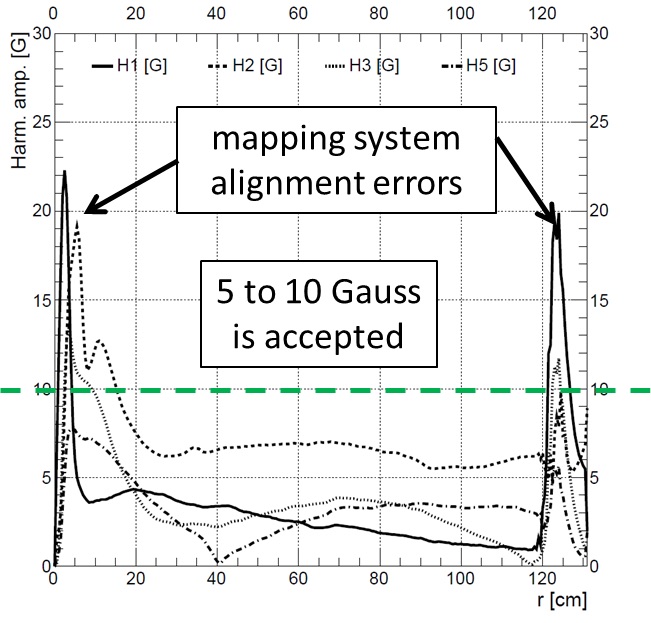}
\caption{Amplitudes of the (most important) harmonic field errors in the measured field map of the IBA C70 cyclotron.}
\label{fig:orbit_centre1}
\end{center}
\end{figure}

We note that the same resonance is used in the synchro-cyclotron for beam extraction (regenerative extraction).
\Figure[b]~\ref{fig:orbit_centre1} shows the first few harmonic errors, as measured in the  IBA C70 cyclotron (discussed in
\Sref{c70_example}). The large peaks observed in the cyclotron centre and near the pole radius are artefacts
due to small to small radial alignment errors of the mapping system. Such errors produce a large effect in the regions where
the radial gradients are high. Often, in general, harmonic errors smaller than 5--$10\UG$ are considered acceptable
in such industrial $\EH^-$ cyclotrons.

\subsection{The revival of the synchro-cyclotron}

In an isochronous cyclotron, vertical focusing is generated by the flutter of the magnetic field.
This flutter can be created by the iron of the magnet (not by the solenoidal coils). The maximum achievable
field modulation with the iron is about $2\UT$. If the average magnetic field is pushed too far up (using superconducting coils), the flutter will steadily decrease and, at a certain point, insufficient vertical focusing will be achieved. In a synchro-cyclotron,
this problem does not occur. Thus, in a synchro-cyclotron one can fully exploit the potential offered by superconductivity.

\begin{figure}
\begin{center}
\includegraphics[width=12cm]{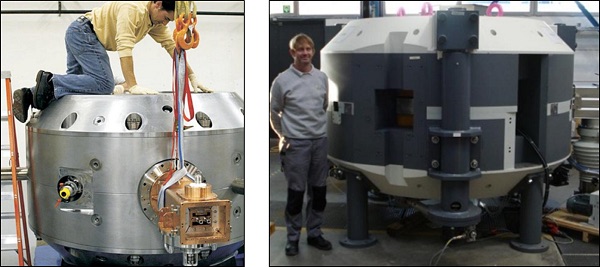}
\caption{Left: Monarch S250 synchro-cyclotron from Mevion (Couresy Mevion medical systems \Bref{Mevion}). Right: superconducting synchro-cyclotron (S2C2) from IBA.}
\label{fig:synchrocyclos}
\end{center}
\end{figure}

In 2007, the company Still River Systems (today, Mevion Medical Systems, Inc.) began manu\-facturing a
superconducting synchro-cyclotron for proton therapy based on the patent of Dr T. Antaya from the
Massachusetts Institute of Technology. This accelerator (left panel of \Fref{fig:synchrocyclos})
has a central magnetic field
of $9\UT$; this high field is obtained with a $\ENb_3\ESn$ super\-conducting coil, cooled by cryo-coolers.
The unique feature of this extremely compact cyclotron is that it is mounted on a gantry rotating around the patient.
The proton beam is extracted at a fixed energy of $250\UMeV$. As with other superconducting magnets,
the large magnetic forces acting on the superconducting coil impose the presence of a special former around the coil.
The total consumed power is about $120\UkW$.

In 2008, IBA began development of the compact superconducting synchro-cyclotron S2C2\cite{Kleeven2013}
(see right panel of \Fref{fig:synchrocyclos}).
With a superconducting $\ENb\ETi$ coil, the magnetic field in the centre of the cyclotron is $5.7\UT$ and the size of the
cyclotron is reduced to a diameter of $2.5\Um$. The total weight of the S2C2 is about 45 tonnes and the beam energy is
constant at $230\UMeV$. The beam from the S2C2 is pulsed at $1\UkHz$ and pulses are about $10\Ums$ long.
This results from the synchro-cyclotron concept by which the RF frequency is reduced synchronously
with the accelerated proton beam. The first S2C2 has been installed at the Centre de Protonth\'erapie Antoine Lacassagne in Nice,
France and is commissioned at the time of writing (end of 2015).

\section{Injection into a cyclotron}\label{injection}

In this section, the main design goals for beam injection are explained and special problems related to a central region
with internal ion source are considered. The principle of a PIG source is addressed.
The issue of vertical focusing in the cyclotron centre is briefly discussed. Several examples of numerical simulations are given.
Axial injection is also briefly outlined.

The topic of cyclotron injection has already been covered in earlier CAS proceedings
of the general accelerator physics course \cite{Heikkinen1994}, as well as in CAS proceedings of
specialized courses \cite{Mandrillon1996,Kleeven2005}. An overview of issues related to beam transport from the ion source
into the cyclotron central region has been given by Belmont at the 23rd ECPM \cite{Belmont2003}.
Since then, not so many substantial changes have occurred in the field,
especially if one only considers small cyclotrons that are used for applications. For this reason, an approach is chosen where the accent is less on completeness and rigorousness (because this has already been done) but more on explaining and illustrating the main principles that are used in medical cyclotrons. Sometimes a more industrial viewpoint is taken. The use of complicated formulae is avoided as much as possible.

Two fundamentally different injection approaches can be distinguished, depending on the position of the ion source.
An internal ion source is placed in the centre of the cyclotron, where it constitutes an integrated part of
the RF accelerating structure. This may be a trivial case, but it is the one that is most often implemented
for compact industrial cyclotrons, as well as for proton therapy cyclotrons.
The alternative is the use of an external ion source, where some kind of injection line
with magnets for beam guiding and focusing is needed, together with some kind of inflector to kick the beam
onto the equilibrium orbit. This method is used for higher-intensity isotope production cyclotrons and also in the proposed
IBA C400 cyclotron for carbon therapy.

\subsection{Design goals}

Injection is the process of particle beam transfer from the ion source, where the particles are created,
into the centre of the cyclotron, where the acceleration can start. When designing an injection system for a cyclotron,
the following main design goals must be identified.

\begin{enumerate}
\item Horizontal centring of the beam with respect to the cyclotron centre. This is equivalent to placing the beam on the correct equilibrium
orbit given by the injection energy.
\item Matching (if possible) of the beam phase space with respect to the cyclotron eigenellipse (accept\-ance).
\item Vertical centring of the beam with respect to the median plane.
\item Longitudinal matching (bunching), \ie compressing the DC beam from the ion source into shorter packages
at the frequency of the RF.
\item Minimization of beam losses and preservation (as much as possible) of the beam quality between the ion source and the
cyclotron centre.
\end{enumerate}

\begin{figure}
\begin{center}
\includegraphics[width=7cm]{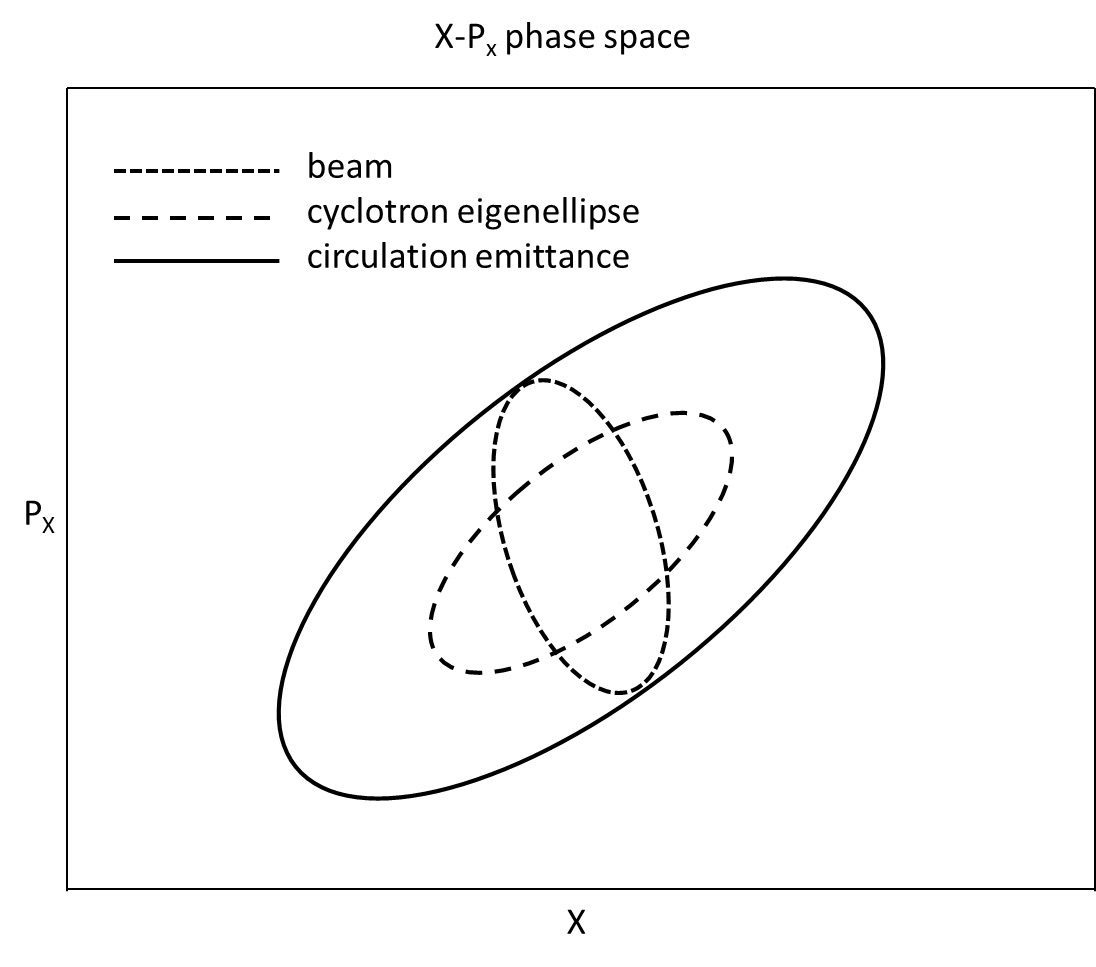}
\caption{Mismatch between the cyclotron eigenellipse and the beam ellipse injected into the machine leads, after many turns,
to an increase of the circulating emittance.}
\label{fig:mismatch}
\end{center}
\end{figure}

The requirement of centring of the beam with respect to the cyclotron centre is equivalent to requiring that the beam
is well positioned on the equilibrium orbit corresponding with the energy of the injected particles.
The underlying physical reasons for the first three requirements are the same. A beam that is not well centred or is badly
matched will execute coherent oscillations during acceleration. In the case of off-centring, these are beam centre-of-mass oscillations.  In the case of    phase-space mismatch, these are beam envelope oscillations.  After many turns , these coherent
oscillations smear out and directly lead to an increase in~the circulating beam emittances (see \Fref{fig:mismatch}).
Consequently, beam sizes will be larger, the beam is more sensitive to harmful resonance, the extraction
will be more difficult, and the beam quality of the extracted beam will be lower.

The last two requirements directly relate to the efficiency of injection into the cyclotron.
Longi\-tudinal matching requires a buncher, which compresses the longitudinal DC beam coming from the ion source into RF buckets.
A buncher usually contains an electrode or small cavity that oscillates at the same RF frequency as the cyclotron dees.
It works like a longitudinal lens that introduces a velocity modu\-lation of the beam. After a sufficient drift,
this velocity modulation transfers into longitudinal dens\-ity modulation. The goal is to obtain an RF bucket
phase width smaller than the longitudinal cyclo\-tron accept\-ance.  For a compact cyclotron without flat-top dees,
the longitudinal acceptance is usually around 10--15\%. With a simple buncher, a gain of a factor of two to three can
easily be obtained. However, at increasing beam intensities, the gain starts to drop, owing to longitudinal space charge forces
that counteract the longitudinal density modulation. The issue of beam loss minimization also occurs,
for example, in the design of the electrodes of an axial inflector.
Here, it must be ensured that the beam centroid is well centred with respect to the electrodes. This is not a
trivial task, owing to the complicated 3D orbit shape in an inflector. An iterative process of 3D electric field simulation and orbit tracking is required.

\subsection{Constraints}

It should be kept in mind that the design of the injection system is often constrained by requirements at a higher level
of full cyclotron design. Such constraints can be determined, for example by:

\begin{enumerate}
\item the magnetic structure:
        \begin{enumerate}
        \item magnetic field value and shape in the centre;
        \item space available for the central region, inflector, ion source, \etc
        \end{enumerate}
\item the accelerating structure:        \begin{enumerate}
        \item the number of accelerating dees;
        \item the dee voltage;        \item the RF harmonic mode.
        \end{enumerate}
\item the injected particle:
        \begin{enumerate}
        \item charge-to-mass ratio of the particle;
        \item number of internal ion sources to be placed (one or two);        \item injection energy.
        \end{enumerate}
\end{enumerate}

\subsection{Cyclotrons with an internal ion source}

The use of an internal ion source is the simplest and certainly also the least expensive solution for in\-jection into a cyclotron.
Internal ion sources are used in proton therapy cyclotrons as well as in iso\-tope production cyclotrons. The internal ion source is
also used in high-field (6--$9\UT$) super\-conducting synchro-cyclotrons for proton therapy (see, for example, \Bref{Kleeven2013}).
Besides the elimination of the injection line, a main advantage lies in the compactness of the design.
This opens up the possibility of placing two ion sources in the machine simultaneously. In many small PET cyclotrons, an $\EH^{-}$
and a $\ED^{-}$ source are included, to be able to accelerate and extract both protons and deuterons.
These two particles are sufficient to produce four well-known and frequently used PET isotopes
$^{11}\EC$, $^{13}\EN$, $^{15}\EO$, and $^{18}\EF$.
However, an internal ion source brings several limitations: (i) often only low to moderate beam intensities
can be obtained; (ii) only simple ion species such as, for example, $\EH^{+}$, $\EH^{-}$, $\ED^{-}$, $^{3}\EHe$, or $^{4}\EHe$ can be obtained;
(iii) injected beam manipulation, such as matching or bunching is not possible; (iv) there is a direct gas leak into the cyclotron, which might be especially limiting for the acceleration of negative ions
because of vacuum stripping; (v) high beam quality is more difficult to obtain;  and
(vi) source maintenance requires venting and opening the cyclotron.

\subsubsection{The Penning ionization gauge ion source}

\begin{figure}
\begin{center}
\includegraphics[width=\textwidth]{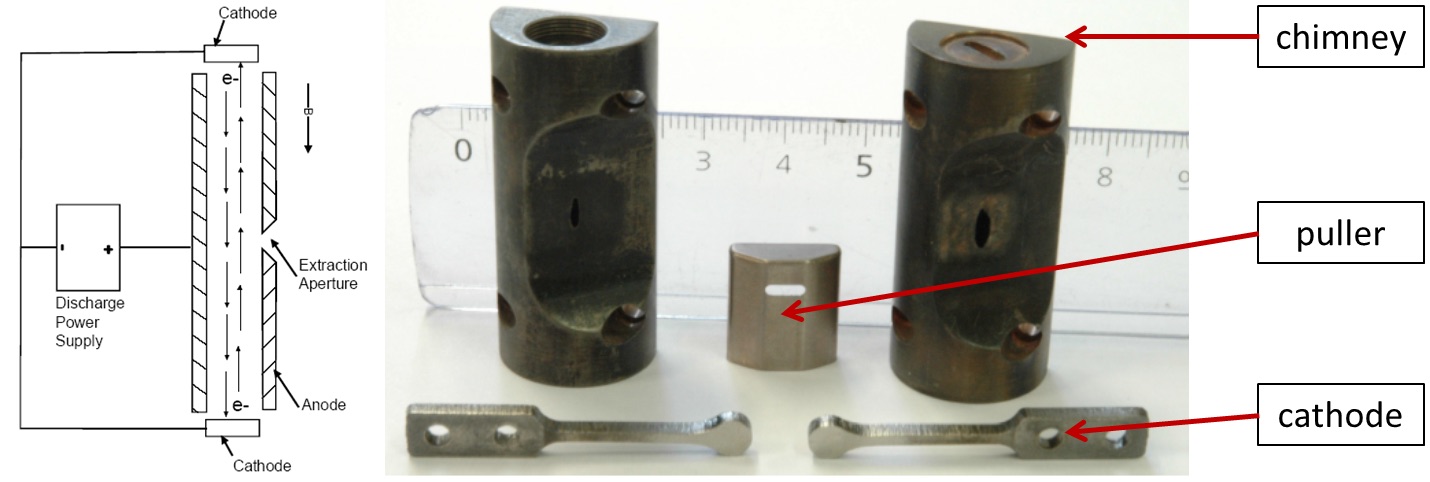}
\caption{Left: cold-cathode Penning ion gauge source. Right: two chimneys, two cathodes, and a puller. The chimney on the right shows an eroded slit.}
\label{fig:pig}
\end{center}
\end{figure}

\begin{figure}
\begin{center}
\includegraphics[width=\textwidth]{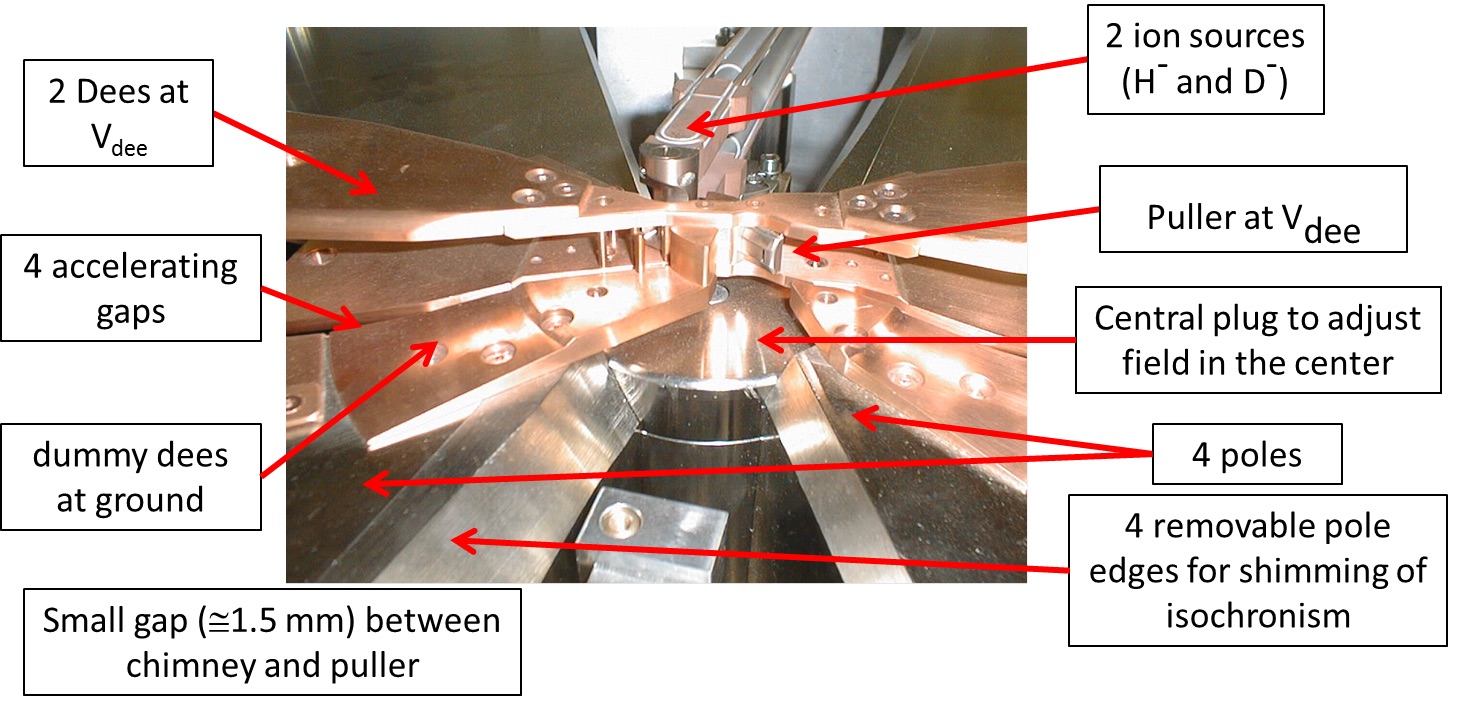}
\caption{Central region of the IBA C18/9 cyclotron showing the dees and dummy dees.
One of the two ion sources has been removed to show the puller.
The figure also shows the four hill sectors. The removable circular disc underneath the central region is the central plug;
it is used to fine tune the magnetic field bump in the centre.}
\label{fig:cregion1}
\end{center}
\end{figure}

A cold-cathode PIG\cite{Gavin1989} ion source is often used as an internal source.
The PIG source contains two cathodes that are placed at each end of a cylindrical anode (\Fref{fig:pig}).
The cathodes are at negative potential relative to the anode (the chimney). They emit electrons that are needed
to ionize~the hydrogen gas and create~the plasma.  The cyclotron magnetic field must be along the axis of the anode.
This field is essential for the functioning of the source, as it enhances confinement of the electrons in the plasma
and therefore the level of ionization of the gas. The electrons oscillate up and down as they are reflected between the
two cathodes and spiralize around the vertical magnetic field.  To initiate the arc current,
the cathode voltage must initially be raised to a few~kilovolts. Once a plasma is formed, the cathodes are self-heated
by ionic bombardment and the arc voltage will decrease with increasing arc current.
Usually, an operating voltage of a few hundred volts is obtained. This is sufficient to ionize the gas atoms.
The ions to be accelerated are extracted from the source via a small aperture called the slit. This extraction is
obtained by the electric field that exists in between the chimney and the so-called puller.
This puller is at the same electric potential as the RF accelerating structure, as it is
mechanically connected to the dees (see \Fref{fig:cregion1}).

The right panel of \Fref{fig:pig} shows chimneys and cathodes used in compact IBA cyclotrons. The chimneys are
made of a copper-tungsten alloy, which has~good thermal properties and good machining properties.
The cathodes are fabricated from tantalum, because of its good thermal properties and its low work function for electron emission.

\subsubsection{Some guidelines for central region design}

\Figure[b]~\ref{fig:cregion1} shows a typical design of a central region for a compact PET-cyclotron with two internal
ion sources; one for $\EH^-$ and one for $\ED^-$.

The design of such a central region with an internal ion source is a tedious task that requires precise numerical calculation
and often many iterations before good beam centring, vertical focusing, and longitudinal matching are obtained.
Some general guidelines for such a design process can be given.

\begin{enumerate}

\item Start with a crude model and refine it step by step. Begin with a uniform magnetic field and assume a hard-edge
uniform electric field in the gaps. Initially, only consider orbits in the median plane. Try to find the approximate position
of the ion source and the accelerating gaps that will centre the beam with respect to the cyclotron centre and
centre the beam RF phase with respect to the accelerating wave (longitudinal matching).
This may be done by drawing circular orbit arcs by hand, using analytical formulae\cite{Mandrillon1996}, or even using a pair of compasses (for drawing circular arcs between gaps) and a protractor (for estimating RF phase advance between gaps).
Transit time effects should be taken into account, especially in the first gap\cite{Reiser1962}.
The starting phase for particles leaving the source should be roughly between  $-40^\circ$ and $-10^\circ$,
assuming that the dee voltage is described by a cosine function.

\item Once an initial gap layout has been found, the model should be further refined by using an orbit integration program.
Here, the electric field map can still be generated artificially by assuming, for each gap, an electric field shape
with a Gaussian profile that only depends on the coordinate that is normal to the gap and not on the coordinate
that is parallel to the gap. Empirical relations may be used to find the width of the Gauss function in
terms of the gap width and the vertical dee gap\cite{Hazewindus1974}. For the first gap between the source and the puller,
a half-Gauss function should be used. The advantage of this intermediate step is that the layout can be easily generated
and modified. At this stage, the vertical motion can be taken into account.

\item Create a full 3D model of the central region and solve the Laplace equation to calculate the
3D electric field distribution. An electrostatic map can be used, as long as the wavelength of the RF is much larger than
the size of the central region. Several 3D codes exist, such as RELAX3D from TRIUMF\cite{Houtman1994},  or the commercial code
TOSCA,  from Vector Fields. The latter is a finite-element code that allows modelling of very fine details
as part of a larger geometry without the need of very fine meshing everywhere (see \Fref{fig:cregion2}). This enables modelling of the full accelerating structure and orbit tracking from the source to the extraction.
If possible, the model should be fully parametric, to allow for fast modifications and optimizations.
With the availability of 3D computer codes, it is no longer necessary to measure electric field distribution
as has been achieved in the past by electrolytic tank measurements\cite{Reiser1968,Liukkonen1979} or a magnetic analogue model\cite{Hazewindus1974}.

\begin{figure}
\begin{center}
\includegraphics[width=9cm]{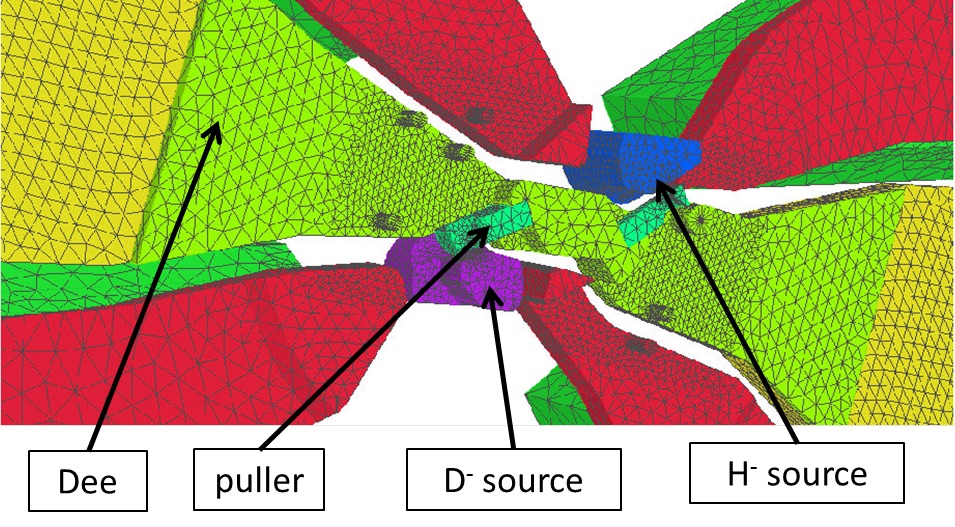}
\caption{Example of a 3D finite-element model of the IBA C18/9 central region. Fine meshing is used in regions with small
geometrical details, such as the source--puller gap. The graded mesh size allows modelling of the full dee-structure.
Complete parameterization of the model is used for fast modification and optimization.}
\label{fig:cregion2}
\end{center}
\end{figure}

\item Track orbits in the calculated electric field and in the realistic magnetic field
(obtained from field mapping or from 3D calculations). Fine tune the geometry further for better centring,
vertical focusing, and longitudinal matching (RF phase centring). An example of such a calculation is given in \Fref{fig:orbits}.

\begin{figure}
\begin{center}
\includegraphics[width=9cm]{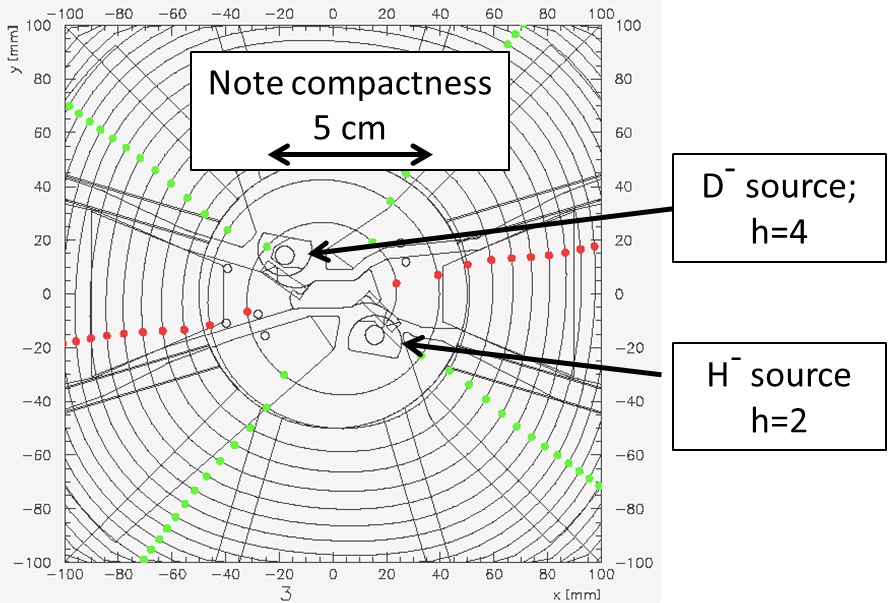}
\caption{Calculated orbits in the IBA C18/9 central region. The $\ED^-$ source is shifted farther outward because its orbit
is larger. Note that parts of the $\ED^{-}$ chimney have been cut away, to give sufficient clearance for the $\EH^{-}$ beam. Assessment of longitudinal matching is illustrated:
the red dots and the green dots give the particle position when the dee voltage is zero and a maximum, respectively.
Ideally, the red dots should be on the dee centre line. Electric fields are obtained from Opera. Magnetic fields from Opera
or from a measured map.}
\label{fig:orbits}
\end{center}
\end{figure}

\item Track a full beam (many particles) to find beam losses and maximize the beam transmission in the central region
(see, for example, \Bref{Kleeven2003}).

\end{enumerate}

The energy gain per turn in the cyclotron is given by:

\begin{equation}
\Delta E_k = qV_{\mathrm{dee}} N\sin\left(\frac{h\alpha}{2}\right)\cos\Phi_{\mathrm{RF}}\ .
\end{equation}

\begin{figure}
\begin{center}
\includegraphics[width=10cm]{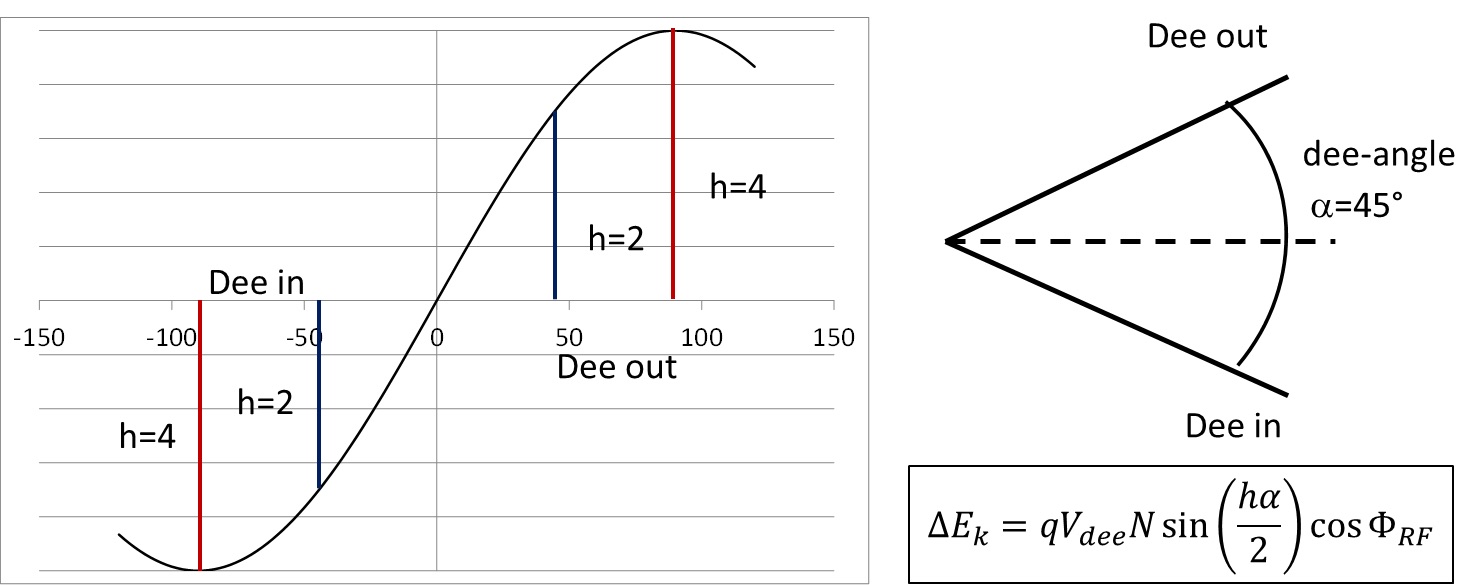}
\caption{Energy gain per turn depends on several parameters, such as the dee voltage $V_{\mathrm{dee}}$, the number of dees, $N$, the harmonic
mode, $h$, the dee angle, $\alpha$, and the RF phase $\Phi_{\mathrm{RF}}$. For a dee angle of 45$^\circ$,
the harmonic mode $h=4$ is most efficient.}
\label{fig:accel}
\end{center}
\end{figure}

\noindent Here, $q$ is the charge of the particle, $V_\mathrm{dee}$ is the dee voltage, $N$ is the number of accelerating gaps,
$h$ is the harmonic mode, $\alpha$ is the dee opening angle and $\Phi_\mathrm{RF}$ is the RF phase of the particle.
Maximum acceleration is obtained if the RF phase advance between the dee entrance and
the dee exit is just 180$^\circ$ ($h\alpha=\pi$). This is illustrated in \Fref{fig:accel}. For
example, if the dee angle is 45$^\circ$ and $h=4$, the energy gain is 100\%, but for $h=2$, it is only 71\%.

\subsubsection{Vertical focusing in the cyclotron centre}

The azimuthal field variation goes to zero in the centre of the cyclotron; therefore, this resource for
vertical focusing is lacking. There are two remedies to restore the vertical focusing.

\begin{enumerate}

\item Add a small magnetic field bump (a few hundred gauss) in the centre. The negative radial gradient of this bump provides
some vertical focusing. The bump must not be too large, to avoid too large an RF phase slip.
In small IBA PET cyclotrons, the bump is fine tuned by modifying the thickness of the central plug (see \Fref{fig:cregion1}).
\item Fully exploit the electrical focusing provided by the first few accelerating gaps.

\end{enumerate}

\begin{figure}
\begin{center}
\includegraphics[width=14cm]{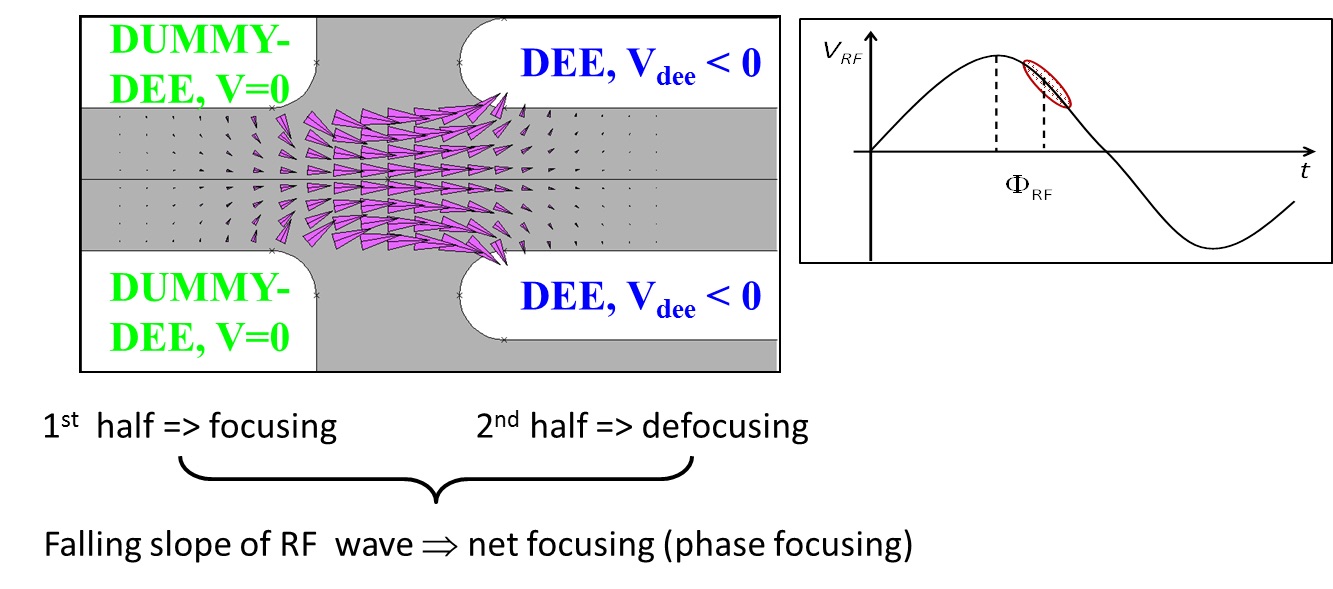}
\caption{Vertical electrical focusing in the cyclotron accelerating gap.
This focusing is important only in the cyclotron centre.}
\label{fig:zfocus1}
\end{center}
\end{figure}

If an accelerating gap is well positioned with respect to the RF phase, it may provide some elec\-trical focusing.
\Figure[b]~\ref{fig:zfocus1} illustrates the shape of the electric field lines in the accelerating gap between a dummy dee (at ground potential) and the dee.
The particle is moving from left to right and is accelerated in the gap. In the first half of the gap,
the vertical forces point towards the median plane and this part of the gap is vertically focusing.
In the second half of the gap, the vertical forces have changed sign and this part of the gap is vertically defocusing.
If the dee voltage were DC, there would already be a net focusing effect of the gap for two reasons:
(a) a focusing and de-focusing lens, one behind the other, provide some net focusing in both planes and
(b) the defocusing lens is weaker because the particle has a higher velocity in the second part of the gap.
This is comparable to the focusing obtained in an Einzel lens.
However, the dee voltage is not DC but varying in time and this may provide an additional focusing term that
is more important than the previous two effects (phase focusing). This is obtained by letting the particle cross the gap at the
moment that the dee voltage is falling (instead of accelerating at the top). In this case, the defocusing effect
of the second gap half is additionally decreased. To achieve this, the first few accelerating gaps
must be properly positioned azimuthally. This forms part of the central region design.

\begin{figure}
\begin{center}
\includegraphics[width=10cm]{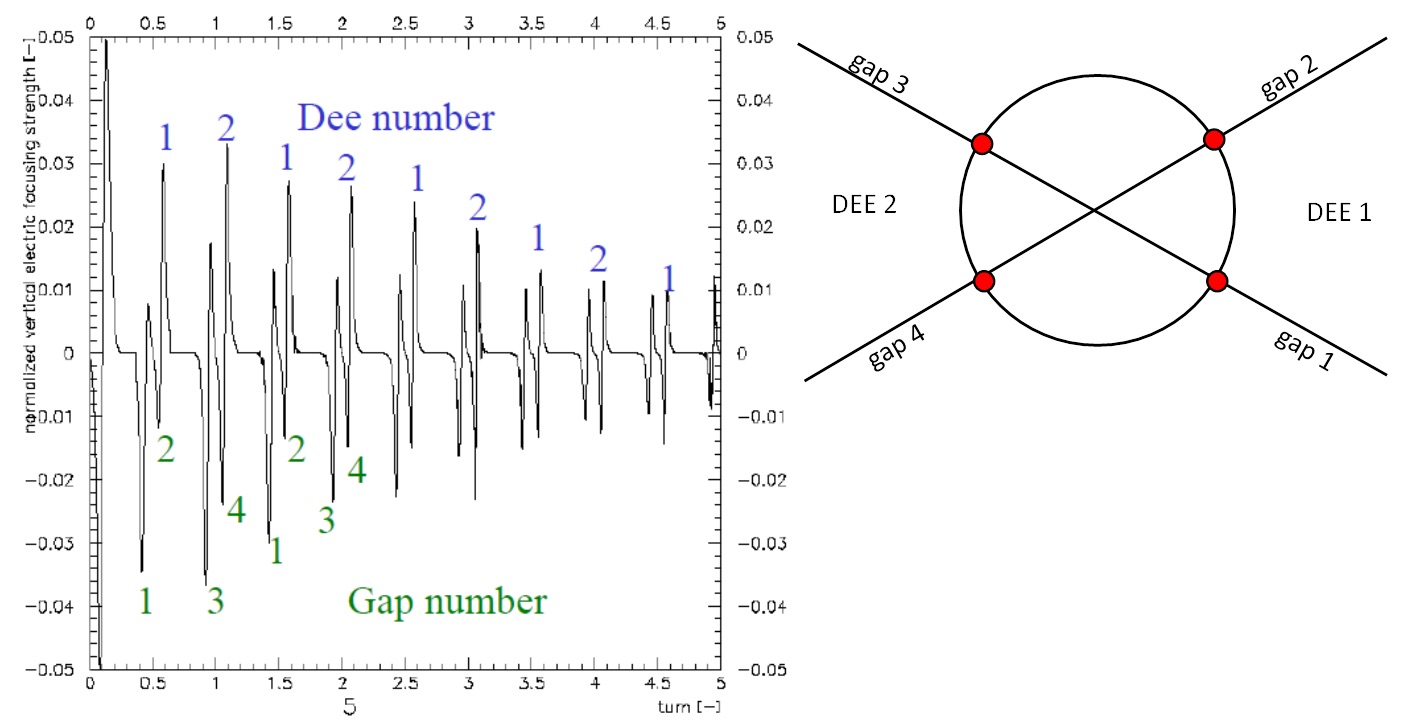}
\caption{Normalized vertical electrical forces obtained from orbit tracking of a particle during five turns in a central region
with two dees (four accelerating gaps). Dee crossings are indicated at the top of the curve (in blue) and gap crossings (two per dee) at the bottom of the curve (in green). Each gap is focusing at the entrance and defocusing at the exit.}
\label{fig:zfocus2}
\end{center}
\end{figure}

Vertical electrical focusing is also illustrated in \Fref{fig:zfocus2}.
This figure shows the (normalized) vertical electrical force
acting on a particle during the first five turns in the cyclotron (IBA C18/9). A minus sign corresponds with focusing
(force directed towards the median plane). The dee crossings (two dees) as well as the gap crossings (four gaps) are indicated.
It can be seen that each gap is focusing at the entrance and defocusing at the exit.
It can also be seen from the amplitude of the force that the electrical focusing rapidly falls with increasing beam energy.
However, after a few turns, the magnetic focusing becomes sufficient.

\subsubsection{The central region of a superconducting synchro-cyclotron}

\begin{figure}
\begin{center}
\includegraphics[width=12cm]{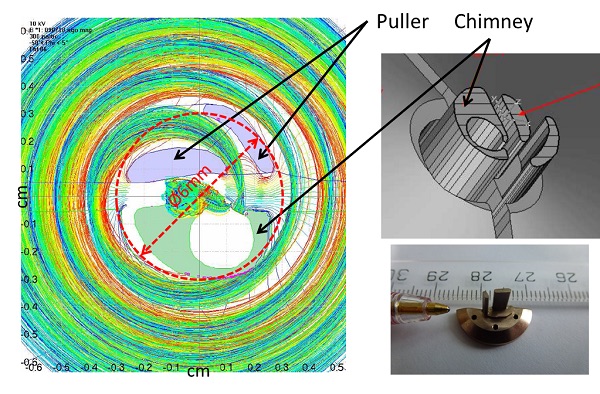}
\caption{Left: calculated orbits in the central region of the IBA S2C2. Right: 3D view of the Penning ion gauge source and puller used in this
central region.}
\label{fig:cregion3}
\end{center}
\end{figure}

The internal cold-cathode PIG ion source is also used in superconducting synchro-cyclotrons for proton therapy.
In such a cyclotron, the magnetic field in the centre is very high (5--$9\UT$) and the energy gain per turn is low, with
a dee voltage of about $10\UkV$. In such a case, the central region
needs to be very compact. This is illustrated in \Fref{fig:cregion3}, which shows the central region of the IBA S2C2.
The source diameter is <$5\Umm$ and the diameter of the first turn in the cyclotron is ${\approx}6\Umm$. The vertical dee gap
in the centre is only $6\Umm$. The first 100 turns are within a radius of only $30\Umm$.

\begin{figure}
\begin{center}
\includegraphics[width=11cm]{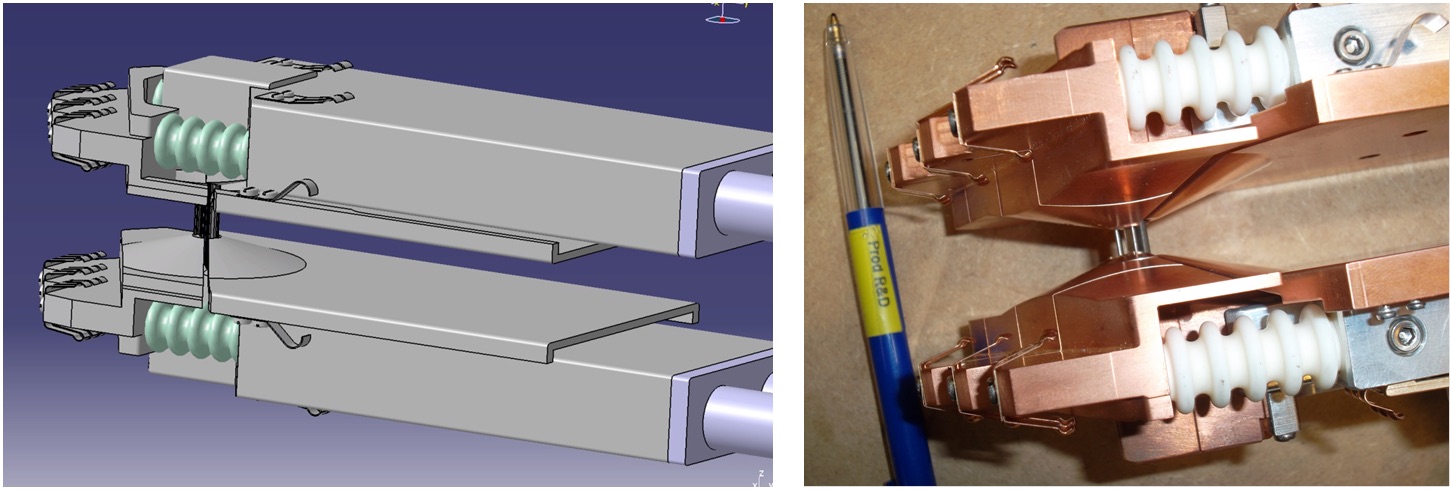}
\caption{The ion source and central region can be extracted from the cyclotron as one assembly, for easy mainten\-ance
and precise repositioning.}
\label{fig:cregion4}
\end{center}
\end{figure}

The precise position of the ion source in a synchro-cyclotron is of the utmost importance, in order to obtain very good
beam centring and the highest beam quality at the extraction.
The central region and the ion source of the S2C2 can be removed as one subsystem for easy maintenance and precise alignment and
realignment after reassembly (see \Fref{fig:cregion4}). To suppress the multipactor, both the dee and the counter
dee are biased at a DC voltage of $1\UkV$.

\begin{figure}
\begin{center}
\includegraphics[width=9cm]{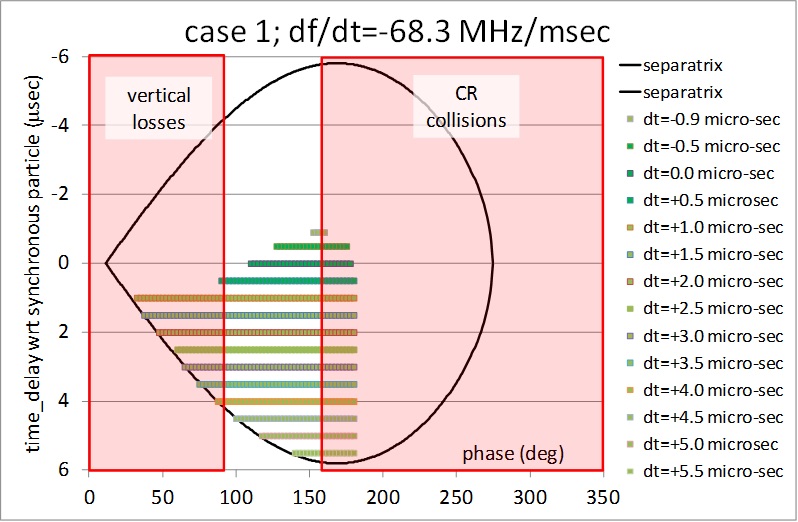}
\caption{Simulation of the beam capture process in the S2C2 central region (CR).}
\label{fig:capture1}
\end{center}
\end{figure}

As mentioned in \Sref{another}, in a synchro-cyclotron, there is only a short time in which the beam can be
captured into phase-stable orbits\cite{Bohm1947}.
Simulation of the beam capture requires a combined study of the orbit dynamics in the cyclotron
central region and the subsequent acceleration. Here, particles are started at the ion source at different time points and at
different RF phases. Only a subset of the started particles are captured. Particles outside of the acceptance window fall
out of synchronism with respect to the RF and are decelerated back towards the ion source, where they are lost.
Super\-imposed on that, there are the usual additional transverse (radial or vertical) losses due to the collisions with
the geometry of the central region. This is illustrated in \Fref{fig:capture1}.

\subsubsection{Burning paper}

\begin{figure}
\begin{center}
\includegraphics[width=7cm]{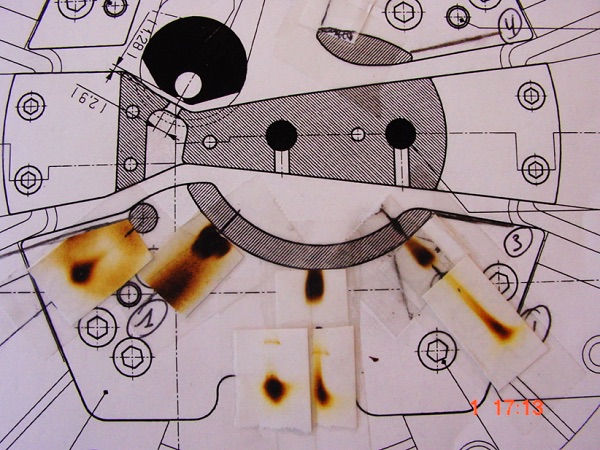}
\caption{Small bits of thin paper are placed in the pole gap and burned by the beam, to find the beam position
and size during the first few turns.}
\label{fig:burn}
\end{center}
\end{figure}

When a new central region has been designed and is being tested in the machine, it does not always immediately function correctly
and it may happen that the beam is lost after a few turns. Owing to space limitations, it is not always possible to have
a beam diagnostic probe that can reach the centre of the cyclotron and it may be difficult to find out why and where
the beam is lost. In such cases, it may help to put small bits of thin paper in the median plane; they will change colour, owing to the interaction with the beam. This is illustrated in \Fref{fig:burn} , which shows the central region layout of the IBA
self-extracting cyclotron. Seven bits of burned paper have been fixed on the central region design drawing.
In this way, the position of each turn and the corresponding beam sizes are nicely indicated.

\subsection{Cyclotrons with an external ion source}

In many cases, the ion source is placed outside the cyclotron. There may be different reasons for this choice:
(i) higher beam intensities are needed, which can only be produced in a more complex and larger ion source than the simple PIG source,
(ii) special ion species, such as heavy ions or highly stripped ions, are required, or
(iii) a good machine vacuum is needed (for example, $\EH^{-}$ acceleration).
External ion sources are used in high-intensity isotope production cyclotrons but, for example, also in the proposed
IBA C400 cyclotron for carbon therapy.
Of course, the external ion source is a more complex and more expensive solution, since it requires an injection line
with all related equipment such as magnetic or electrostatic beam guiding and focusing elements,
vacuum equipment, beam diagnostics, \etc

\subsubsection{Different methods of injection}

There are a few different ways to inject into a cyclotron.

\begin{enumerate}

\item Axial injection: this case is most relevant for small cyclotrons. The beam travels along the vertical symmetry axis
of the cyclotron towards the cyclotron centre. In the centre, the beam is bent through $90^{\circ}$ degrees from vertical
to horizontal into the median plane. This is achieved using an electrostatic or magnetostatic inflector.
\item Horizontal injection: the beam is travelling in the median plane from the outside towards the cyclo\-tron centre.
Generally speaking, this type of injection is more complicated than axial injection, owing to the vertical magnetic field
exerting a horizontal force on the beam, thereby trying to bend it in the horizontal plane.
It has been attempted to cancel this force with electrical forces from an electrode system installed
near the median plane\cite{Beurtly1965}.
It has also been attempted to tolerate this force and to let the beam make a spiral motion along the hill--valley pole edge
towards the cyclotron centre. This is called trochoidal injection and is illustrated in \Fref{fig:trochoidal}.
In the centre, an electrostatic deflector places the particle on the correct equilibrium orbit.
Both methods are very difficult and therefore are no longer used.

\begin{figure}
\begin{center}
\includegraphics[width=7.5cm]{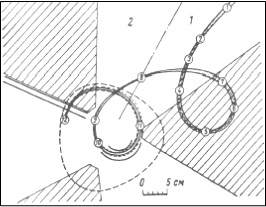}
\caption{Horizontal (trochoidal) injection. The beam travels along the hill--valley pole edge. In the centre, an
electrostatic device places the beam on the equilibrium orbit. Figure taken from \Bref{Heikkinen1994}}
\label{fig:trochoidal}
\end{center}
\end{figure}

\item Injection into a separate sector cyclotron: this must be qualified as a special case.
Much more space is available in the centre to accommodate magnetic bending and focusing devices.
Injection at much higher energies (in the megaelectronvolt range) is possible. This topic is considered as out of the scope of (small) medical accelerators.

\item Injection by stripping: a stripper foil positioned in the centre changes the particle charge state and its local
radius of curvature so that the particle aligns itself on the correct equilibrium orbit.
This method is mostly applied for separate sector cyclotrons, where the beam is injected horizontally.

\end{enumerate}

\subsubsection{Inflectors for axial injection}

The electrical field between two electrodes bends the beam $90^{\circ}$ from vertical to horizontal. The presence of the
cyclotron magnetic field creates a complicated 3D orbit; this makes the inflector design difficult.
Four different types of electrostatic inflectors are known.

\begin{figure}
\begin{center}
\includegraphics[width=12cm]{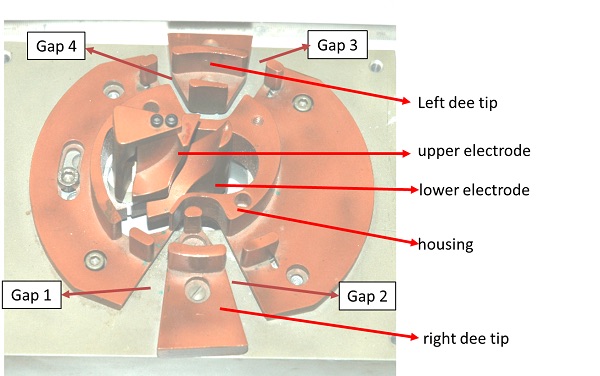}
\caption{1:1 model of the spiral inflector and central region of the IBA Cyclone 30 cyclotron}
\label{fig:inflector1}
\end{center}
\end{figure}

\begin{enumerate}

\item The mirror inflector: two planar electrodes are placed at $45^{\circ}$ with respect to the vertical beam direction.
In the upper electrode is an opening for beam entrance and exit (see \Fref{fig:mirror}).
The advantage of the mirror inflector is its relative simplicity. However, because the orbit is not following
an equipotential surface, a high electrode voltage (comparable to the injection voltage) is needed.
At the entrance, the particle is decelerated and at the exit it is re-accelerated. Furthermore, to obtain a reasonable electrical field distribution between the electrodes, a wire grid is needed across the entrance and exit
opening in the upper electrode. Such a grid is very vulnerable and is easily damaged by the beam.

\begin{figure}
\begin{center}
\includegraphics[width=15cm]{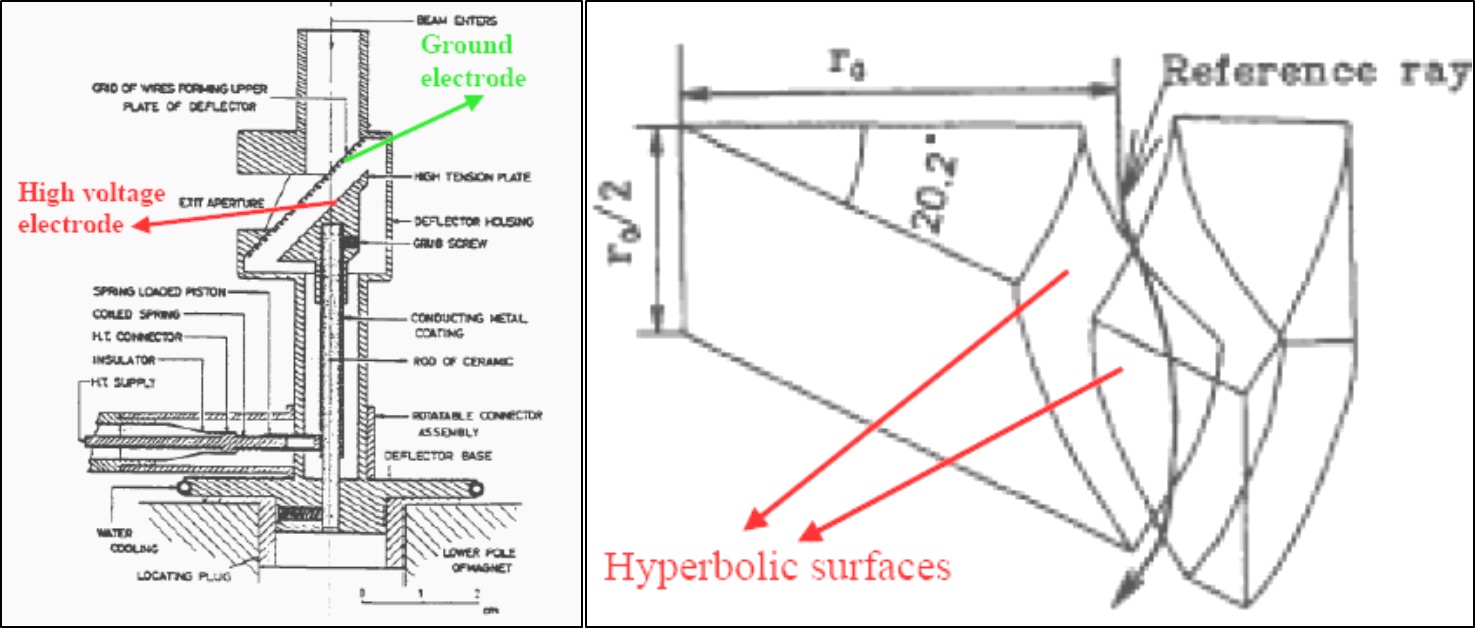}
\caption{Left: mirror inflector. Right: hyperboloid inflector. Figure taken from \Bref{Heikkinen1994}}
\label{fig:mirror}
\end{center}
\end{figure}

\item The spiral inflector: this is a cylindrical capacitor that is gradually twisted to take
into account the spiralling of the trajectory, induced by the vertical cyclotron magnetic field. The design is such
that the electrical field is always perpendicular to the velocity vector of the central particle and the orbit is therefore positioned on an equipotential surface. The electrode voltage can be much lower for a mirror inflector.
A simple formula for the electrode voltage is:

\begin{equation}
 V =2\cdot\frac{E}{q}\cdot\frac{d}{A}\ ,
\end{equation}

where $V$ is the electrode voltage, $E$ is the injection energy,  $q$ is the particle charge, $d$ is the electrode spacing,
and $A$ is the electric radius of the inflector (which is almost equal to the inflector height).
It can be seen that the ratio between electrode voltage and injection voltage is equal to twice the ratio of the electrode spacing
and the height of the inflector. An important advantage of the spiral inflector is that it has two free design parameters
that can be used to place the particle on the correct equilibrium orbit. These two parameters are the electrical radius, $A$,
and the so-called tilt parameter, $k^\prime$. This second parameter represents a gradual rotation of the electrodes around
the particle moving direction by which a horizontal electric field component is obtained that is proportional
to the horizontal velocity component of the particle. Varying the tilt parameter , $k^\prime$,  is, therefore, equivalent
(as far as the central trajectory is concerned) to varying the cyclotron magnetic field in the inflector volume.
Another advantage of the spiral inflector is its compactness. However, the electrode surfaces are complicated
3D structures, which are difficult to machine. Fortunately, with the wide availability of computer
controlled five-axis milling machines, this is not really a problem anymore. \Figure[b]~\ref{fig:inflector1} shows a 1:1 model of
the spiral inflector used in the IBA C30 cyclotron.

\item The hyperboloid inflector: the electrodes are hyperboloids with rotational symmetry around the vertical $z$-axis
(see \Fref{fig:mirror}).
As for the spiral inflector, the electrical field is perpendicular to the particle velocity and a relatively low electrode voltage
can be used. However, for this inflector, no free design parameters are available. For given particle charge and mass,
injection energy and magnetic field, the electrode geometry is fixed and it is more difficult to inject the particle
in the correct equilibrium orbit. Furthermore, this inflector is quite large compared with the spiral inflector.
However, owing to the rotational symmetry, it is easier to machine.

\item The parabolic inflector: the electrodes are formed by bending sheet metal plates into a parabolic shape.
This inflector has the same advantages and disadvantages as the hyperboloid inflector: relatively low voltage
and ease of construction, but no free design parameters and relatively large geometry.
\end{enumerate}

\subsubsection{Example: axial injection in the IBA C30HC high-intensity cyclotron}

Nowadays, the spiral inflector is almost always used for axial injection.
Analytical formulae exist for central orbits in a spiral inflector placed in a homogeneous
magnetic field\cite{Belmont1966,Belmont1986,Root1972,Baartman1993}.
However, the field in the cyclotron centre is certainly not uniform, owing to the axial hole needed for axial injection.
In practice, the inflector design requires extensive numerical effort, which can be broken down into three main parts:
(1) 3D modelling of the electrical fields of the inflector and central region,
(2) 3D modelling of the magnetic field in the central region, and
(3) orbit tracking in the central region.
The complete process is tedious and requires many iterations.
First, the central trajectory has to be defined and optimized.
There are three main requirements, namely that the injected orbit is nicely on the equilibrium orbit,
correctly placed in the median plane and well centred with respect to the inflector electrodes.
After an acceptable electrode geometry has been obtained, for which these requirements are fulfilled,
the beam optics must be studied. Here the main requirement is that reasonable matching into the cyclotron
eigenellipse can be achieved, so that large emittance growth in the cyclotron  is avoided\cite{Kleeven1993}.

\begin{figure}
\begin{center}
\includegraphics[width=\textwidth]{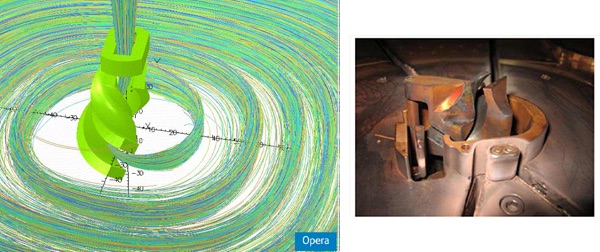}
\caption{Left: axially injected 3D orbits calculated with the cyclotron tracking code AOC are imported in the Opera-3d
finite-element model of the IBA C30XP cyclotron. Right: central region of the IBA C70XP cyclotron, showing the inflector
with an additional electrode at its exit that provides a radial kick, which is needed to centre particles with different $q/m$ ratios
on their respective equilibrium orbits.}
\label{fig:inflector2}

\end{center}
\end{figure}

It may be necessary to calculate several inflectors of different height, $A,$ and tilt parameter, $k^\prime$, to optimize
this process. At IBA, both the 3D magnetic field computations as well as the 3D electrical field computations are done
using the Opera-3d software package from Vector Fields\cite{Cobham2015}. Often, the models are completely parameterized, for
quick modification and optimization. \Figure[b]~\ref{fig:inflector2}
shows such a model of the central region and the inflector.
The inflector model uses the following parameters:  the electrode width and spacing (both may vary along the inflector),
the tilt parameter, $k^\prime$, and the shape of the central trajectory itself, in terms of a list of points
and velocity vectors.

\begin{figure}
\begin{center}
\includegraphics[width=11.5cm]{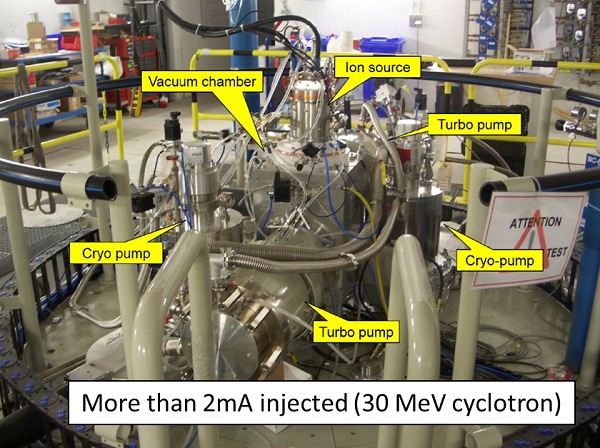}
\caption{ C30HC injection line installed on top of the cyclotron. The ion source is on top of the vacuum box.
This box is pumped by two turbo pumps (front and back).}
\label{fig:injax1}
\end{center}
\end{figure}

Recently, the IBA C30 has been upgraded to a new high-current version (C30HC)\cite{Kleeven2011}.
For this purpose, a new ion source, a new injection line, and a new central region have been installed.
A new final amplifier provides $100\UkW$ of RF power as needed for beam acceleration.
The new source, as shown in \Fref{fig:injax1}, is the D-pace DC volume cusp source, providing $15\UmA$ of $\EH^-$ beam
within a 4 rms beam emittance of $110\pi\Umm\, \UmradZ$ at an injection energy of $30\UkeV$\cite{Denhel}.
This performance approximately doubles the injected beam current as compared with the standard C30 ion source.

\begin{figure}
\begin{center}
\includegraphics[width=\textwidth]{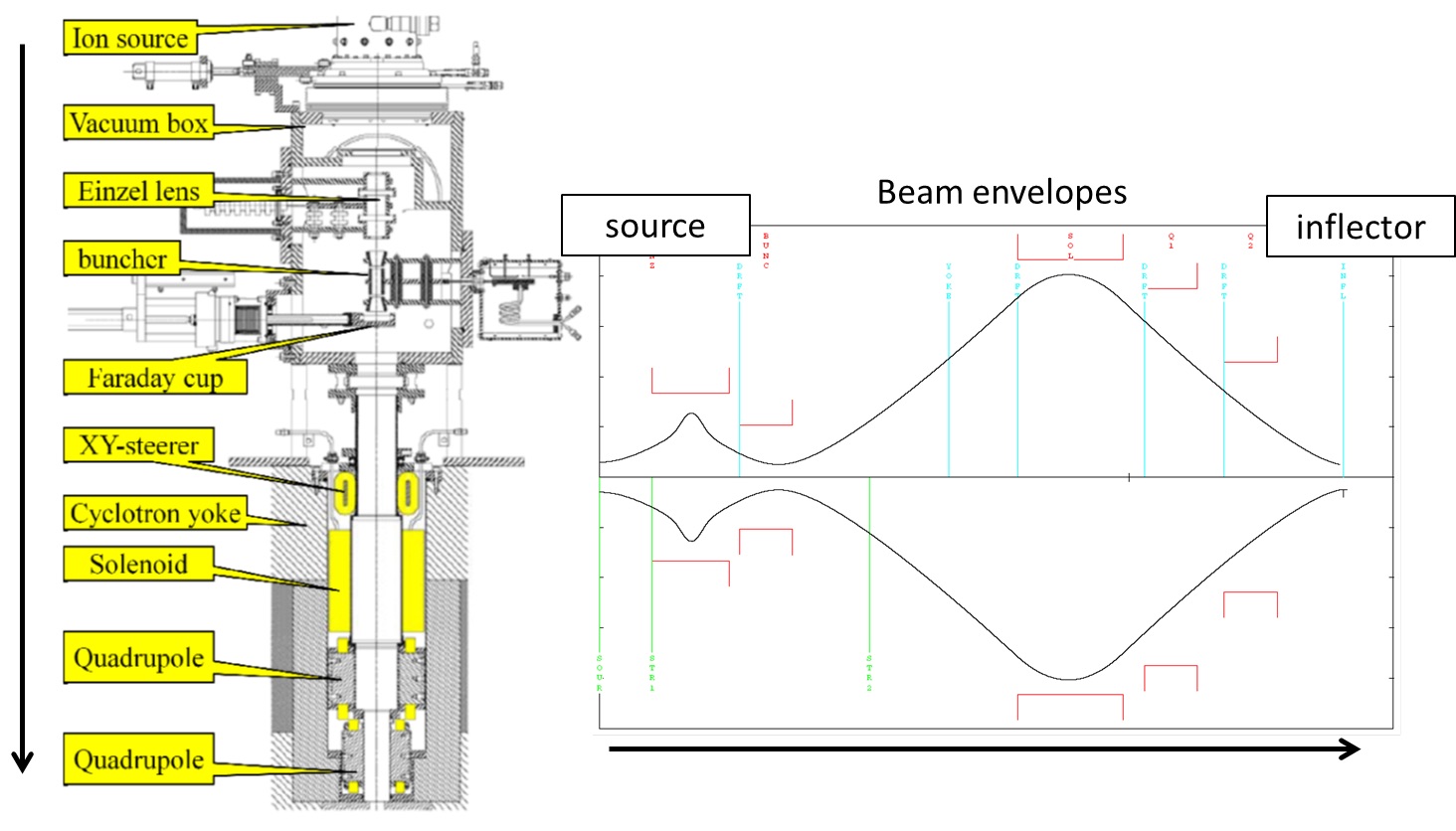}
\caption{Left: layout of the C30HC injection line. Right: beam envelopes in this beam line, calculated with TRANSPORT.}
\label{fig:injax2}
\end{center}
\end{figure}

The new injection line layout is shown in \Fref{fig:injax2}. The ion source is mounted on top of the
vacuum box, which contains an Einzel lens, a buncher, and a Faraday cup.
The axial bore of the cyclotron contains a solenoid, two small quadrupoles, and an $xy$-steering magnet.
With this steering magnet and a second pair at the exit of the ion source, the beam at the inflector can be adjusted
to the correct position and direction. The design has been optimized to maximize $\EH^-$ beam injection.
The beam line is compact (short) to minimize stripping on the residual gas.
Differential pumping is applied with a first turbo pump, which acts directly below the ion source, and a second turbo pump,
which pumps the second separated part of the vacuum box and the downstream part of the beam line.
Both pumps are magnetically shielded from the cyclotron stray field.
The elements in the cyclotron bore
are contained at atmospheric pressure around the beam transport tube. This reduces outgassing.
The cyclotron bore diameter has been increased to allow sufficient space for these elements.
Opera-3d calculations have confirmed that this does not compromise the magnetic field in the median plane.
The cyclotron iron is used as return yoke for the solenoid; the quadrupoles have their own return yoke.
The beam is focused by the Einzel lens to a small size at the centre of the buncher, as shown in the calculation
of the beam envelopes along the beam line (right panel of \Fref{fig:injax2}). In this way, the spread in
transit time factor due to the finite beam size is minimized and the bunching efficiency is maximized\cite{Kleeven1992b}.
Finally, the beam envelope naturally increases at buncher exit, to a maximum value in the solenoid, as
permitted by the solenoid bore. This enables the smallest possible beam size to be obtained at the inflector
entrance, through the focusing action of the solenoid. The relative positions of the Einzel lens,
buncher, and solenoid have been optimized to obtain this condition. The two quadrupoles enable the shape of the beam to be adjusted asymmetrically so that better matching is obtained with respect to the
cyclotron acceptance ellipse\cite{Kleeven1993}. For the same reason, the two quadrupoles can be rotated around their axis.

The initial central region was designed almost 30 years ago, when advanced 3D programs for electromagnetic modelling
were not yet available. With the better design tools available today, it could be improved and a new inflector
and central region was designed. Optimization was conducted for good beam centring in the inflector and in the cyclotron,
good vertical focusing, and large beam capture efficiency. The following stepwise approach was used.

\begin{enumerate}

\item Construct precise Opera-3d models of the cyclotron magnet and the accelerating structure and obtain the 3D magnetic field
in and around the inflector volume.
\item Optimize the azimuthal opening angle of the dees in the centre to maximize the capture efficiency.
\item Determine the accelerated equilibrium orbit by backtracking from high energy towards the centre.
\item Choose an injection point on the accelerated equilibrium orbit, allowing a good position of the inflector relative to the first dee gap.
\item Obtain an initial estimate of the inflector central orbit, passing through the injection point.
A special tracking mode is used, simulating the inflector by an electric field that is always perpendicular to the orbit.
\item Construct the real 3D inflector electrodes around the estimated reference orbit using Opera-3d.
\item Verify orbit centring with respect to the inflector electrodes and modify the fringe field parameters if needed.
\item Verify the design by orbit tracking in the real 3D fields and modify the inflector position slight, if necessary, for beam horizontal or vertical centring.

\end{enumerate}

An inflector bending radius of $A=29.5\Umm$ was chosen. For this case, a tilt of $k^\prime = -1.0$ was needed to centre the beam.
The inflector gap is $8\Umm$ and the aspect ratio equals 2. \Figure[b]~\ref{fig:inflector2} shows the 3D model of the inflector, developed in Opera-3d.
With this new central region and injection line design, a beam current of $>2\UmA$ was obtained at the $1\UMeV$ beam stop in the cyclotron centre.

\section{Extraction from cyclotrons}

Different solutions for beam extraction are treated. These include extraction by stripping, resonant extraction using a deflector,
and the regenerative extraction used in synchro-cyclotrons. The different methods of creating a turn separation are explained.
The purpose of different types of extraction device, such as harmonic coils, deflectors, and gradient corrector channels, are outlined.
Several illustrations are given, in the form of photographs and drawings.

The topics of cyclotron extraction have already been covered in earlier CAS proceedings in the framework of the
general accelerator physics course\cite{Heikkinen1994},  as well as in the framework of specialized courses\cite{Botman1996,Kleeven2005}.
Since then, not so many substantial changes have occurred in the field, especially if one only considers small cyclotrons
that are used for applications. For this reason, it was decided to choose an approach where the accent is less on
completeness and rigorousness (because this has already been done) but more on explaining and illustrating
the main principles that are used in small cyclotrons.
Sometimes a more industrial viewpoint is taken. The use of complicated formulae is avoided as much as possible.

Extraction is the process of beam transfer from an internal orbit to a target that is placed outside of the magnetic field.
There are three main reasons why extraction is considered as difficult.

\begin{enumerate}
\item The magnetic field itself behaves as a kind of trap. When the particles are accelerated into the falling
fringe field they will run out of phase with respect to the RF wave; if the phase slip is more than 90$^\circ$,
they will be decelerated instead and move inward. This may be considered as a kind of reflection of the beam on
the radial pole edge of the magnet.
\item In a cyclotron, the turn separation is inversely proportional to the radius ($R \approx \sqrt{E}$). Because of this, the turns pile up
closely together near the extraction radius. Therefore, it is difficult to deflect the last orbit,
without influencing the inner orbits and without important beam losses.
\item  During extraction, the beam has to cross the fringe field. This is an area where there are very large gradients
and non-linearities in the magnetic field. Special precautions have to be made to avoid substantial beam losses,
beam blow-up, or loss of beam quality.
\end{enumerate}

There are a few different ways to solve the problem of extraction.

\begin{enumerate}
\item No extraction at all: avoid the problem by using an internal target. This can be done for isotope production, but is a little
bit dirty.
\item Extraction by stripping ($\EH^-$ or $\EH_2^+$ cyclotrons): this is often applied in isotope production cyclo\-trons.
\item Use one (or more) electrostatic deflectors (ESDs) that peel off the last orbit: this is, for example, done in the proton
therapy cyclotrons of Varian, IBA, and SHI.
\item Regenerative extraction, as used in synchro-cyclotrons: this is, for example, done in the proton therapy synchro-cyclotrons of
Mevion and IBA.
\item Self-extraction: this requires a suitable and precise shaping of the magnetic field. IBA has made one such prototype
cyclotron for isotope production.
\end{enumerate}

\noindent Cases (3) and (4) require some method to increase the turn separation between the last and penultimate orbits.
The different methods will be discussed in some more detail next.

\subsection{The use of an internal target}

\begin{figure}
\begin{center}
\includegraphics[width=14cm]{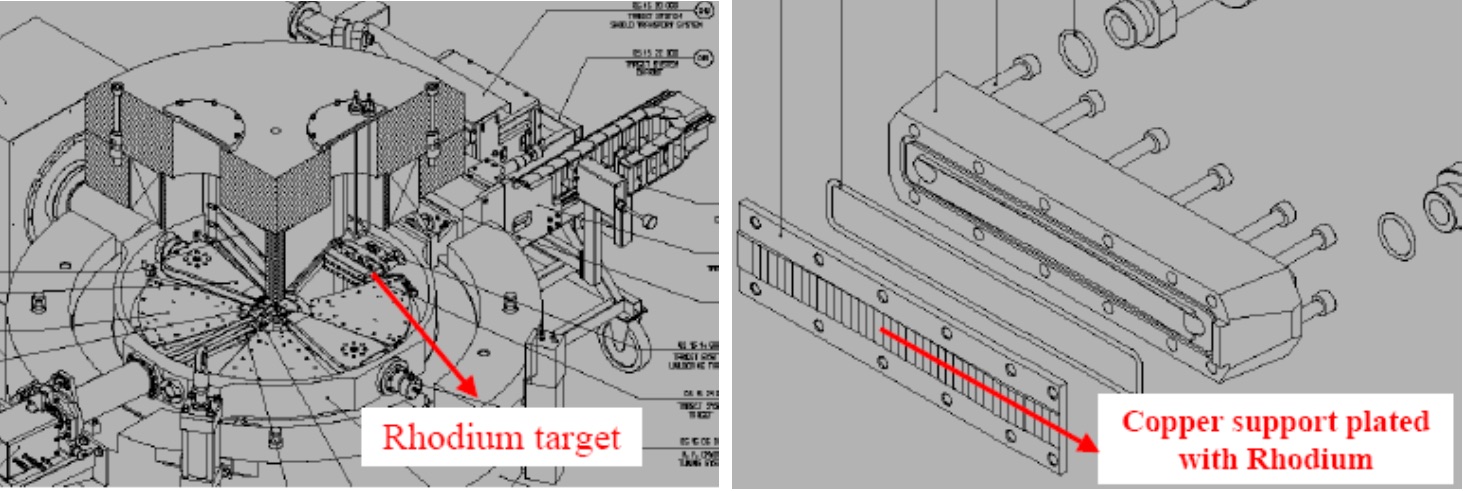}
\caption{Left: IBA C18+ cyclotron with an internal rhodium target for the production of palladium.
Right: detail of an internal target showing the profiling of the target surface, optimized
to maximize the beam spot and minimize beam reflection on the target.}
\label{fig:internal}
\end{center}
\end{figure}

The target is placed between the poles of the cyclotron at a radius where the field is still isochronous.
The method was applied quite frequently for the production of radioisotopes, such as $^{103}\EPd$ or $^{201}\ETl$.
The method is relatively simple and also non-expensive.
The energy can be selected by choosing the correct radius of the target in the cyclotron.
However, it is a rather dirty way of working because of radioactive contamination of the cyclotron.
If the target is perpendicular to the beam, the beam spot is very small and local heating will pose a problem.
To avoid this, the target is placed at a small grazing angle with respect to the beam.
In this case, however, a certain fraction of the incoming beam will be reflected from the target surface, as a result of multiple scattering.
This in turn will activate the cyclotron.
The right panel of \Fref{fig:internal} shows an IBA C18+ cyclotron with an internal rhodium target, which is used for the
production of $^{103}\EPd$.
The target can be handled fully remotely. The $^{103}\EPd$ was used for brachytherapy (in the treatment of prostate cancer).
Sixteen of these machines have been sold to one customer.
The left panel of \Fref{fig:internal} shows some detail of the internal Rh/Pd target.
The target is heavily water-cooled, so that it can take a beam power of $30\UkW$ ($2\UmA$/$15\UMeV$).
The target surface is profiled, which is optimized to maximize the beam spot and minimize the beam reflection.

More strict legal rules concerning the prevention of radioactive contamination will certainly elimin\-ate
this method in future.

\subsection{Stripping extraction}

\begin{figure}
\begin{center}
\includegraphics[width=12cm]{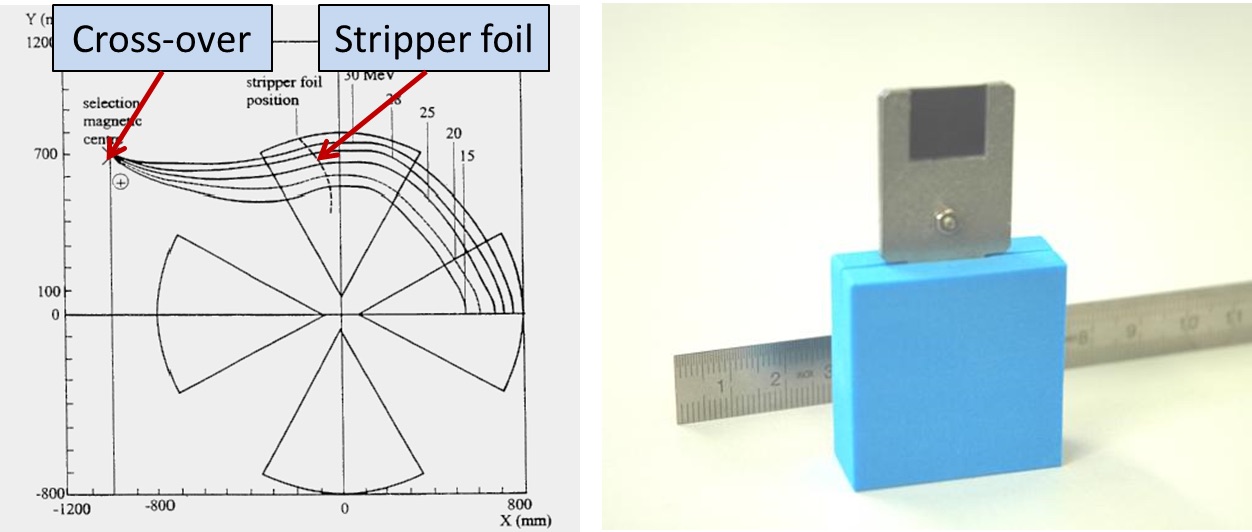}
\caption{Left: $\EH^{-}$ extraction by stripping. Energy is selected by moving the stripper foil to the correct radius.
Right: a simple carbon stripper foil is used to extract the beam (typically 50--$200\Uug/\UcmZ^2$).}
\label{fig:stripping1}
\end{center}
\end{figure}

To extract the beam, the particles pass a thin stripper foil (see the right panel of \Fref{fig:stripping1}),
by which one or more electrons are removed from the ion.
Because of this, there is an instantaneous change of the orbit local radius of curvature.
The relation between the local radius before ($\rho_\mathrm{i}$) and after ($\rho_\mathrm{f}$) stripping is given by:

\begin{equation}
\rho_\mathrm{f}=\frac{Z_\mathrm{i}}{Z_\mathrm{f}}\frac{M_\mathrm{f}}{M_\mathrm{i}}\rho_\mathrm{i}\ ,
\end{equation}

\noindent where $M_\mathrm{i}$ and  $M_\mathrm{f}$ are the particle mass before and after stripping, respectively.
As an example, for $\EH^{-}$, we have $\EH^-$ $\Rightarrow$  $\EH^+ + 2 \mathrm{e}^-$
and the local radius of curvature practically only changes sign ($\rho_\mathrm{f}=-\rho_\mathrm{i}$) because $M_{\EH^{+}} \simeq M_{\EH^{-}}$.
As a result, the stripped particle is immediately deflected outward,
away from the cyclotron centre. This is illustrated in the left panel of \Fref{fig:stripping1}.

\begin{figure}
\begin{center}
\includegraphics[width=12cm]{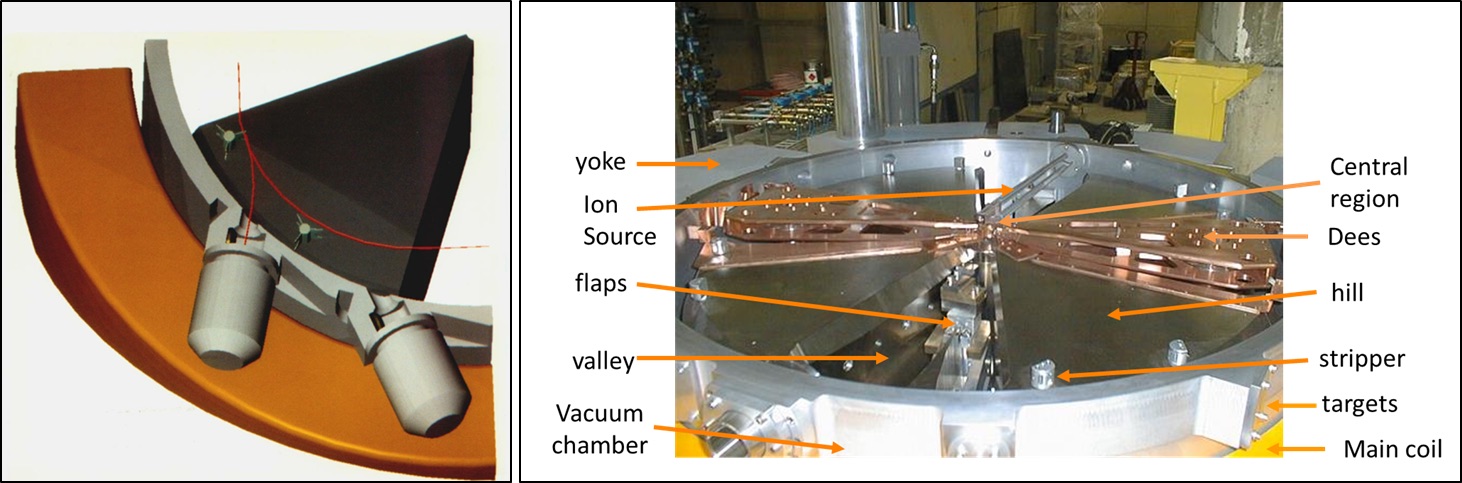}
\caption{Left: $\EH^{-}$ extraction by stripping. Several targets can be placed around the machine.
Right: median plane of a typical IBA isotope production cyclotron, showing the most important subsystems.}
\label{fig:stripping2}
\end{center}
\end{figure}

Multiple targets can be placed around the machine. This is illustrated in the left panel of \Fref{fig:stripping2}.
A given target is selected by rotating the corresponding stripper foil
into the beam.  $\EH^{-}$ extraction is applied in many commercial isotope production cyclotrons, fabricated,
for example, by IBA (Cyclone 30, C18/9, C10/5), Advanced Cyclotron Systems (TR30, TR13), or General Electric (PETtrace).
The right panel of \Fref{fig:stripping2} shows a view on the median plane of a typical IBA isotope production cyclotron. In
this cyclotron, eight different target ports are available.

The most important features and advantages of stripping extraction are:

\begin{enumerate}
\item very simple extraction device;
\item 100\% extraction efficiency;
\item variable energy;
\item the ability to place several targets around the machine;
\item simultaneous dual beam extraction;
\item good beam optics.
\end{enumerate}

The energy can easily be varied by moving the radial position of the stripper probe (see \Fref{fig:stripping1}).
By proper azimuthal positioning of the stripper foil, all orbits come together in the same cross-over point outside
the magnetic field. Here, a combination magnet can be placed that deflects the beam into the beam line.
Simultaneous dual beam operation is made possible by positioning two stripper foils at an azimuth of 180$^\circ$ with respect
to each other. Some fine tuning of the turn-pattern is necessary to distribute the total beam current
precisely between the two stripping foils. This may be achieved by fine adjustment of the dee voltage, or by using first-harmonic coils.
Since the extracted beam crosses the radial pole edge at an angle that is close to $90^{\circ}$,
the large (de-)focusing effects of the fringe field are avoided and the beam quality remains intact. $\EH^-$ stripping is an
ideal solution for low- and medium-energy industrial cyclotrons.

There is a serious limitation of an $\EH^{-}$ cyclotron, owing to magnetic stripping that may occur during acceleration.
Because of this, the magnetic field cannot be high and to obtain high energy the pole  radius of the machine must be increased.
A well-known example is the TRIUMF cyclotron\cite{Dutto1993}, which accelerates $\EH^{-}$ up to $520\UMeV$.
The average magnetic field is only $0.3\UT$ (in the cyclotron centre), resulting in a magnet diameter of $18\Um$
and a magnet weight of 4000 tonnes. The $\EH^{-}$ is also stripped on the vacuum rest gas and, to limit
 beam losses, good vacuum pumping (expensive)  and an external ion source is required (IBA Cyclone-30, ACS TR30).
The $\EH^-$-cyclotron is good for isotope production, but not for proton therapy.

Another example is the acceleration of molecular hydrogen $\EH_2^+$  and related extraction by stripping:
$\EH_2^+$ $\Rightarrow$ 2$\EH^+ + \mathrm{e}^-$. In this case, the radius of curvature does not change sign but is practically divided by
two ($\rho_\mathrm{f} = \rho_\mathrm{i}/2$). In this case, extraction is more difficult, because the beam initially remains in the machine,
since the particle is deflected inward immediately after stripping.
The method, therefore, only works when the flutter is large enough.
Note that in this case, the $K$-value describing the bending power of the cyclotron magnet\cite{Livingood1961} must be
four times higher than $K$-value for the cyclotron accelerating $\EH^{-}$ ions to the same energy.

\subsection{Extraction by means of electrostatic deflector devices}

\subsubsection{Turn separation in a cyclotron}

Some qualitative aspects of orbit separation are explained, to illustrate the general effects. In a cyclotron, the
position of a particle with a given energy is determined by a betatron oscillation relative to the equilibrium orbit:

\begin{equation}
r(\theta) =r_0(\theta) + x(\theta)\sin(\nu_r\theta+\theta_0)\ . \label{eq:betatronosc}
\end{equation}

\noindent Here $r$ and $\theta$  are the polar coordinates of the particle, $r_0$ is the equilibrium orbit
for a given energy, $x$ is the amplitude of
the betatron oscillation, $\nu_r$ is the radial betatron oscillation frequency, and $\theta_0$ is an arbitrary offset angle.
The equilibrium orbit can be ideally centred and shaped (in the case of a perfectly symmetric magnetic field),
or it can be displaced with respect to the centre of the cyclotron (when there is a first-harmonic field
perturbation in the field). The betatron oscillation is quasi-sinusoidal but the oscillation amplitude may slightly
depend on $\theta$, owing to the AVF characteristic of the magnetic field. For the present purpose, this effect is not important.
\Eq[b]~\ref{eq:betatronosc} describes the oscillation of a single particle. However, it can also be used to describe a coherent
oscillation of the centre of the beam. The latter case is relevant for the study of turn separation.
We can evaluate the radius $r(\theta)$ at a fixed azimuth $\theta_n$ but for successive turns $n$.
It is easily derived that in this case\cite{Botman1996}:

\begin{equation}\label{eq:tsep}
\begin{split}
\Delta r(\theta_n) = \Delta r_0(\theta_n) +& \Delta x\sin(2\pi n(\nu_r-1)+\theta_0)\\
                                          +& 2\pi(\nu_r-1)x\cos(2\pi n(\nu_r-1)+\theta_0)\ ,
\end{split}
\end{equation}

\noindent where $\Delta r$ is the radial increase between two successive turns. In this equation, there are three different terms.
They relate to three different methods that can be used to generate a turn separation. The first term, $\Delta r_0$, represents
an increase in the radius of the equilibrium orbit and is related to the energy increase $\Delta E_0$ per turn:

\begin{equation}
\frac{\Delta r_0}{r_0} \approx \frac{1}{2} \frac{\Delta E_0}{E_0}\ ,
\end{equation}

\noindent where $E_0$  is the kinetic energy at the radius $r_0$. Thus, the relative radial increase is only half the
relative energy increase.  However, for a given cyclotron, the turn separation $\Delta r_0$ will double when the dee voltage is doubled.

\begin{figure}
\begin{center}
\includegraphics[width=12cm]{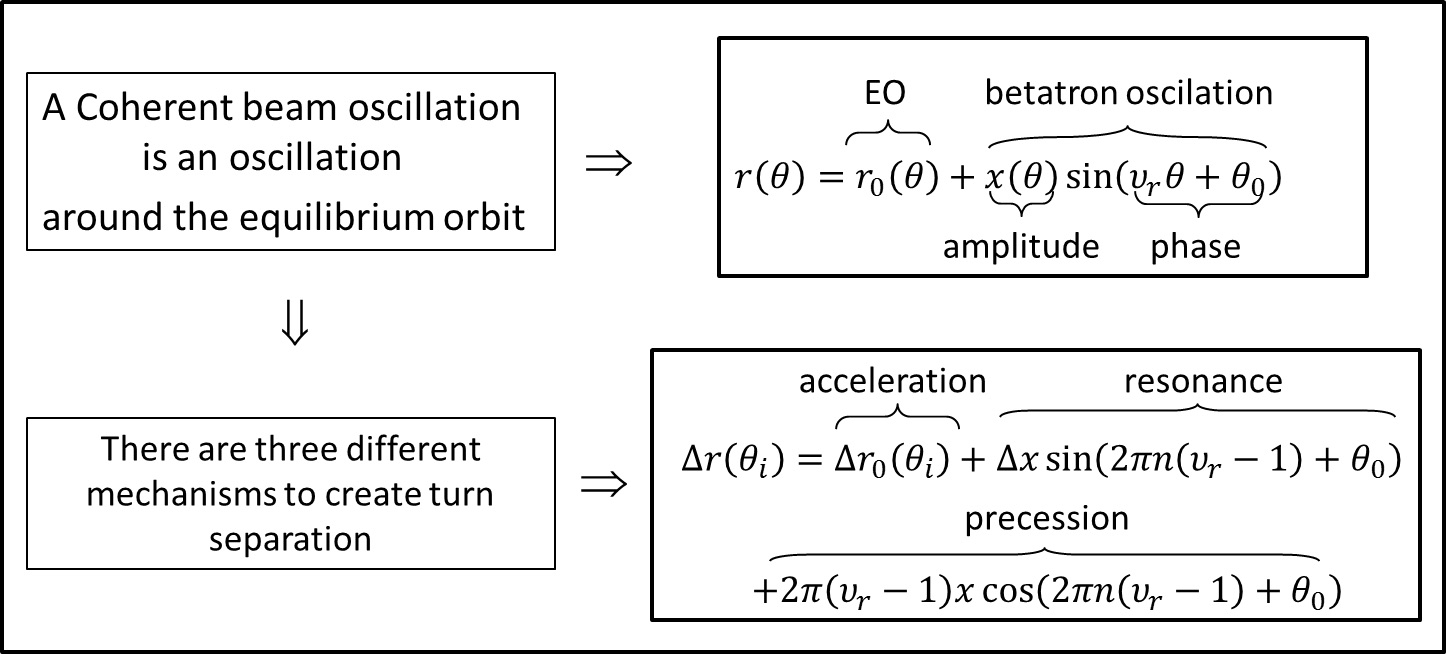}
\caption{Mechanisms that may contribute to an increase of the turn separation in a cyclotron}
\label{fig:turnsep}
\end{center}
\end{figure}

The second term in \Eref{eq:tsep} relates to a turn separation due to an increase in the betatron oscil\-lation
amplitude between two successive turns. This is, in general, what happens in a resonance.
The resonances that are important for extraction from a (small) cyclotron are the $\nu_r=1$ resonance and the $2\nu_r=2$ resonance.
The $\nu_r=1$ resonance is a first-order integer resonance (displacement of the beam) and is driven by a first-harmonic dipole bump.
The $2\nu_r=2$ resonance is a second-order half-integer resonance (exponential growth in the stop band) and is driven by
a second-harmonic gradient (quadrupole) bump.
The third term in \Eref{eq:tsep} describes the case where a coherent amplitude already exists,
but the turn separation arises from the fact that the phase of the oscillation has advanced between two successive turns.
The different mechanisms that create turn separation are illustrated in \Fref{fig:turnsep}.

\subsubsection{Different extraction methods relying on turn separation}

The following classification\cite{Joho1971} has been made for the different methods of extraction that rely on
an increase in the turn separation.

\begin{enumerate}
\item Extraction by acceleration:  no means other than acceleration is used to increase the turn separation. This requires
a sufficiently high dee voltage and is done, for example, in the IBA/SHI C235 proton therapy cyclotron.

\item Brute force extraction: the beam is extracted during the build-up of the resonance. Owing to this, there is an
increase in the amplitude of coherent oscillation in between the last and the penultimate turn.
This oscillation is created on or close to the $\nu_r=1$ resonance by applying a first-harmonic dipole bump. At the same time,
the acceleration gives an additional important contribution to the turn separation.
The IBA self-extracting cyclotron\cite{Kleeven2003} can probably be best classified in this group.

\item Resonant extraction: here, some coherent beam oscillation is created just before extraction. Two different schemes
can be distinguished.
\begin{enumerate}

\item Precessional extraction: this is more subtle\cite{Hagedoorn1966,Hagedoorn1963,Gordon1963}.
Here, a coherent oscillation is also created with
a first-harmonic dipole bump on the $\nu_r=1$ resonance (or alternatively the oscil\-lation may be
created by off-centring the beam at the injection). However, after passing this resonance,
the beam is further accelerated into the fringe field of the cyclotron. Here, the value of $\nu_r$ drops below 1
(typically to 0.6--0.8). As a result, the betatron phase advance between the last two turns is substantially
different from $360^{\circ}$ and a turn separation is obtained, which is proportional to the oscillation amplitude
and to $\nu_r-1$. The number of turns in the fringe field should not be too large, to avoid a
too large RF phase slip. This extraction method is used in the Varian SC cyclotron for proton therapy.

\item Regenerative extraction: this is also subtle. Here the beam is also extracted during resonance build up.
The $2\nu_r=2$ resonance is used. This resonance is driven by a second-harmonic gradient bump of the field.
This means that the second-harmonic bump should show as a function of radius a quadrupole-like dependence (linear increase with radius).
This shape is more critical than for the previous methods. If the gradient is large enough, then the resonance
will lock the real part of $\nu_r$ to a value of one. There is also an imaginary part of $\nu_r$ that will cause exponential growth
of the betatron oscillation.
Regenerative extraction is used in modern superconducting synchro-cyclotrons, such as the IBA S2C2 and the Mevion Monarch.
\end{enumerate}

\end{enumerate}

\subsubsection{General layout for resonant extraction}

The extraction process for resonant extraction using an electrostatic deflector can generally be subdivided into four steps.

\begin{enumerate}
\item Use harmonic coils (placed at a radius where $\nu_r=1$) to push the beam, create a coherent oscil\-lation,
and create the required turn separation. Instead of harmonic coils, the field bump may also be created by properly placed
and shaped iron bars. Then obtain turn separation by precession.
\item Use an electrostatic deflector to peel off the last turn and provide an initial radial kick to the beam.
\item Use gradient correctors or focusing channels to guide the beam through the cyclotron fringe field. The primary goal is
to reduce the magnetic field locally and to control the field gradients in order to avoid too much optical damage to the beam.
\item Place external focusing elements (quadrupole doublets) as close as possible to the radial pole edge
(if necessary, in the cyclotron return yoke), to handle the large beam divergences in the extracted beam.
\end{enumerate}

\subsubsection{Harmonic extraction coil}

\begin{figure}
\begin{center}
\includegraphics[width=6cm]{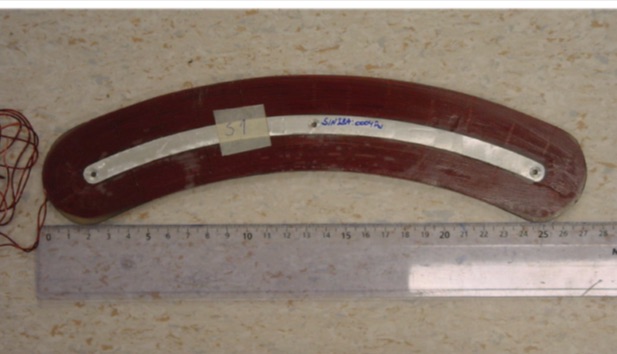}
\caption{Harmonic coil used in the IBA self-extracting cyclotron (note that the ruler measures centimeters)}
\label{fig:hcoil}
\end{center}
\end{figure}

A harmonic extraction coil  may be placed at a radius where $\nu_r$ is close to one to create a coherent beam oscillation.
The principle is simple: if $\nu_r=1$, the radial kicks that are given to the beam are all in phase and the oscillation
amplitude increases linearly with turn number. In reality, this may be more difficult, however. The coil covers a certain radial width
(or better, a certain energy range) and $\nu_r$ will not be equal to one over the full range. Depending on the number of turns
that are seen by the harmonic coil, a situation may very quickly occur where the oscillation amplitude that was already created
is lost again because new radial kicks are out of phase with the oscillation built up so far. This occurs when the
beam follows the shifted equilibrium orbit adiabatically (too slow acceleration).
To avoid this situation, it is important to limit the radial width of the coil. If the shape of the coil follows the
local shape of the equilibrium orbit, the particle energy range covered by the coil is small and a non-adiabatic effect may be
obtained.
The coil should be placed in the pole gap, to use the field amplification produced by the iron.
However, this may complicate the design, owing to vertical space limitations.  Also, the heat load will be important
because water cooling is often not possible.  \Figure[b]~\ref{fig:hcoil} shows a photograph of the harmonic coil that is used in
the IBA self-extracting cyclotron.

\subsubsection{Electrostatic deflector}

\begin{figure}
\begin{center}
\includegraphics[width=7cm]{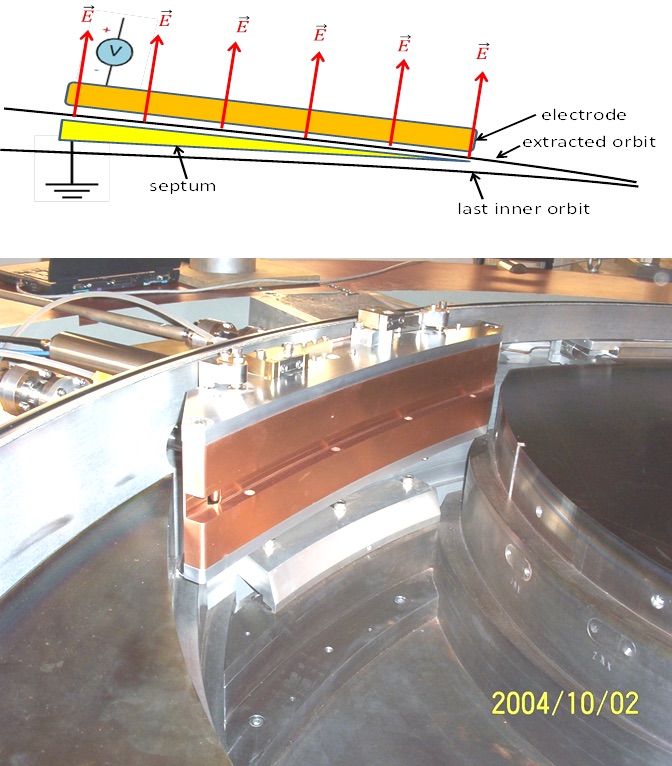}
\caption{Top: principle function of an electrostatic deflector. Bottom:
 electrostatic deflector installed in the IBA C235 cyclotron.}
\label{fig:esd1}
\end{center}
\end{figure}

\begin{figure}
\begin{center}
\includegraphics[width=10cm]{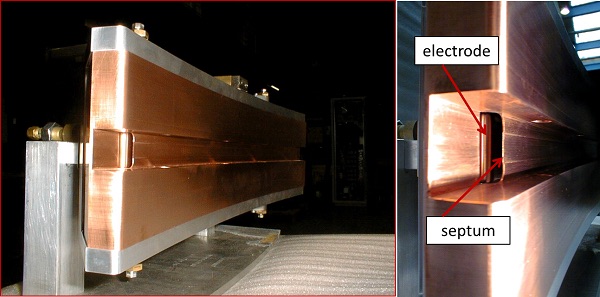}
\caption{Close-up of C235 ESD, showing the septum and the electrode}
\label{fig:esd2}
\end{center}
\end{figure}

The ESD creates an outwardly directed DC electric field between two electrodes.
The goal is to give an initial angular deflection to the beam. The inner electrode (called the septum) is placed in
between the last and penultimate turn. The septum is at ground potential so that the inner orbits in the cyclotron are not affected.
At the entrance, the septum is knife-thin (of the order of $0.1\Umm$) in order to peel off the last orbit and
minimize beam losses on the septum itself. To better distribute the heat due to beam losses, the beginning of
the septum is often V-shaped. The septum is water-cooled. The heat loss on the septum usually determines the maximum
current that can be extracted from the cyclotron. The outer electrode is on a negative potential (assuming extraction of
positively charged particles). Of course, the shape of the electrodes must follow the shape of the extracted orbit.
\Figure[b]~\ref{fig:esd1} illustrates the principle and shows the electrostatic deflector that
is used in the IBA C235 cyclotron. \Figure[b]~\ref{fig:esd2} shows the septum and the electrode.

\subsubsection{A gradient corrector channel}

The goal of a gradient corrector channel is to guide the beam through the fringe field,
to reduce the magnetic field on the extraction path, and to reduce the
vertical or increase the radial focusing through the fringe field. Often, more than one magnetic channel is
needed along the extraction path.  Different types are used:

\begin{enumerate}
\item passive channel: made of soft iron bars that are magnetized by the cyclotron magnetic field;
\item active channel:  using coils or permanent magnets.
\end{enumerate}

\begin{figure}
\begin{center}
\includegraphics[width=12cm]{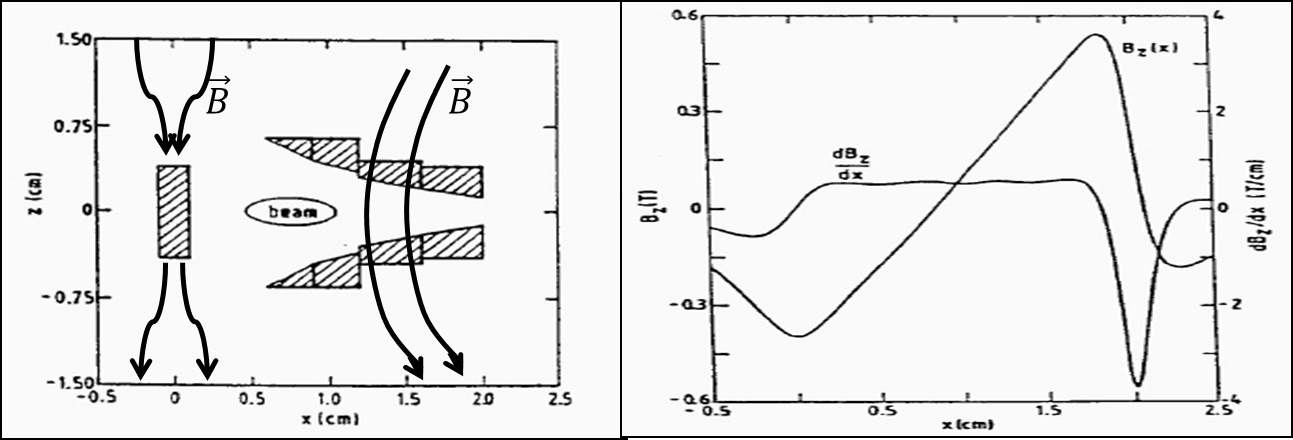}
\caption{Left: vertical cross-section of the passive gradient corrector channel used in the small ILEC cyclotron at Technical University Eindhoven. Right: calculated magnetic field and field gradient. The field is increasing with
radius, increasing the radial focusing. Figure taken from \Bref{Botman1996}}
\label{fig:channel1}
\end{center}
\end{figure}

The design always includes an effort to reduce the adverse effect of the channel on the internal orbits. \Figure[b]~\ref{fig:channel1}
shows a vertical cross-section of the passive focusing channel that was used in the small ILEC cyclotron at Eindhoven
University\cite{Kleeven1988}. Here, the poles are shaped to provide a smooth and constant radial gradient at the location of
the beam. Near the inner bar, the field decreases because the field lines are sucked into the iron. In between the two outer bars,
the field increases because there, the effective pole gap decreases.

\subsubsection{Extraction in the IBA C235 cyclotron}

\begin{figure}
\begin{center}
\includegraphics[width=7cm]{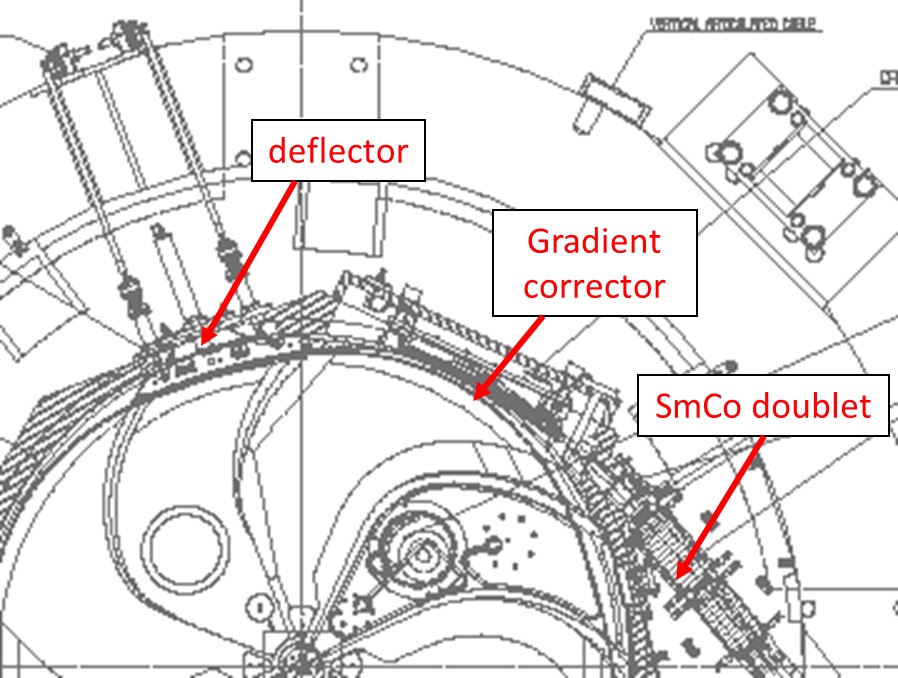}
\caption{Extraction scheme used for the IBA C235 proton therapy cyclotron}
\label{fig:extraction1}
\end{center}
\end{figure}

\Figure[b]~\ref{fig:extraction1} shows the extraction scheme that is used in the IBA C235 proton
therapy cyclotron. In this cyclotron,
turn separation is created by acceleration only. There are no harmonic extraction coils.
The only extraction elements are the deflector, the gradient corrector, and the permanent magnet quadrupole doublet
that is placed in the return yoke. This cyclotron is a special case because the beam can be accelerated very close
to the radial pole edge of the machine. This is achieved by using a pole gap with an elliptical shape.
Furthermore, the RF cavities are designed such that there is a strong increase in the dee voltage at large radii
so that a larger turn separation is obtained at extraction.

\begin{figure}
\begin{center}
\includegraphics[width=12cm]{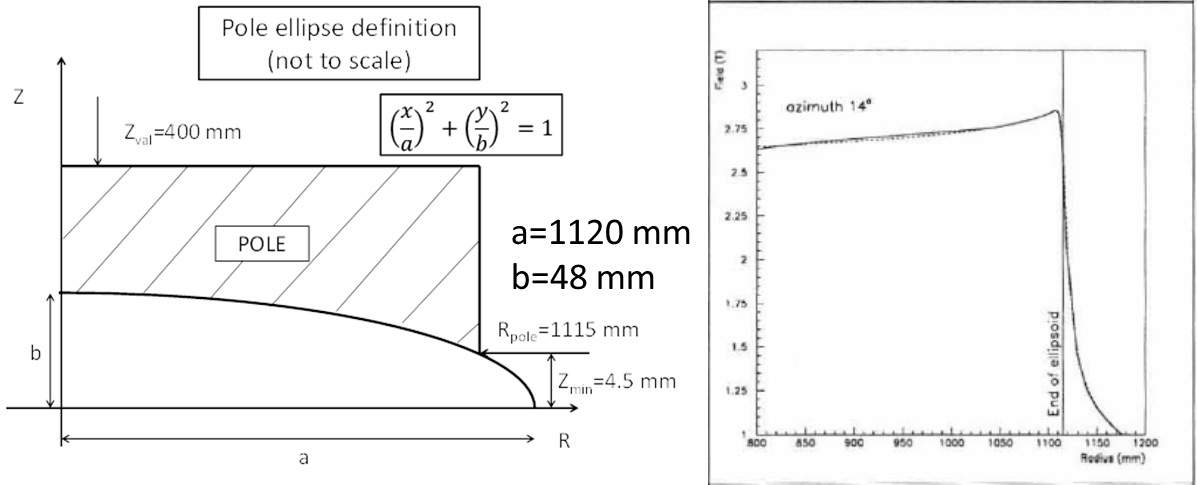}
\caption{Left: definition of the elliptical pole gap used in the IBA C235 cyclotron. Right: the magnetic field
shows a very sharp cut-off near the radius of the pole.}
\label{fig:ellipse}
\end{center}
\end{figure}

The parameters of the elliptical pole gap are illustrated in the left panel of \Fref{fig:ellipse}.
It can be seen in the right part of this figure that a good field region is obtained even very close
to the radius of the pole, enabling particle acceleration (with isochronous field) very close to the pole radius. Beyond this stable radius, the field
decreases very sharply. Because of this feature, only a small kick (provided by the deflector) is needed to extract the beam. The orbit
is extracted in about one-quarter of a turn.

\begin{figure}
\begin{center}
\includegraphics[width=12cm]{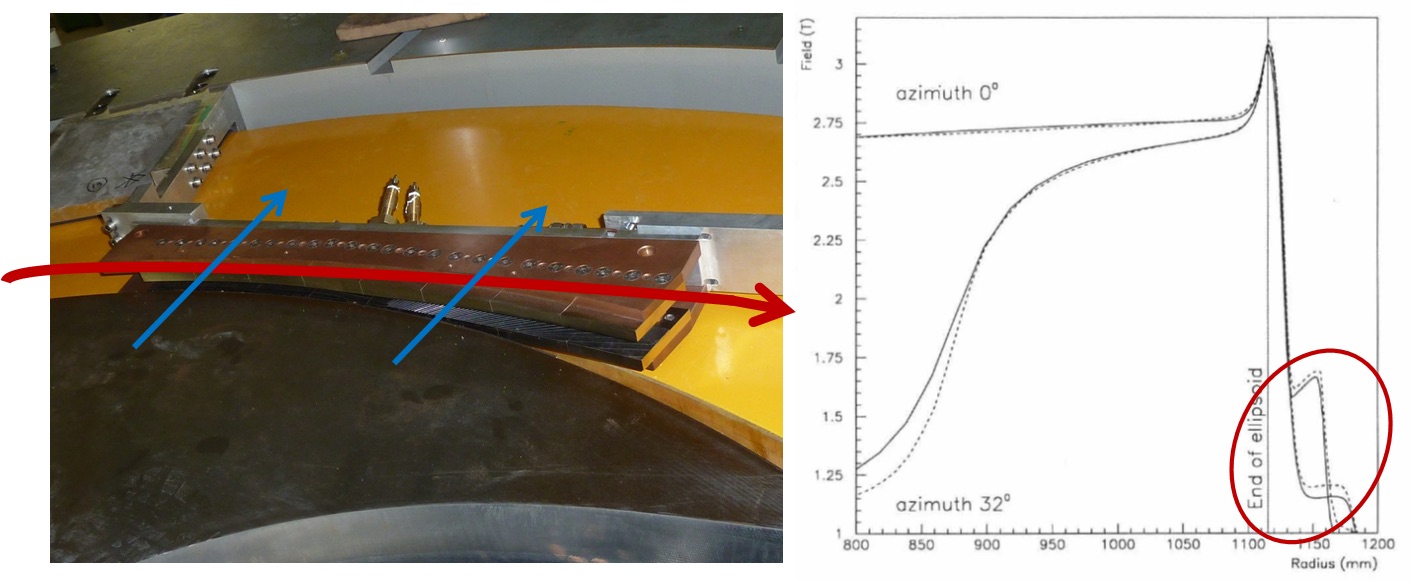}
\caption{Left: gradient corrector installed in the C235 cyclotron. Right: magnetic field at two different
azimuths, showing the field profile experienced by the extracted beam.}
\label{fig:gradcorr}
\end{center}
\end{figure}

The beam leaves the cyclotron by crossing the radial pole edge.
Here there is a very strong nega\-tive radial magnetic field gradient that would completely defocus the beam horizontally.
The gradient corrector creates a kind of plateau with descending magnetic field value.
\Figure[b]~\ref{fig:gradcorr} shows the installation of this gradient corrector in the C235 cyclotron.
It is a passive
system made of soft iron that is magnetized by the cyclotron magnetic field and placed between the two main coils, very close
to one of the hills of the cyclotron magnet. The gap in the gradient corrector is profiled to obtain a steady magnetic field
drop along the direction of the beam but, at the same time, a radial gradient which is well under control.
This is shown in the right panel of \Fref{fig:gradcorr}, where two field profiles are shown in the median plane and perpendicular
to the particle orbit, but at two different azimuthal positions along the trajectory (indicated by the blue arrows).

\begin{figure}
\begin{center}
\includegraphics[width=12cm]{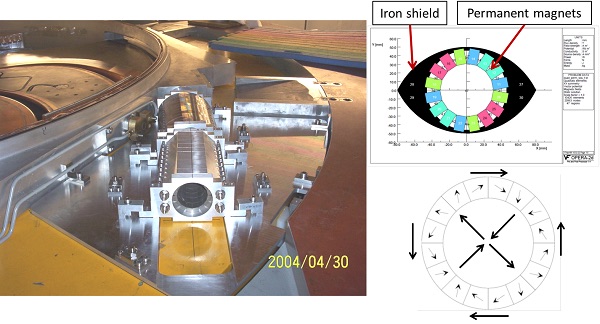}
\caption{Left: permanent magnet quadrupole doublet, installed in the IBA C235 cyclotron. Right: layout and polarity of the
permanent magnet, producing the high-quality quadrupole field.}
\label{fig:doublet}
\end{center}
\end{figure}

\Figure[b]~\ref{fig:doublet} shows the installation of the $\ESm\ECo$ permanent magnet doublet in the C235.
These magnets are placed immediately
at the exit of the vacuum chamber but still in the beam exit penetration of the return yoke. The layout of the permanent magnets
is shown in the upper-right panel of \Fref{fig:doublet}. For the first quadrupole, an iron housing is placed around the permanent
magnets to shield them from the external cyclotron magnetic field, which is still rather high at this position. The lower-right panel shows the polarity of the individual permanent magnets and the global circulation of the magnetic flux.

\subsection{Self-extraction}

In a cyclotron, the average magnetic field starts to decrease when approaching the maximum pole radius.
This limits the maximum energy that can  be achieved in the cyclotron. There are, in fact, two limits.

\begin{enumerate}
\item There is a limit of radial stability. This limit is reached on the equilibrium orbit for which the
radial betatron frequency $\nu_r$ has fallen to zero ($\nu_r\downarrow  0$). Note that in a rotational symmetric magnetic
field this corresponds to the situation where the field index equals $-1$,

\begin{equation}
n=\frac{r}{B}\frac{\mathrm{d}B}{\mathrm{d}r} = -1\ .
\end{equation}

This occurs at the radius where the magnetic rigidity $ p/q = B \cdot r$ reaches its maximum.

\item There is a limit of acceleration that occurs as a result of the loss of isochronism.
This limit is achieved when the RF phase has slipped $90^{\circ}$. Of course, it depends on the  dee voltage.
\end{enumerate}

\begin{figure}
\begin{center}
\includegraphics[width=12cm]{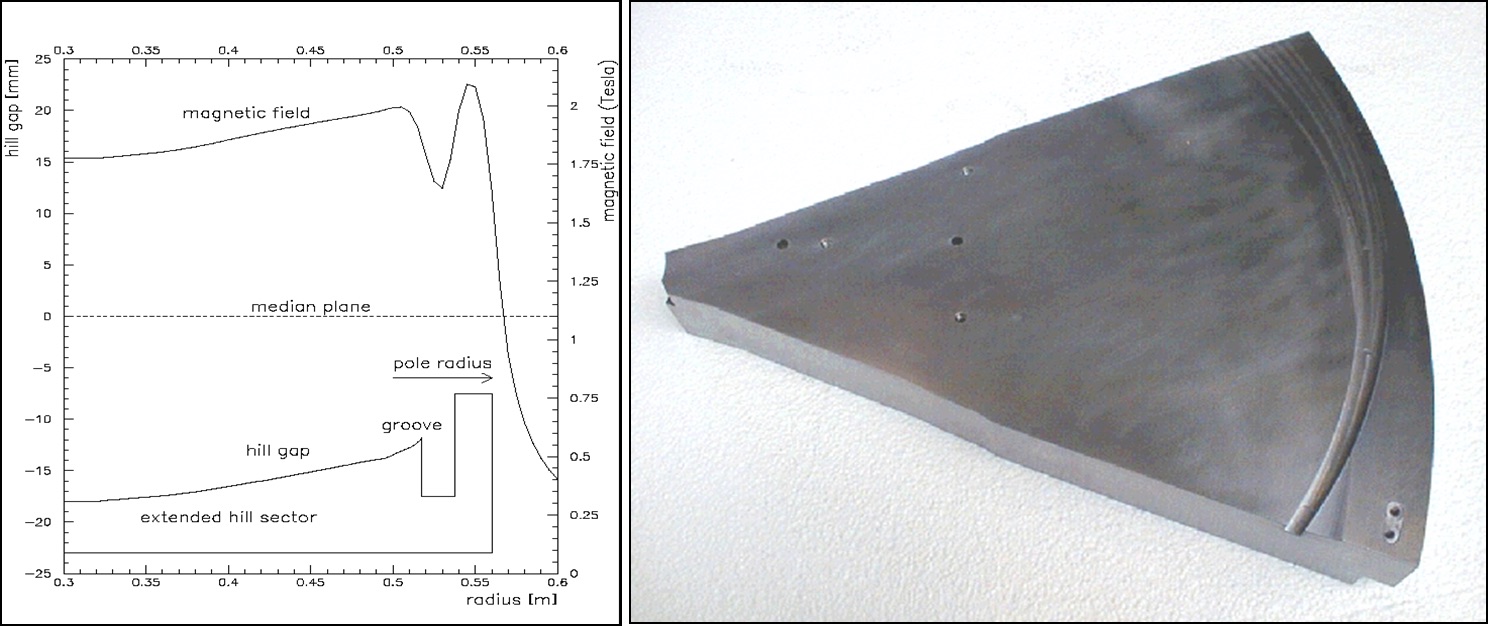}
\caption{ In the IBA self-extracting cyclotron, a groove is machined in one of the poles (right), to create an extraction channel. This channel provides a magnetic septum at the entrance and,
at the same time, gradient correction and focusing. An elliptical pole gap is used, allowing for sharp
radial gradients in the magnetic field near the pole edge.}
\label{fig:se2}
\end{center}
\end{figure}

If the vertical pole gap is much larger than the radial gain per turn (as is the case for many cyclo\-trons that
have been built so far), the second limit is achieved earlier than the first. However, when the pole
gap is small and, furthermore, when this gap is elliptically shaped (and the dee voltage is not too small),
the first limit is achieved first and the beam is self-extracted. However, in this case, the particles
will come out at all possible azimuths and there is not really a well-defined coherent extracted beam.
To achieve this, an elliptical pole gap is used, which makes enables the realization of very sharp
magnetic gradients close to the pole radius. At the same time, a groove or plateau is machined in one of the cyclotron
poles to provide an extraction channel. This channel simultaneously serves as magnetic septum and gradient corrector. This is illustrated in \Fref{fig:se2}.

In principle, the scheme for self-extraction is quite similar to the usual scheme for resonant pre\-cessional extraction:

\begin{enumerate}
\item harmonic coils create a coherent oscillation;
\item the beam is accelerated into the fringe field where $\nu_r\approx 0.6$;
\item the groove creates a kind of magnetic septum and at the same time provides for a gradient corrector channel;
\item a permanent magnet doublet is placed within the vacuum chamber; this doublet continues the ex\-traction
path and focuses the beam in both directions.
\end{enumerate}

\begin{figure}
\begin{center}
\includegraphics[width=12cm]{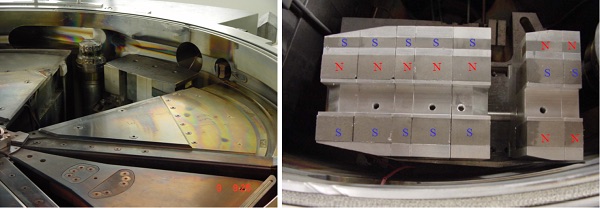}
\caption{Left: extraction elements in the IBA self-extracting cyclotron. The harmonic coils
are placed underneath the pole covers.
Right: lower part of the permanent magnet corrector used in the IBA self-extracting
cyclotron. The corrector is placed close to the vacuum chamber visible in the lower part of the photo.
The polarity of the magnets is indicated.
The polarities are inverted in the upper part of the corrector.}
\label{fig:se1}
\end{center}
\end{figure}

The left panel of \Fref{fig:se1} shows the interior of the IBA self-extracting
cyclotron\cite{Jongen1995,Kleeven2003}. The part of the beam
that is not well extracted is intercepted by a low-activation water-cooled beam dump. With this machine,
a beam intensity of $2\UmA$ has been extracted with an efficiency of 80\%.
The horizontal beam emittance (2$\sigma$) is about $300\pi\Umm\Umrad$.

The right panel of \Fref{fig:se1} shows the gradient corrector that is used.
This is an active channel made of samarium-cobalt permanent magnets. It acts like a quadrupole doublet, but with the first doublet
being longer than the second. The quadrupole-like field shape is obtained by using two opposite dipole fields that
are radially displaced a few centimetres with respect to each other. Some additional small magnets are placed to minimize the adverse effect of the gradient corrector on the internal orbits.

\subsection{Two different extraction schemes in one cyclotron}

\subsubsection{The IBA C70XP cyclotron}

The IBA C70XP cyclotron was discussed in \Sref{iso_cyclo}.
In this cyclotron, two independent extraction systems are installed:

\begin{enumerate}
\item an ESD for the extraction at fixed energy of positively charged particles ($^{4}\EHe^{2+}$ and $\EH_2^+$);
\item a variable-energy stripper system for negatively charged particles ($\EH^-$ and $\ED^-$).
\end{enumerate}

\begin{figure}
\begin{center}
\includegraphics[width=10cm]{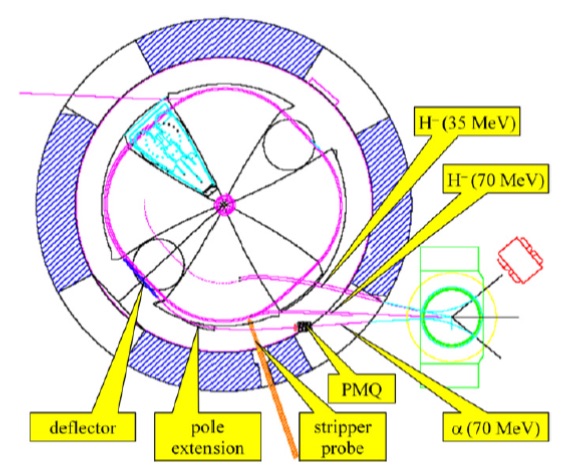}
\caption{Dual extraction system (stripping and electrostatic deflector) used in the IBA C70XP-cyclotron: PMQ, permanent magnet quadrupole.}
\label{fig:extraction2}
\end{center}
\end{figure}

The stripping extraction is implemented on two opposite poles, allowing simultaneous dual beam extraction.
On one side, the two independent extraction modes are linked by the strong
geometrical constraint that both beams converge to the centre of one common switching magnet,
as illustrated in \Fref{fig:extraction2}.
This is obtained by having both extractions taking place at the same pole.
The position of the stripped beam at the switching magnet is controlled by the position of the stripping probe
on the pole. The position of the beam extracted by the ESD is determined by the
optimized design and length of the radial pole extension and also depends on the deflector voltage.
The radial pole extension locally modifies the falling gradient in the fringe field and facilitates
good optical beam transport in this critical region.

\begin{figure}
\begin{center}
\includegraphics[width=10cm]{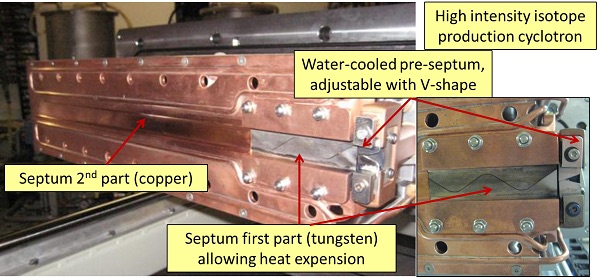}
\caption{High-power electrostatic deflector used in the IBA C70XP-cyclotron}
\label{fig:esd3}
\end{center}
\end{figure}

\Figure[b]~\ref{fig:esd3} shows the ESD.
A pre-septum made of tungsten is placed just in front of the actual septum
of the ESD. It is water-cooled and protects the septum from overheating. The septum itself consists of
two parts: the first part is made of two (upper and lower) tungsten foils that are
brazed to their copper supports. This allows for a good heat evacuation. The two tungsten foils can expand
independently, avoiding an increase in effective septum thickness due to thermal expansion.
The second part of the septum is made of copper and is machined from a solid copper block.
Extraction efficiencies higher than 80\% have been obtained with this ESD.

\subsubsection{The IBA C400 cyclotron}

\begin{figure}
\begin{center}
\includegraphics[width=12cm]{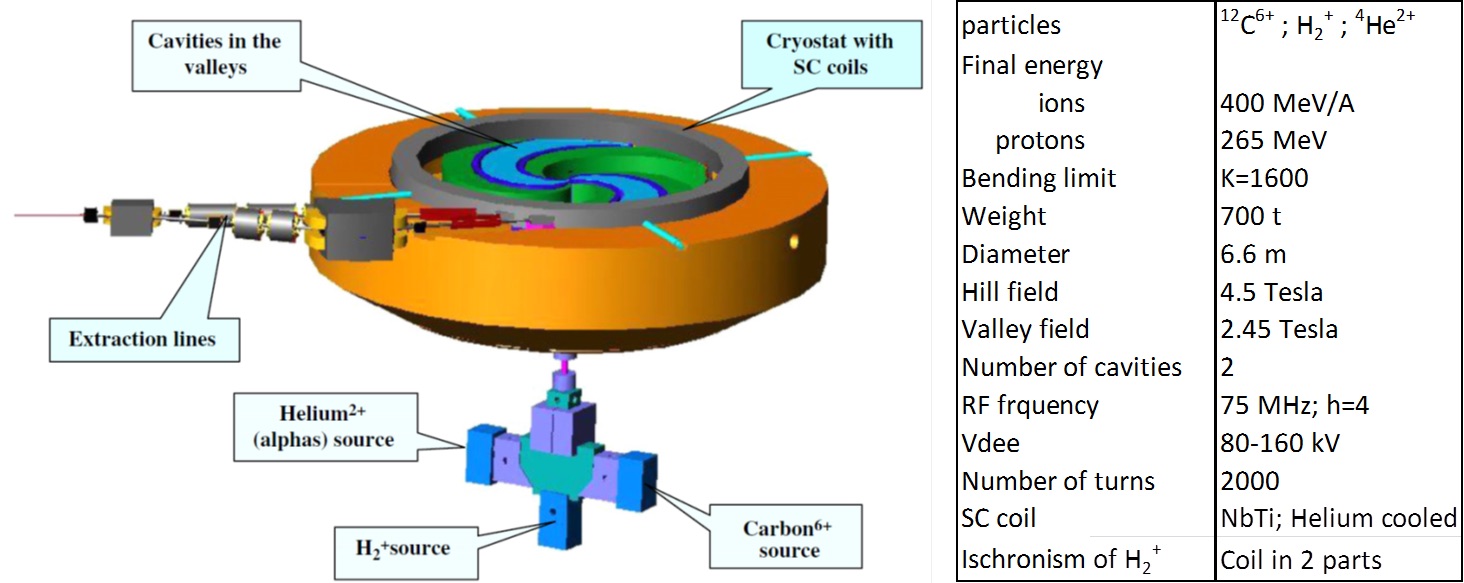}
\caption{C400 superconducting cyclotron and main parameters, proposed by IBA for carbon and proton therapy: SC, superconducting.}
\label{fig:c400_1}
\end{center}
\end{figure}

A design study of the compact superconducting isochronous cyclotron C400\cite{Jongen2010a,Jongen2010b}.
has been made by IBA in collaboration with the JINR at Dubna. When built, it will be the first cyclotron in the world capable
of delivering protons, carbon, and helium ions for cancer treatment. The $^{12}\EC^{6+}$ and the $^{4}\EHe^{2+}$ will be accelerated
to $400\UMeV/\UAZ$ and extracted by an electrostatic deflector. $\EH_2^+$ ions will be accelerated to an energy of $265\UMeV/\UAZ$ and extracted
by stripping. The left panel of \Fref{fig:c400_1} shows an artist's impression of the cyclotron. The table on the right lists the
main parameters. The magnet yoke has a diameter of $6.6\Um$ and the total magnet weight is about 700 tonnes. The maximum magnetic
field in the hills is $4.5\UT$.  Three ion sources are mounted on a switching magnet placed in the injection line below the cyclotron.

\begin{figure}
\begin{center}
\includegraphics[width=12cm]{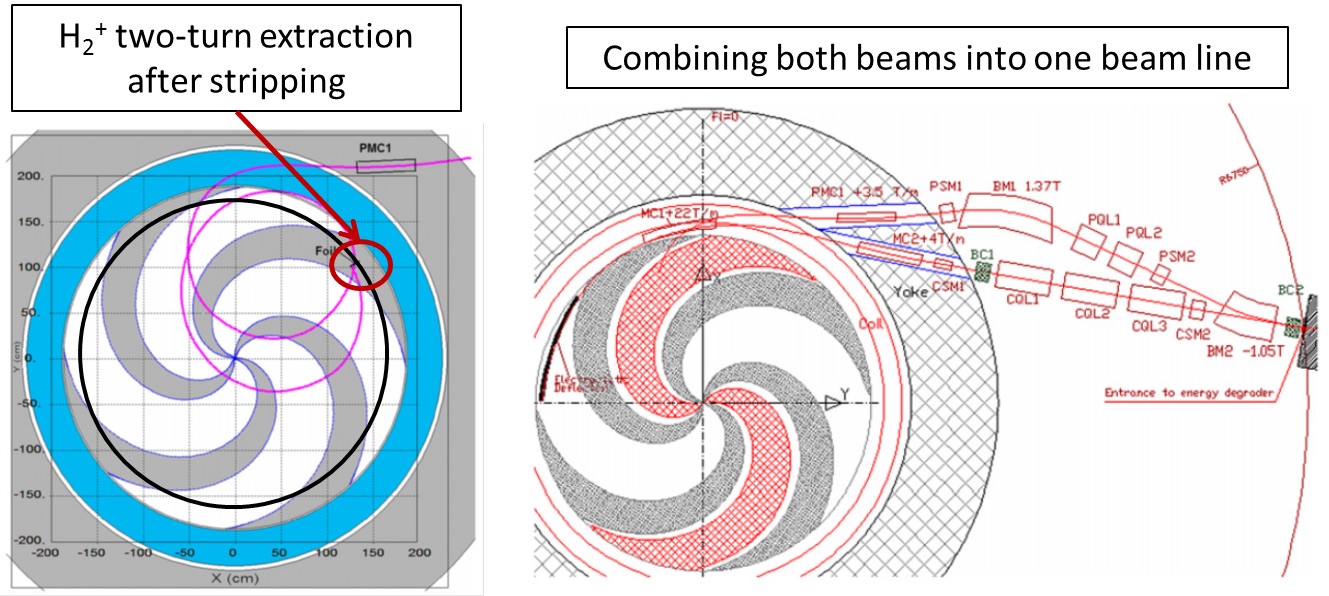}
\caption{Left: in the IBA C400 cyclotron, protons are extracted by stripping the accelerated $\EH_2^+$ ion by a thin stripping foil.
Right: $^{12}\EC^{6+}$ is extracted with an electrostatic deflector. Both particles, protons and carbon, are combined into one beam line.}
\label{fig:c400_2}
\end{center}
\end{figure}

Extraction of protons is done with a stripping foil.
The minimum proton energy that can be obtained is $320\UMeV$, for single-turn
extraction, and $265\UMeV$, for two-turn extraction (see \Fref{fig:c400_2}).
The second solution was chosen in order to be closer to the usually applied energy for the proton beam treatment.

Electrostatic deflection extraction is used for the $\EC$ and $\EHe$ beams. A single ESD
located in a valley is used, with an electric field of ${\approx}15\UMV/\UmZ$. From the right panel of \Fref{fig:c400_2}, it can be
seen that the beams from both extraction systems do not exit in the same direction and position:
they are combined by two external beam lines onto the common degrader. Downstream, both beams travel in the same beam line.

\subsection{Regenerative extraction in the IBA S2C2}

\begin{figure}
\begin{center}
\includegraphics[width=12cm]{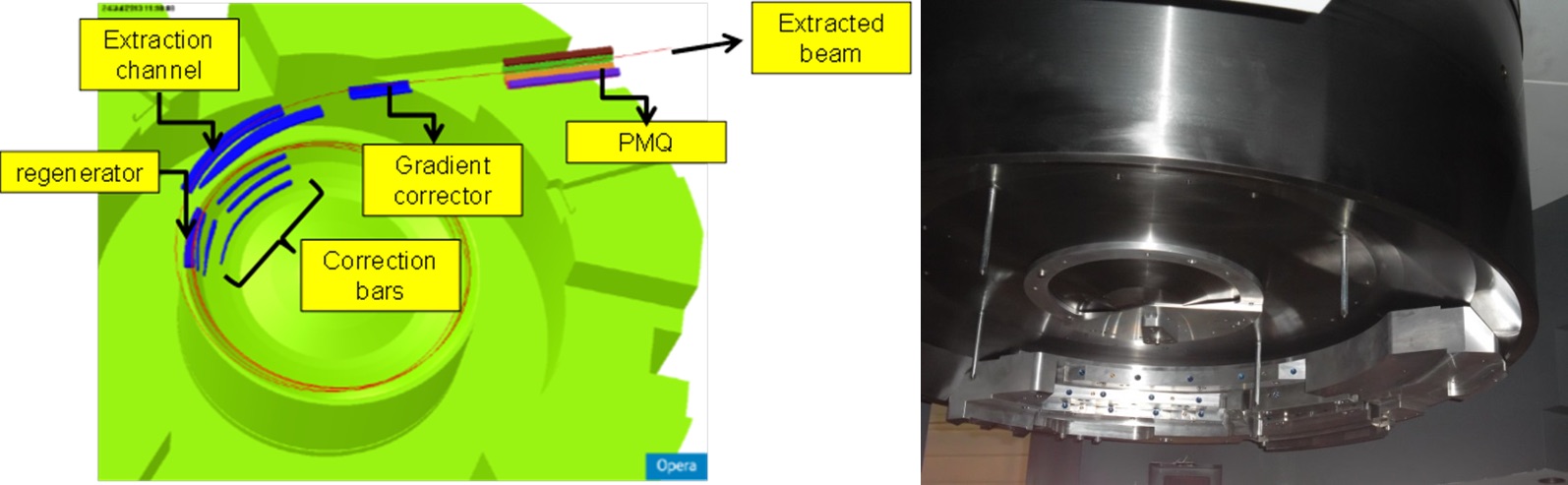}
\caption{Passive extraction system of the IBA superconducting synchro-cyclotron S2C2: PMQ, permanent magnet quadrupole.}
\label{fig:extraction3}
\end{center}
\end{figure}

\begin{figure}
\begin{center}
\includegraphics[width=12cm]{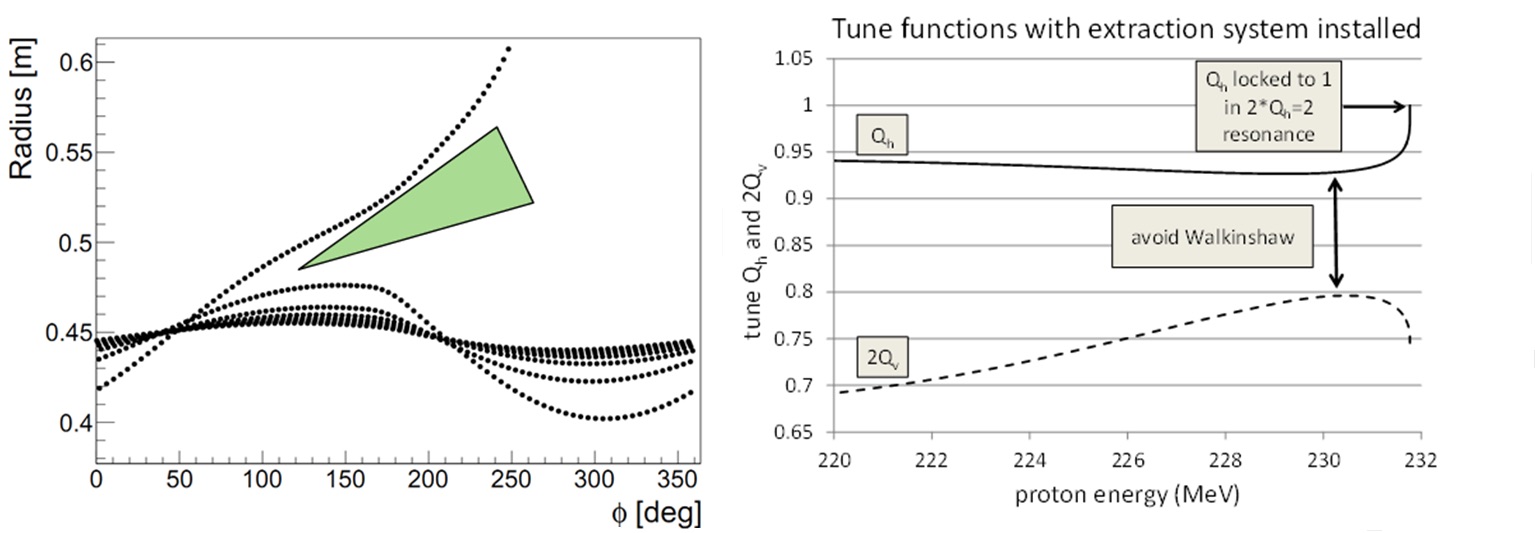}
\caption{Regenerative extraction: a strong quadrupole bump increases $\nu_r$ and locks it to one. A steady shift of the beam towards
the extraction channel is built up. The Walkinshaw resonance ($\nu_r=2\nu_z$) must be avoided.}
\label{fig:extraction4}
\end{center}
\end{figure}

In the IBA superconducting synchro-cyclotron (S2C2), extraction with an electrostatic deflector cannot be used,
because the turn separation at extraction is far too small.
Instead, the regenerative method of extraction based on the $2\nu_r=2$ resonance is used.
The extraction system is fully passive: only soft iron elements are used. The layout is shown in \Fref{fig:extraction3}.
The regenerator creates a strong magnetic field bump, of which the quadrupole component increases the radial
focusing and locks the horizontal betatron frequency $\nu_r$ to one. The orbit becomes unstable and is pushed towards
the extraction channel by the first-harmonic component of the magnetic field, also produced in the bump. It is essential
to avoid the Walkinshaw resonance in this process, as illustrated in \Fref{fig:extraction4}. When the extraction sets in,
the displacement of the beam towards the extraction channel steadily and exponentially builds up. The magnetic septum of the
extraction channel separates the extracted orbit from the last internal orbit. The locking of the  betatron frequency $\nu_r$ to one
can be observed in the left panel of \Fref{fig:extraction4} by the fact that the subsequent azimuth of the nodes of the
oscillation is fixed for multiple turns.

A series of iron correction
bars is needed to compensate the magnetic field undershoots towards inner radii that are produced by the regenerator and the extraction
channel. Farther downstream, between the main coils, a three-bar gradient corrector is placed, which reduces the radial defocusing
in the fringe field of the pole. Finally a permanent magnet quadrupole is used to further match the extracted beam
to the external beam line.

\section{Some aspects of cyclotron magnetic design}

A good overview of considerations and methods applied for the design of compact cyclotron magnets has already been made
in previous topical CAS courses by Jongen and Zaremba\cite{Jongen1994,Zaremba2005}. Therefore, after some
general considerations, we limit ourselves in this section to some recent examples and some new methods that have been developed.

\subsection{Some general considerations}

In industrial practice, many design choices of the cyclotron magnet are often already fixed before the start of any design
calculations. Such choices are determined by more general considerations, such as:

\begin{enumerate}
\item the application of the cyclotron, which will determine the particles to be accelerated and their
required maximum kinetic energy;
\item from the maximum particle rigidity, the choice of maximum $B$-field will determine the pole radius, or vice versa;
\item the choice of the cyclotron type: isochronous or synchro-cyclotron;
\item the coil technology: superconducting or normally conducting;
\item the type of extraction to be applied: stripping extraction, ESD, or regenerative extraction;
\item when choosing, for example, an isochronous solution, the following parameters are often already determined beforehand:
        \begin{enumerate}
        \item number of sectors, pole gap, pole angle, valley depth, pole spiral, \etc;
        \item number of dees, dee voltage, harmonic mode, and RF frequency;
       \item for injection, the use of an internal or external ion source.
        \end{enumerate}
\end{enumerate}

Other choices can be made with some rough back-of-the-envelope calculations: here are a few examples.
\begin{enumerate}
\item The maximum magnet rigidity is obtained from the particle maximum kinetic energy as follows:

        \begin{equation}
        B\rho=\frac{\sqrt{T^2+2TE_0}}{300 Z}\ ,
        \end{equation}

        where $Z$ is the charge ionization number of the particle. The rest-energy, $E_0$, and the kinetic energy, $T$,
        are expressed in MeV and the rigidity $B\rho$ in Tm.

\item For a magnet that is far from magnetic saturation, the relation between the magnetic field in the pole gap and
        the number of ampere turns in the coil can be approximated as:

        \begin{equation}
        \oint\vec{H}\cdot \mathrm{d}\vec{l} = (NI)_\mathrm{tot} \approx \frac{1}{\mu_0}\left(gB_\mathrm{gap}\right)+\frac{1}{\mu_0\mu_\mathrm{iron}}\left(L_\mathrm{iron}B_\mathrm{iron}\right)\ .
        \end{equation}

        Here, the integration is made over a closed loop that crosses the gap and closes via the return yoke around
        the coils of the magnet. Further $(NI)_\mathrm{tot}$ is the total number of ampere turns in the coils,
        $g$ is the pole gap, $L_\mathrm{iron}$ is the length of the loop in the iron, $B_\mathrm{gap}$
        and $B_\mathrm{iron}$ are the magnetic fields in the gap and iron, respectively, $\mu_0$ is the magnetic permeability in vacuum and
        $\mu_\mathrm{iron}$ is the relative magnetic permeability in iron. In practice, for a conventional magnet far from saturation,
        the contribution of the iron is smaller than a few percent.

\item Assuming a hard-edge approximation for the fields produced from the iron pole sectors, the aver\-age field $\bar{B}$
        can be estimated as:

        \begin{equation}
        \bar{B} \approx \alpha B_\mathrm{h} + (1-\alpha)B_\mathrm{v}\ .
        \end{equation}

        Here, $\alpha$ is the fraction of the hill angle on one symmetry period
and $B_\mathrm{h}$ and $B_\mathrm{v}$ are the magnetic field values in the hill and valley, respectively.

\item To estimate the width of the return yoke, the total magnetic flux, $\Phi$, produced between the poles must be roughly
        equal to the flux guided in the yoke:

        \begin{equation}
        \Phi \approx 2\pi R_\mathrm{p}^2 \bar{B} \approx B_{\mathrm{ret}}A_{\mathrm{ret}}\ .
        \end{equation}

        Here $R_\mathrm{p}$ is the (effective) pole radius, $B_{\mathrm{ret}}$ is the magnetic field in the return
        yoke, and $A_{\mathrm{ret}}$ is the total surface cross-section of the return yoke.

\item The required increase in average magnetic field, necessary for isochronous operation, may be estimated from the relativistic
        mass increase of the particles as follows:

        \begin{equation}
        \bar{B}(r) \approx B_0\gamma_{\mathrm{rel}}\ ,
        \end{equation}

        where $B_0$ is the magnetic field in the centre and $\gamma_{\mathrm{rel}}$ is the relativistic gamma.

\item The magnetic flutter may be obtained from the hard-edge approximation, as given in \Eref{flutje}.

\item The betatron frequencies may be related to the flutter and the pole spiral angle, as given in Eqs.~(\ref{nuz}) and (\ref{nur}).

\end{enumerate}

\subsection{Tools for magnetic modelling in Opera}

Almost all magnet designs at IBA are done using Opera simulation software from Cobham (VectorFields)\cite{Cobham2015}.
The main packages used are as follows.
\vspace{0.5cm}

\subsubsection{Opera-2d} This is the 2D version of the software, which can be used to model magnets with rotational symmetry. It is a
perfect choice for an initial design of a synchro-cyclotron magnet (without the extraction system). However, it can also be useful
for an initial design of a 3D isochronous cyclotron magnet. In this case, stacking factors can be used to take into account
the hill--valley structure of such a magnet. This yields a preliminary idea of the average magnetic field (as
a function of radius),
the stray field around the cyclotron, and the dimensions of the return yoke.
\subsubsection{Opera-3d---modeller interface} This module allows for full 3D simulations. It is easy to use and easy to include fine geometrical
details. The 3D finite-element mesh is generated automatically. The most commonly used is the tetrahedral mesh. This mesh is
not always very regular, which may result in some noise in the calculated magnetic field.
\subsubsection{Opera-3d---preprocessor interface} This module also allows for full 3D simulations, but it is more difficult to use
and even more
difficult to include fine geometrical details. However, the 3D finite-element mesh is fully created by the user and therefore
fully under control. A hexahedral mesh is used and less noisy magnetic fields can be obtained. This module may be useful,
for example, for the precise calculation of magnetic forces.

\subsection{Design approach using Opera-3d modelling}

The cyclotron magnet design approach used at IBA can be characterized with the following features.

\begin{enumerate}
\item The 3D models are fully parameterized.
\item Models are automatically generated using macros. Together with the previous feature, this allows some parameters to be changed easily, so that the corresponding new model is quickly obtained.
\item The macros for all different types of IBA cyclotron (such as S2C2, C230, C19/9-family, C30-family, C70-family) have
similar structures. This enables the quick creation of new macros for new machines if needed.
\item The magnetic properties of the iron (the $BH$-curves) are verified with an in-house permeability meter.
\end{enumerate}

The following types of property are parameterized:

\begin{enumerate}
\item all important dimensional parameters and pole profiles;\item main coil settings;
\item material properties, such the different $BH$-curves for different iron parts;
\item the sizes and other properties of the finite-element mesh;\item \ solver tolerances;\item switches for the selection or de-selection of individual subsystems;\item switches for filling individual subsystems with air or  iron.\end{enumerate}

\begin{figure}
\begin{center}
\includegraphics[width=13cm]{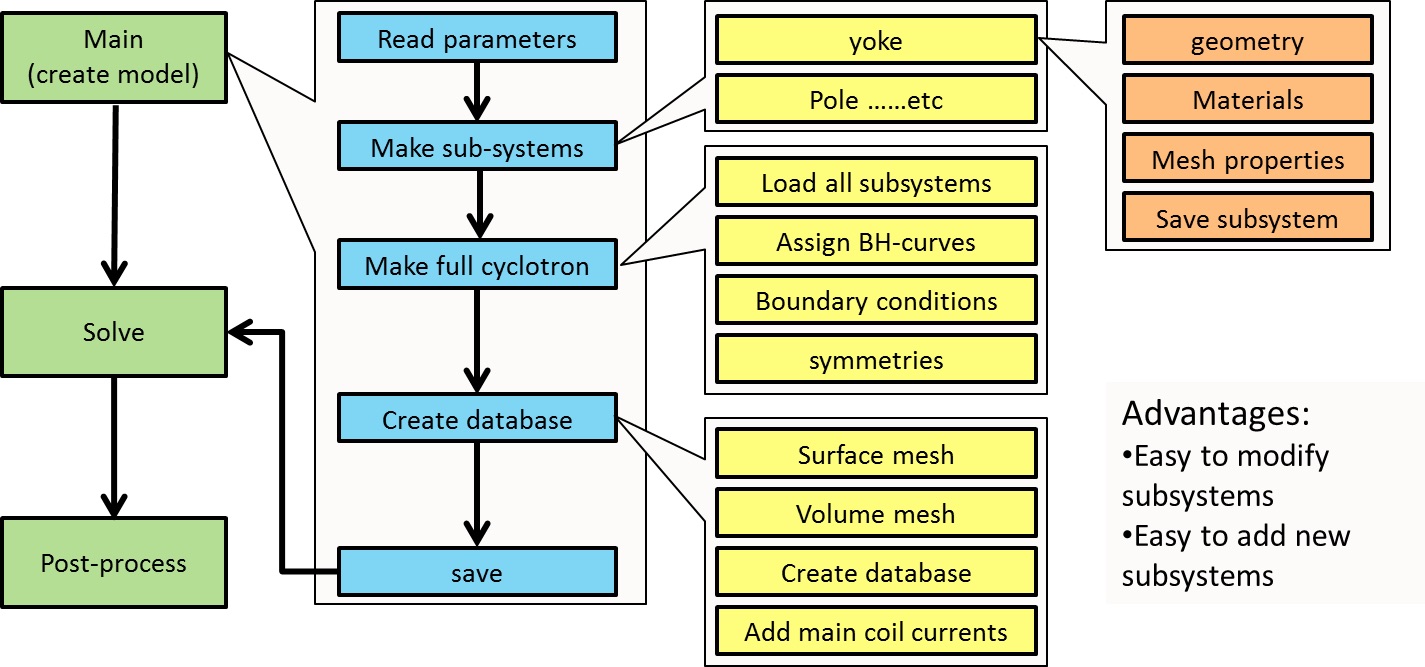}
\caption{Logical structure of a typical input file used for the fully parameterized and automated
creation of an Opera-3d model of a cyclotron.}
\label{fig:opera4}
\end{center}
\end{figure}

The main structure of a cyclotron modelling macro is illustrated in \Fref{fig:opera4}. The first column (blocks in green)
shows the main steps, namely: (i) creation of the model in the Opera-3d modeller, (ii) solving the model with the
Opera-3d solver, and (iii) analysing the result with the Opera-3d post\-processor. The second column (blocks in blue) illustrates
how a model is created in the modeller: (i) read from a file all the parameters that are needed to create the model, (ii) create
individual subsystems, (iii) assemble these subsystem into one model that represents the full cyclotron, (iv) create and save the database
that serves at input for the solver. The third column (yellow blocks) shows some details of some of the blue blocks. Subsystems
can be, for example, the yoke (upper, return), pole, or main coils, but also, for example, a passive extraction system,
a three-bar gradient corrector, or a permanent magnet quadrupole. Creation of the full cyclotron model includes steps like (i) loading
all subsystems, (ii) defining material properties, and (iii) defining boundary conditions or symmetries. The creation of the database
requires surface and volume meshing but also enables the inclusion of different cases to be solved, such as, a
list of different main coil currents. The last column (in orange) shows some details for the creation of one subsystem, such as (i)
creation of the geometry, attaching material labels and mesh-properties to different cells in the subsystem, and saving.

\subsection{Some examples}

\begin{figure}
\begin{center}
\includegraphics[width=6cm]{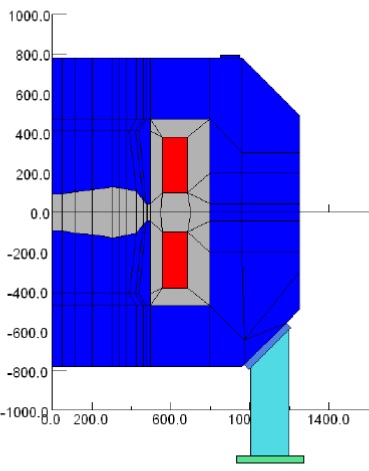}
\caption{The Opera-2d model of the S2C2}
\label{fig:opera1}
\end{center}
\end{figure}

\begin{figure}
\begin{center}
\includegraphics[width=12cm]{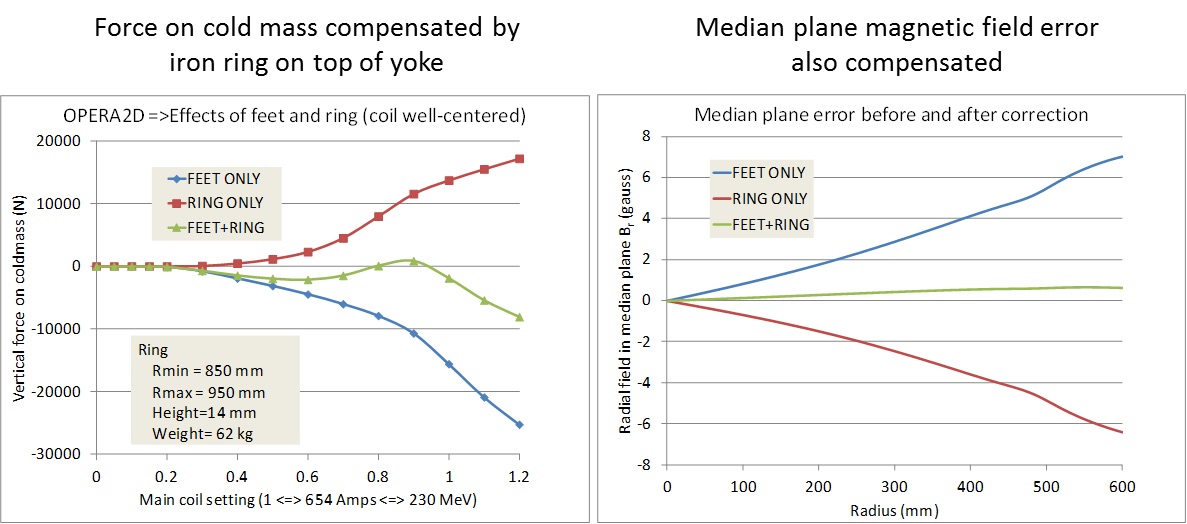}
\caption{Opera-2d simulation of the median plane error in the S2C2, as produced by the feet underneath the cyclotron,
and compensation by an iron ring placed on top of the yoke.}
\label{fig:opera2}
\end{center}
\end{figure}

In this section, we consider as an example the recent development of the IBA superconducting synchro-cyclotron S2C2.
\Figure[b]~\ref{fig:opera1} shows a model of the S2C2 made in Opera-2d. The vertical axis is the axis of rotational symmetry.
The main coils are shown in red and the magnet iron in deep blue.
The iron of the S2C2 magnet is heavily saturated; therefore, additional iron placed around the yoke might introduce a
median plane error or increase the vertical forces on the cold mass.
The goal of this study was to calculate the vertical asymmetry
introduced by the cyclotron support feet. These are shown in light blue. The feet do not obey rotational symmetry
(there are four of them), but a stacking factor was used to approximate their average effect.
To compensate for the effect of the feet, an iron ring is placed on top of the yoke. \Figure[b]~\ref{fig:opera2}
shows some results of these simulations. The left panel shows the total vertical force acting on the
cold mass during ramp-up of the main coil current. Three cases are shown: (i) for the blue curve, the feet were present in the
model but the ring was not; (ii) for the red curve, the ring was present but the feet were not; and (iii) for the green curve,
both the feet and ring were present. The dimensions of the ring were optimized to minimize the total vertical force
during ramp-up. It can be seen that the total asymmetry due to the feet (about $25000\UN$ at full current) is reduced to about
$8000\UN$ by placing the ring.
The right panel shows the magnetic median plane error (the radial field in the median plane) as a function of radius for the same
three cases at the maximum excitation current. It can be seen that for the same optimized dimensions of the ring, the maximum
field error is reduced from $7\UG$ to <$1\UG$. It is noted that in such a synchro-cyclotron, median plane
errors must be avoided as much as possible, because of the weak vertical focusing and also because of
the large number of turns in such a machine.
Based on these calculations, the described method has successfully been used in the actual machine.

\begin{figure}
\begin{center}
\includegraphics[width=12cm]{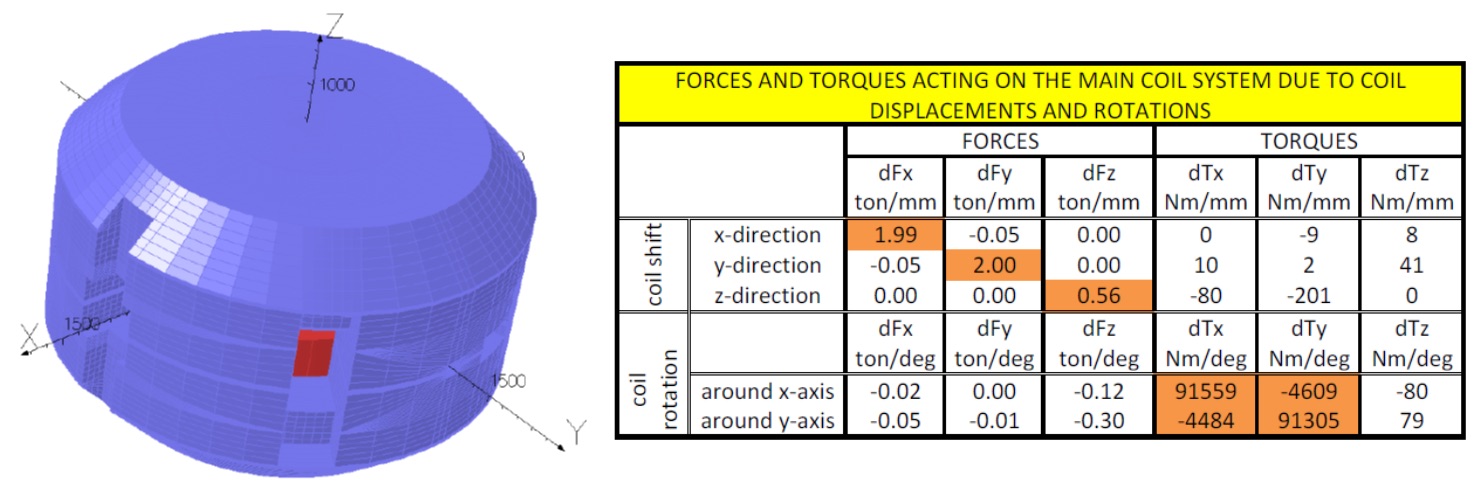}
\caption{Opera-3d preprocessor model of the S2C2, to estimate forces and torques due to displacement  acting
on the cold mass  with respect to the geometrical centre of the cyclotron.}
\label{fig:opera3}

\end{center}
\end{figure}

The left panel of \Fref{fig:opera3} shows a model of the S2C2 that was created using the Opera-3d pre\-processor. The typical
hexahedral mesh is clearly visible in this model. The goal of this model was to calculate precisely the forces and torques
on the cold mass as a function of its position and rotation. An under\-standing of these forces is crucial for the design of
the tie rods by which the cold mass is suspended in the cryostat. To have little heat flow from the cold mass to the exterior,
the diameter of these tie rods should be as small as possible. Conversely, the diameter should be large enough to withstand
the magnetic forces acting on the cold mass. The differential forces on the cold mass due to translations or rotations
can be calculated with better precision in the preprocessor. Results of these calculations are summarized in the table on the right
of \Fref{fig:opera3}. It was found that all forces and torques vary linearly with displacement or rotation. Furthermore, it
can be seen that all coil movements are unstable: a given displace\-ment will create a force in the
same direction as the displacement. Horizontal displacements result in a force of about 2 tonnes per~mm. For vertical
displacements this is about 0.5 tonnes per~mm.

The optimization of the magnetic circuit of the S2C2 has been a long process, carried out by three main contributors:
IBA, AIMA Developpement, and ASG (the company that made the coil and the cryostat). Many aspects were considered, such as:

\begin{enumerate}
\item optimization of the pole-gap profile;
\item definition of the pole radius and the $230\UMeV$ extraction radius;
\item optimization of coil current density and dimensions to assure a margin with respect to the critical surface of
the used superconducting material and to allow operation up to $250\UMeV$;
\item dimensions of the yoke to balance outside stray fields reasonably;
\item dimensions and placement of all horizontal and vertical yoke penetrations (ports are needed for RF system, ion source, vacuum, beam exit, cryo-coolers, 3 horizontal and $2\times 3$ vertical tie rods);\item optimization of the extraction system;
\item shielding required for external systems, such as the rotating capacitor and the cryogenic coolers;
\item the influence of the external iron systems on the accelerated beam;
\item the influence of the fringe field on the external beam line;
\item median plane errors introduced by the vertical asymmetry in the magnetic design and com\-pen\-sation of these errors;
\item magnetic forces acting on the return yoke, the coils, the extraction system elements, external com\-ponents, \etc;
\item compensation of first-harmonic field errors;
\item the influence of the cyclotron feet and the yoke lifting system.
\end{enumerate}

For the design of the superconducting coil, the transient behaviour of the magnet with eddy cur\-rents and AC losses needs to be studied in detail, along with  the quenching behaviour.

\begin{figure}
\begin{center}
\includegraphics[width=9cm]{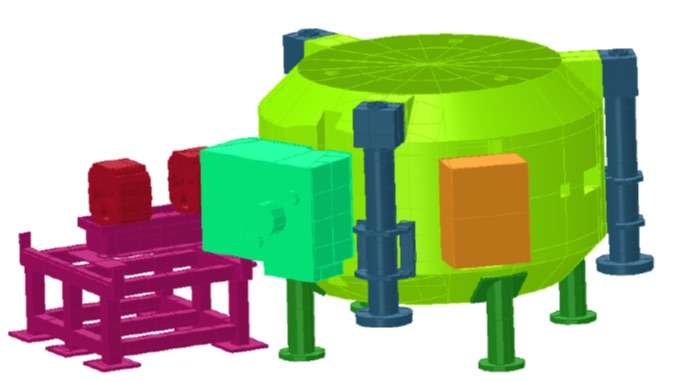}
\caption{A complete and detailed Opera-3D model was made of the IBA S2C2}
\label{fig:opera5}
\end{center}
\end{figure}

To make this optimization, magnetic finite-element models were produced, using Opera-2d, Opera-3d, and CST.
In Opera-3d especially, very detailed models were made, as illustrated in \Fref{fig:opera5}. Besides the yoke, poles and main coils,
more detailed elements and features were included, such as the yoke penetrations, the extraction system (regenerator and extraction
channel with their first-harmonic corrector bars, gradient corrector, and permanent magnet quadrupole), and external systems, such as the cyclotron feet,
yoke lifting system, cryo-cooler shields, rotating capacitor shield, and external quadrupoles with support structure.

\section{Conclusions}

The three main subjects discussed in this report (magnetic field design, beam injection, and beam ex\-traction)
are the most difficult problems in cyclotron design.
For industrial applications and isotope production, there is generally a need for ever-increasing beam intensity. This
implies, at the same time, a minimization of beam losses during both injection and extraction and also an
understanding of space charge effects in these processes. For particle therapy applications,the importance lies much more
in understanding beam quality, reproducibility, and stability. Except perhaps for Baartman's paper on space charge effects in cyclotrons~\cite{Baartman2013},
nothing really new has appeared in theoretical and analytical methods during the last 5--10 years that was helpful in this aspect.
However, increasing importance is given to the computational tools that are needed to optimize the design.
This concerns 3D finite-element software packages that are used to model the magnetic field as well as electrical field, but
also more specialized orbit tracking codes that can be produced in sufficient detail. The latter are usually
developed in house.

\section*{Acknowledgements}

We would like to thank our colleagues Jarno Van de Walle, Vincent Nuttens, Emma Pearson, and Eric Forton for their help
in preparing this paper.

\end{document}